\documentclass[12pt,preprint,revtex4]{emulateapj}
\usepackage{graphicx}
\usepackage{longtable}
\usepackage{pdflscape}
\usepackage{amsmath}
\usepackage{xcolor}
\usepackage{enumitem}
\usepackage{url} 

\shorttitle{Stellar Abundance Measurement Techniques}
\shortauthors{Hinkel et al.}
\slugcomment{Submitted for publication in the Astrophysical Journal}

\def\teff{{T$_{\text{eff}}$\,\,}}
\def\lg{{$\log$(g)\,}}

\def\aaa{{\,$\text{\AA}\,$}}
\newcommand{\kms}{km~s$^{-1}$}

\def\gtaprx{ \mathrel{ \vcenter{
      \offinterlineskip \hbox{$>$}
      \kern 0.3ex \hbox{$\sim$}    } } }

\def\ltaprx{ \mathrel{ \vcenter{
      \offinterlineskip \hbox{$<$}
      \kern 0.3ex \hbox{$\sim$}    } } }
      
\def\aj{{AJ}}
\def\apj{{ApJ}}
\def\apjs{{ApJS}}
\def\apjl{{ApJL}}
\def\araa{{ARA\&A}}
\def\aap{{A\&A}}
\def\pasp{{PASP}}

\def\mnras{{MNRAS}}

\begin{document}

\title{A Comparison of Stellar Elemental Abundance Techniques and Measurements}

\author{
  Natalie R. Hinkel\altaffilmark{1},
  Patrick A. Young\altaffilmark{1},
  Michael D. Pagano\altaffilmark{1},
  Steven J. Desch \altaffilmark{1},
  Ariel D. Anbar\altaffilmark{1},
  Vardan Adibekyan\altaffilmark{2},
  Sergi Blanco-Cuaresma\altaffilmark{3},
  Joleen K. Carlberg\altaffilmark{4, 5},
  Elisa~Delgado Mena\altaffilmark{2},
  Fan Liu\altaffilmark{6},
  Thomas Nordlander\altaffilmark{7},
  Sergio~G.~Sousa\altaffilmark{2},
  Andreas Korn\altaffilmark{7},
  Pieter Gruyters\altaffilmark{7, 8},
  Ulrike Heiter\altaffilmark{7},
  Paula Jofr\'e\altaffilmark{9},
  Nuno C. Santos\altaffilmark{2, 10},
  Caroline Soubiran\altaffilmark{11},
}
\email{natalie.hinkel@gmail.com}
\altaffiltext{1}{School of Earth \& Space Exploration, Arizona State University,
  Tempe, AZ 85287, USA}
\altaffiltext{2}{Instituto de Astrof\'isica e Ci\^encias do Espa\c{c}o, Universidade do Porto, CAUP, Rua das Estrelas, 4150-762 Porto, Portugal}
\altaffiltext{3}{Observatoire de Gen\`eve, Universit\'e de Gen\`eve, CH-1290 Versoix, Switzerland}
\altaffiltext{4}{NASA Goddard Space Flight Center, Code 667, Greenbelt MD 20771, USA}
\altaffiltext{5}{Department of Terrestrial Magnetism, Carnegie Institution of Washington, 5241 Broad Branch Road, NW, Washington DC 20015, USA}
\altaffiltext{6}{Research School of Astronomy \& Astrophysics, Australian National University, Cotter Road, Weston Creek, ACT 2611, Australia}
\altaffiltext{7}{Department of Physics and Astronomy, Uppsala University, Box 516, 75120 Uppsala, Sweden}
\altaffiltext{8}{Lund Observatory, Department of Astronomy and Theoretical Physics, Box 43, 221 00, Lund, Sweden}
\altaffiltext{9}{Institute of Astronomy, University of Cambridge, Madingley Road, Cambridge CB3 0HA, United Kingdom}
\altaffiltext{10}{Departamento de F\'isica e Astronomia, Faculdade de Ci\^encias, Universidade do Porto, Rua do Campo Alegre, 4169-007 Porto, Portugal}
\altaffiltext{11}{CNRS/Univ. Bordeaux, LAB, UMR 5804, 33270, Floirac, France}

\begin{abstract}
Stellar elemental abundances are important for understanding the fundamental properties of a star or stellar group, such as age and evolutionary history, as well as the composition of an orbiting planet. However, as abundance measurement techniques have progressed, there has been little standardization between individual methods and their comparisons. As a result, different stellar abundance procedures determine measurements that vary beyond quoted error for the same elements within the same stars \citep{Hinkel14}. The purpose of this paper is to better understand the systematic variations between methods and offer recommendations for producing more accurate results in the future. We have invited a number of participants from around the world (Australia, Portugal, Sweden, Switzerland, and USA) to calculate ten element abundances (C, O, Na, Mg, Al, Si, Fe, Ni, Ba, and Eu) using the same stellar spectra for four stars (HD~361, HD~10700, HD~121504, HD~202206). Each group produced measurements for each of the stars using: 1) their own autonomous techniques, 2) standardized stellar parameters, 3) standardized line list, and 4) both standardized parameters and line list. We present the resulting stellar parameters, absolute abundances, and a metric of data similarity that quantifies homogeneity of the data. We conclude that standardization of some kind, particularly stellar parameters, improves the consistency between methods. However, because results did not converge as more free parameters were standardized, it is clear there are inherent issues within the techniques that need to be reconciled. Therefore, we encourage more conversation and transparency within the community such that stellar abundance determinations can be reproducible as well as accurate and precise. 

\end{abstract}

\keywords{solar neighborhood --- stars: abundances --- stars: fundamental parameters -- techniques: spectroscopic -- methods: data analysis -- stars: individual (HD~361, HD~10700, HD~121504, HD~202206)}

\section{Introduction}
\label{introduction}
Like many stars, the Sun is composed overwhelmingly of H and He ($\sim 98.5\%$ by mass), with the
elements O, C, Fe, Ne, Si, N, Mg, S and the rest, in descending order, comprising the remainder
\citep{Lodders:2009p3091}.
How much of a star's mass is in non-H/He elements (termed `metals') and the proportions
of the metals are very important parameters to constrain.
Elemental abundances help astronomers constrain stellar ages \citep{bond_2010_aa, Nissen15} and 
trace Galactic chemical evolution \citep{Timmes:1995p3197,Venn:2004p1483,Soubiran:2005p1496}. 
Most obviously, the abundances of elements in a star---and therefore its protoplanetary 
disk---help determine the composition of planets that form around that star.

In disks with varying chemical make-ups, different minerals will condense out of the
gas, creating unique proportions of solids to form planets  
\citep{Bond:2008p2099, bond_2010_aa}. Systems with more C than O could potentially form planets not of silicates but rather SiC 
\citep{Kuchner05, bond_2010_aa}. That being said, it is worth noting that detailed observations
show little variation from the solar value ${\rm C}/{\rm O} \sim 0.54$ (\citep{Lodders:2009p3091}, such that
most stars have ${\rm C}/{\rm O} < 0.8$ \citep{Nissen14, Teske14}. More subtle mineralogical effects could be more common. The Mg/Si ratio in a planet can substantially change the mineral assemblage and mantle viscosity \citep{Umemoto06, Ammann11} and composition \citep{Santos15}.  

The widespread idea of chemical tagging, among stars with comparable dynamics, seeks to identify a 
subset of stars with very similar element abundance patterns. In this way, there is potential to find stars that
formed in the same stellar cluster \citep[e.g.][]{Mitschang14, Barenfeld13},
in particular stars that may have formed with the Sun \citep[e.g.][]{GonzalezHernandez:2010p7714,Ramirez14b,Nissen15}.
Chemical tagging studies benefit from measuring many elements, but especially those elements that vary the most from star to star \citep{Ramirez14b}. 
Using 11 stars with both physical properties 
and abundance patterns similar to the Sun, or ``solar twins", \citet{Melendez09} looked for patterns
among those refractory elements expected to be trapped in planets. They inferred that the Sun appears to be slightly depleted, relative to these solar twins, in elements trapped within the orbiting planets \citep{Adibekyan12}. Comparatively, \citet{GonzalezHernandez:2010p7714} argued that they did not find such a discrepancy between solar twins that do and do not host planets. For all these reasons it is important to carefully measure elemental abundances in stars \citep{Truitt15}. 

Of all the non-H/He elements, the abundance of Fe is most easily measured in a star's spectrum, 
owing to a large number of absorption lines at optical wavelengths. 
Its abundance (number or molar abundance) in a star's atmosphere is usually reported 
normalized with respect to the abundance of H in the Sun, expressed as 
$[{\rm Fe}/{\rm H}] = \log_{10}\left[ ({\rm Fe}/{\rm H}) / ({\rm Fe}/{\rm H})_{\odot}\right]$.
The iron-content of nearby, disk stars in the Milky Way spans a range, but the vast majority 
lie between $-0.5 < [{\rm Fe}/{\rm H}] < +0.5$, meaning stars have between 1/3 and 3 times
the Sun's abundance of Fe \citep[e.g.][]{Casagrande11}. 
When the Galaxy formed, $\sim 1$ Gyr after the Big Bang, the only dominant elements were H and He. 
But as the Galaxy evolved over time, supernovae, novae, asymptotic giant branch (AGB) stars, etc., 
enriched the interstellar medium with various elements.
The range of metallicities is largely understood as reflecting the time of a star's formation.
Indeed, some of the oldest known stars, with metallicity $< -2.0$ dex, illustrate that the ${\rm C}/{\rm O}$ has not been constant over Galactic history \citep{Frebel15}.
Given the variety of nucleosynthetic sources occurring over the Galaxy's lifetime, 
there is no reason to expect that both C and O, or any element, would be produced in exact proportion
to Fe. It is widely recognized that in the Galaxy the ratio of $\alpha$-elements (intermediate mass elements with even atomic number Z, as would be produced by successive $\alpha$ captures on elements from carbon up to the iron-peak) compared to Fe shows a trend towards super-solar values at low [Fe/H] \citep[e.g.][]{Timmes:1995p3197}. This is a result of differing contributions from core collapse and Type Ia supernovae over time. What has not been widely understood before the modern era of large surveys is the amount of variation of individual elements in stars of similar [Fe/H]. In other words, while the iron-content is easily measured and is even colloquially synonymous with total metallicity, 
it only broadly constrains the abundances of other elements.  
These other elements must be measured and understood in stars. 

Many research groups have attempted to measure multiple elements, not just Fe, in stellar 
atmospheres (including many of the authors of this paper).
Spectroscopic abundance data for 50 elements across $> 3000$ stars were compiled from 84
literature sources by \citet{Hinkel14}, who created the {\it Hypatia Catalog}.
A surprising result of this compilation, similar to that seen in \citet{Torres12}, was that different stellar spectroscopy groups 
infer quite discrepant abundances for the same elements in the same stars.
The variations in elemental abundances often exceed (by factors of 3 or more) the 
formal uncertainties in the measurements.
Discrepancies between varying techniques were large enough, and found often enough, that the lack of agreement represents
a crisis in the field of abundance determination by stellar spectroscopy.

Indeed, a number of recent papers have sought to compare different abundance methodologies in order to understand their inherent variations. For example, \citet{Bruntt10a} compared a variety of direct and indirect techniques used to calculate the fundamental parameters of bright, solar-like stars. The Gaia-ESO \citep{Gilmore12, Randich13} survey team has also analyzed a combination of many methods to remove and understand systematic differences in stellar abundances. \citet{Smiljanic14} conducted a spectroscopic analysis employing 13 different techniques who performed abundance measurements on 1300 FGK-type stars. In addition, \citet{Jofre14, Heiter15a} and \citet{Jofre15} determined the iron-content, effective temperatures and surface gravities, and the abundances for $\alpha$- and iron-peak elements, respectively, for 34 ``Benchmark Stars'' selected as the cornerstone for the Gaia-ESO survey's data calibration. A variety of tests were performed on the spectra of these stars in order to understand the effect that, for example, data resolution, instrument, and local thermodynamic equilibrium (LTE) approximations may have on different techniques used to analyze the data.  

In an attempt to reconcile the discrepancies, a Workshop Without
Walls called ``Stellar Stoichiometry" was held at Arizona State University, April 11-12, 2013, 
supported by the NASA Astrobiology Institute. 
The effects of elemental abundance variations on planet formation, planetary structure, and 
habitability were discussed \citep{Young14}, but the emphasis was on resolving the abundance inconsistencies.
To that end, workshop attendees from stellar spectroscopy groups were asked to participate 
before and after the workshop in a comparative study (affectionately called our ``homework 
assignment" but what we will refer to here as the ``Investigation"), in which they were provided a common set of 4 stellar spectra and asked to derive the abundances of several key elements in a variety of ways.

This paper reports the results of the comparative study, which was elaborated upon due to the findings and discussion from the initial workshop.  
In \S \ref{s.homedescription} we describe the comparative study: we list the participants and the 
numerical methods that each group used and then describe the common data set each group was 
asked to analyze. In \S \ref{s.results} we present the results from the study, first discussing the stellar parameters and then we present the abundances of 10 key elements inferred by each group.
In \S \ref{s.litcomp}, we report on updated findings from the {\it Hypatia Catalog} exemplifying the 
magnitude of the discrepancies in abundance determinations. We also compare the results of the analysis with respect to literature abundances.  
In \S \ref{s.disc} we analyze some of the details of the Investigation, namely the error calculations by each group and the effect of varying line lists on the abundance results. We compare results between methods that used the curve-of-growth technique and those that used spectral fitting. We also offer an overall recommendation to the field in order to establish more consistent results between future stellar abundances determinations. Finally, in \S \ref{s.summary}, we summarize the findings of our Investigation.

\section{Description of the Investigation}
\label{s.homedescription}
We chose four stellar spectra for our analysis:  HD~361, HD~10700, HD~121504, and HD~202206, which are discussed in \S \ref{s.sample}. Each group, introduced in \S \ref{s.people}, was asked to determine the abundances for each star in four different ways (or `analyses'). The end result was comprised of 16 abundance data sets for each group: four analyses for each of the four stars.

\vspace{1mm}
{\it -- Run 1: Autonomous Method} - For the first run, we asked that the analysis be done in the way most typical for the participants, namely using their own line lists, stellar parameters, and customary abundance measuring techniques.

\vspace{1mm}

{\it -- Run 2: Standard Stellar Parameters } - We provided all of our participants with standard stellar parameters (effective temperature \teff, surface gravity \lg, and microturbulent velocity $\xi$) to use for each star during the second run, shown in Table \ref{standardparams}. The standard parameters were chosen via an amalgamation of literature values that were consistent with each spectral type without giving preference to any one methodology. See Appendix \ref{a.standard} for more details. 

\vspace{1mm}

{\it -- Run 3: Standard Line List} - The participants were asked to use the provided line list (Table \ref{standardlines}), which also specified excitation potentials, oscillator strengths, van der Waals damping constant, and the EW of each line. During this analysis, the groups determined their own stellar parameters. The line list was compiled from a variety of references (Column 7) with a similar wavelength range. See Appendix \ref{a.standard} for more details.

\vspace{1mm}

{\it -- Run 4: Standard Stellar Parameters and Line List }- For the final run, we wanted to know whether standardizing both the stellar parameters and line list would force all of the abundance measurements to be nearly identical to within error. Therefore, the groups were told to use Tables \ref{standardparams} and \ref{standardlines} concurrently to determine the stellar abundances.

\vspace{3mm}

Each participating group was asked to document their entire produce and decisions during their analysis. We asked for the equivalent widths (EWs) to be recorded for every line where the curve-of-growth (CoG) method was used, such that we could do a more thorough comparison (\S \ref{s.elemlinebyline}). The line lists employed by the groups when one was not provided are shown in Tables \ref{anulines}-\ref{uppsalalines}.  All of the final abundance outputs are given as ``absolute" abundances, A(X), that were not normalized to the Sun, in order to reduce further cause for dispersion. Therefore, when we compare the results from this Investigation to literature sources (see \S \ref{s.litcomp}), all of the abundances - including the literature sources - will be absolute and without a solar normalization.

\subsection{Stellar Sample}
\label{s.sample}
The stellar spectra for this study were taken by Paul Butler and his team from the Carnegie Institute of Washington and were given to ASU for elemental abundance analysis. The spectra were observed through the Magellan Planet Search Program between December 2002--July 2009. Data were taken using the Magellan Inamori Kyocera Echelle (MIKE) spectrograph \citep{Bernstein03} on the 6.5 meter Magellan II telescope. The spectra have an average resolution of $\Delta \lambda / \lambda \sim50\,000$ in the wavelength range 4700--7100\aaa (red chip) with a S/N varying between 150--300, depending on the quality of the original spectra, with an average of $\sim$200. Individual orders were combined using the \textsc{odcombine} function in IRAF while wavelength calibration, carried out by the iodine absorption method \citep{Marcy92}, was done in IDL. The unified spectra were velocity shifted using \textsc{dopcor} with respect to the 6750.15\aaa\ Fe I line scaled across the spectra. The function \textsc{continuum} was used to normalize the continuum. The calibrated data set was sent to the Investigation participants because it was considered not only typical for both resolution and S/N, but representative for standard stellar spectra (see \S \ref{a.field}) and thus, for determining typical variations.

As an aside, after the stellar abundances were sent to the participants, we discovered that there was a mismatch between the stellar spectra and the corresponding wavelengths in the region below 5050 \aaa. This directly impacted any element that had a line below this region, since the wrong absorption features would have been measured, namely Mg (4730.04 \aaa). Therefore, all groups were instructed to ignore any lines below the 5050 \aaa cutoff. We discuss this more thoroughly in \S \ref{s.cali}.

We chose the four stars, HD~361, HD~10700, HD~121504, and HD~202206, based on a number of considerations. To begin, we wanted the MIKE spectra to be relatively common and ``clean" for all of our stars, in order to reduce additional variation in the abundance determinations between groups.  Despite analyzing a small sample, we sought to cover a range of stellar temperatures and/or stellar types which varied away from solar, but not so far that the stars were no longer comparable to each other.  \citet{Gray06} classified HD~361 as a G1V star while \citet{Morgan73} used HD~10700, or Tau Ceti, as the standard G8V star. Comparatively, HD~121504 is a G2V \citep{Houk75}, although may be considered more of a ``G0.5V" given the systematic differences between the \citet{Houk75} classifications with respect to the \citet{Morgan73} grid in modern use. For similar reasons, while \citet{Houk88} classifies HD~202206 as a G6V, it may more closely resemble a ``G8V." For more information regarding the stellar subtype comparisons, we refer to the breakdown offered by Eric Mamajek\footnote[1]{\url{http://www.pas.rochester.edu/$\sim$emamajek/memo$\_$G2V.html}}.

\begin{figure*}
\begin{center}
\centerline{\includegraphics[height=5.5in]{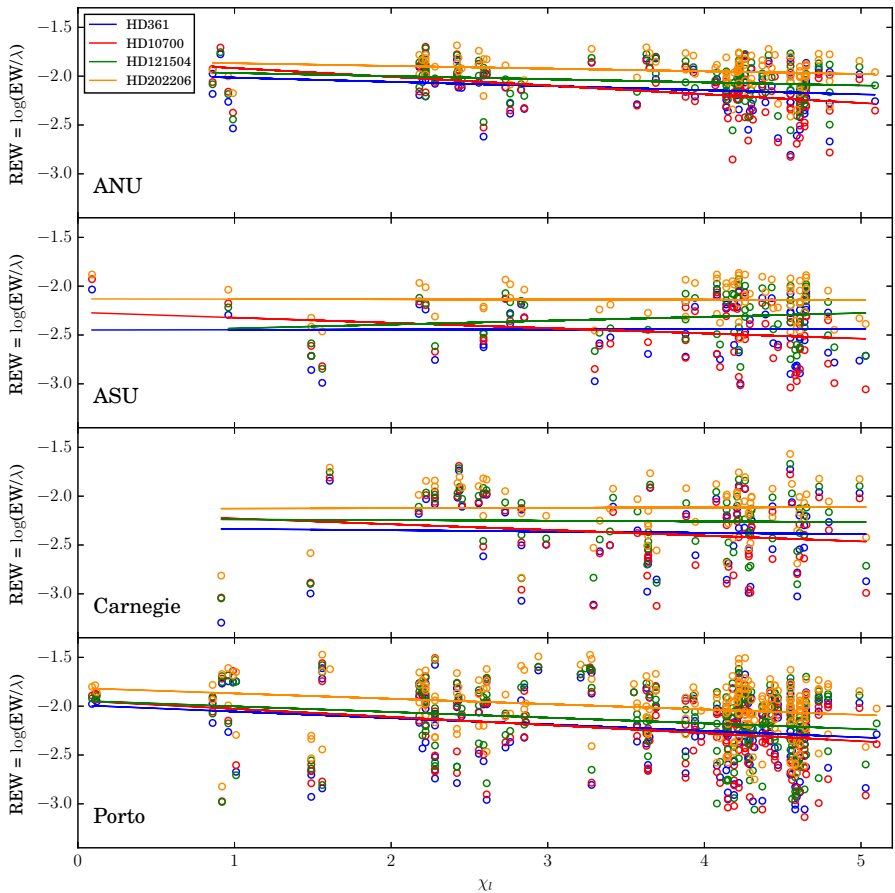}}
\end{center}
\caption{
The reduced equivalent width (REW), which is determined by $\log$(EW/$\lambda$), with respect to the excitation potential, $\chi_l$, for the four CoG groups, as listed in the lower left-hand corner of each panel. Each star is represented by a different color, given by the legend in the top left corner. Linear regression lines are overlaid with respect to each star; the slopes are given in Table \ref{rew}.
}\label{rew-ep}
\end{figure*}

\subsection{Investigation Participants}
\label{s.people}
As part of ASU's NASA Astrobiology Institute-sponsored ``Stellar Stoichiometry" Workshop Without Walls, a number of teams known for their stellar abundance research were contacted about participating in the Investigation. The aim was to accrue as many groups as possible, with as large a variety in stellar abundance techniques as possible. 
The six groups that participated in this Investigation were:  Australian National University (ANU), Arizona State University (ASU), Carnegie Institution of Washington/Department of Terrestrial Magnetism (Carnegie), Observatoire de Gen\`eve (Geneva), Instituto de Astrof\'isica e Ci\^encias do Espa\c{c}o, Universidade do Porto (Porto), and Uppsala University (Uppsala). The names in parenthesis indicate the short-name that we use throughout the paper when referencing each group. We give an in-depth description of the methods used by the participants in Appendix \ref{a.people}. A summary of these methods can be found in Table \ref{models}, which show that both Geneva and Uppsala employ a spectral fitting technique for determining stellar abundances while the rest use CoG. Finally, we note that only two groups participating in this Investigation overlapped with the analysis conducted by the Gaia-ESO team \citep[i.e.][]{Smiljanic14, Jofre14, Jofre15}, namely the Porto and Uppsala groups.

\section{Investigation Results}
\label{s.results}
Each of the 6 groups that participated in the Investigation generated one data set for each of the four stars per run analysis (see \S \ref{s.homedescription}), including line lists, stellar parameters, EWs (in most cases), stellar abundances, and error bars. We have compiled all of this information into a number of figures and tables for better visualization. In this section, we analyze the stellar parameters between Runs 1 and 3, element abundances for 10 elements from Runs 1--4, and conduct a line-by-line analysis for interesting elemental lines. Note that [Fe/H] content was not fixed as part of the standard stellar parameters and will be discussed in \S \ref{s.elements} with the other element abundances.

\subsection{Reduced Equivalent Width vs. Excitation Potential}
\label{s.rew}

When using the CoG method, the excitation and line-strength balance approach can only strictly work if there is no inherent correlation in excitation potential and line strength for the entire iron line sample utilized. If there is a correlation, then the solutions are not determinate without some additional constraint, in which case, solutions and their comparisons are meaningless. Fortunately, reducing or removing the correlation between these two parameters can be handled within the parameter calculation module of MOOG, utilized by all four CoG groups. To verify that the results were indeterminate, we have plotted the reduced equivalent width, or $\log$(EW/$\lambda$), for all Fe I lines used by the four CoG groups (see Table \ref{models}) with respect to the excitation potential, $\chi_l$, shown in Fig. \ref{rew-ep}. Note, that the REW and $\chi_l$ determinations were the same for all groups between Runs 1 and 2, and were provided for Runs 3 and 4, which is why we have only plotted one set of values regardless of run.

We generated a linear fit for the four stars as calculated by each group. The linear fits are overlaid on top of Fig. \ref{rew-ep}, with respective color for each star, and the slopes are given in Table \ref{rew}. All of the slopes are below an absolute value of 0.09, with averages per group below an absolute value of 0.07 (last column). These are acceptable tolerances, especially considering the dispersion in the plots. \citet{Colucci09} note that a small correlation of the REW Fe I lines with excitation potential, on par with those measured here, will not significantly affect the mean [Fe/H] abundance. In general, a correlation between REW and $\chi_l$ is most significant when a technique determines parameters in serial order, for example, constraining \teff, fixing \teff, constraining $\xi$, etc \citet{Sousa14}. None of the CoG groups determine parameters in this way, instead they find the best parameters in parallel where any possible correlations are considered during the minimization process.  
Overall, we found that there is no significant correlation between REW and $\chi_l$, meaning that the resulting abundances are not indeterminate and are suitable for comparison.

\subsection{Stellar Parameters}
\label{s.parameters}
\begin{figure*}
\begin{tabular}{p{8.5cm}p{2.0cm}}
  \includegraphics[width=95mm]{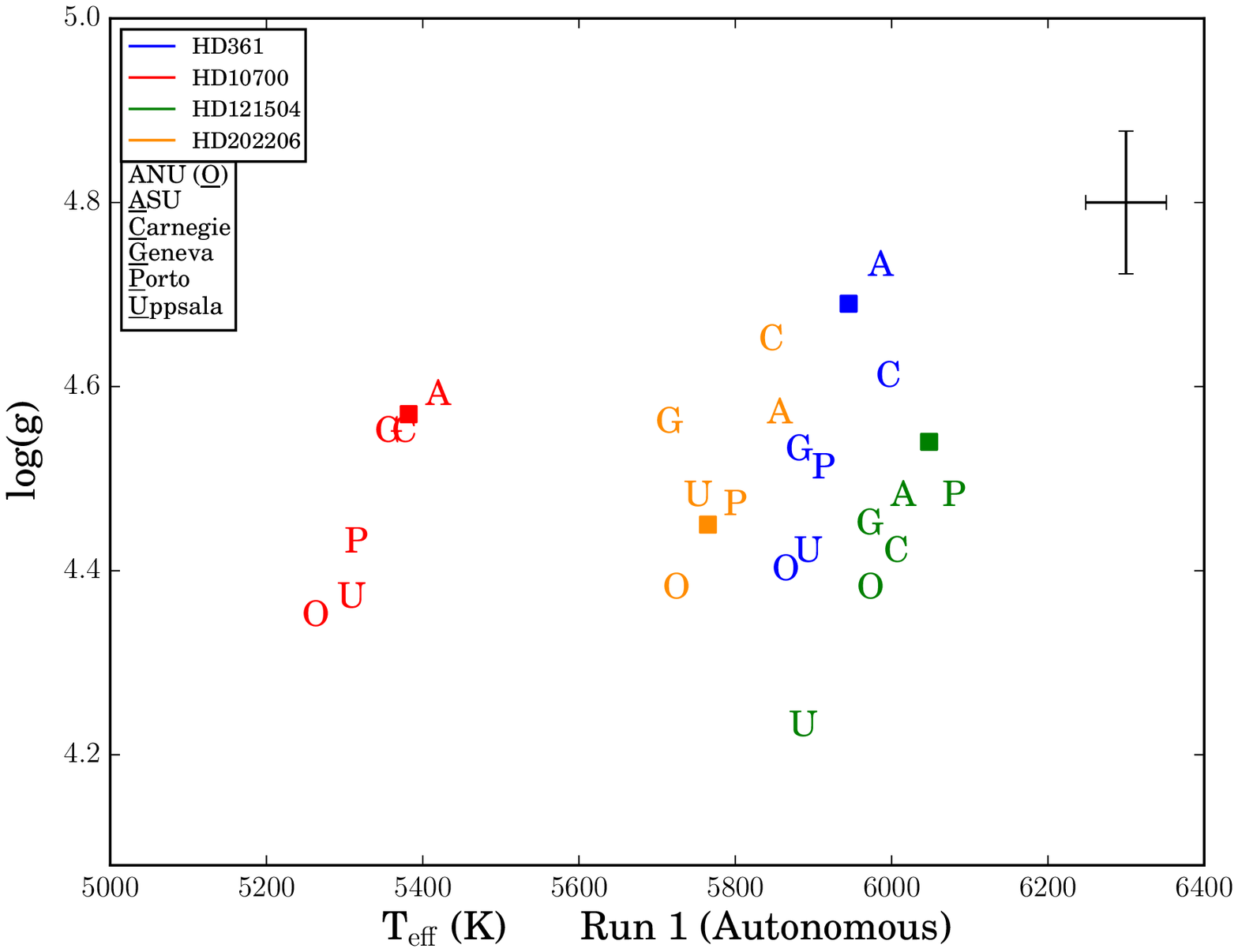} &   \includegraphics[width=95mm]{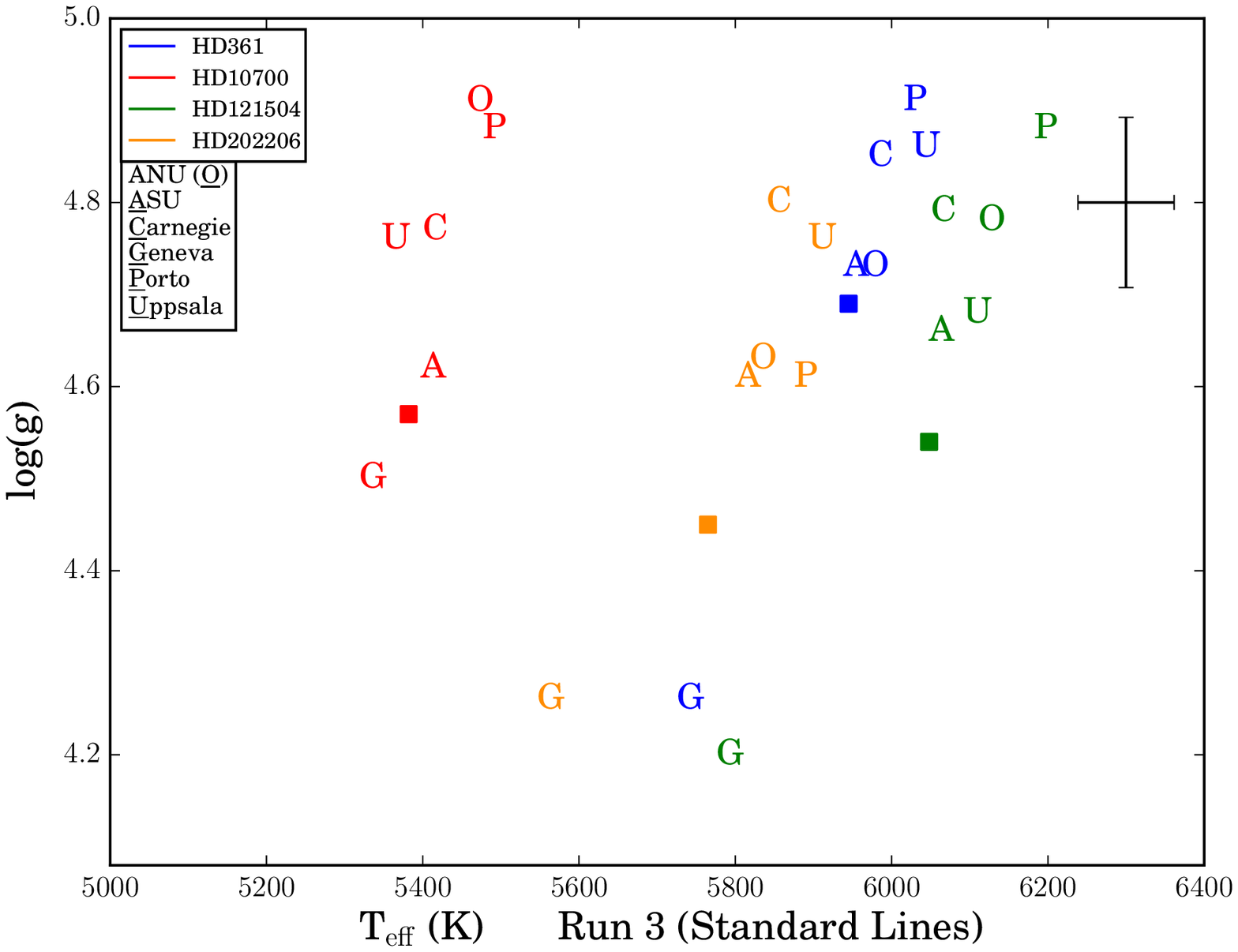} \\
 \includegraphics[width=95mm]{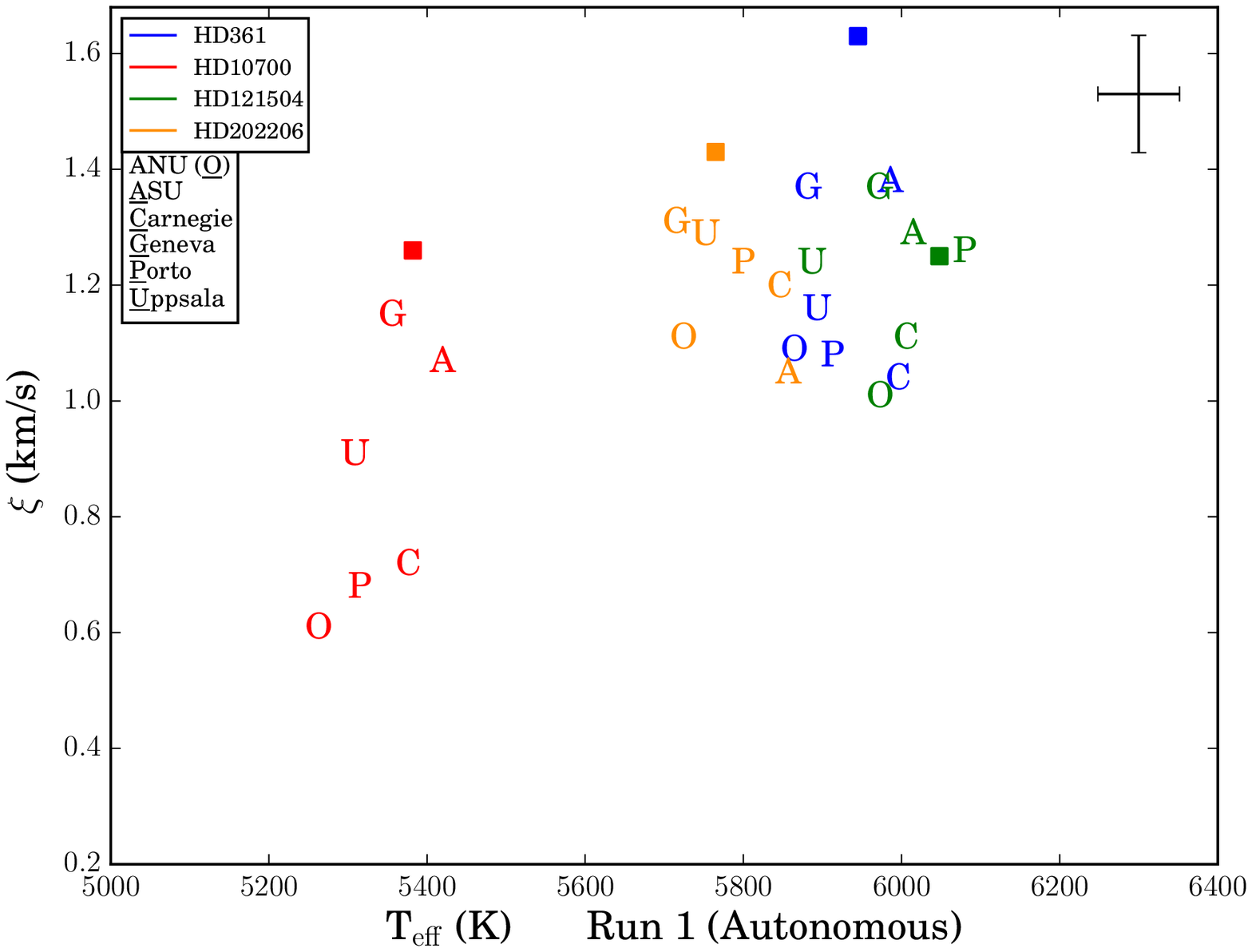} &   \includegraphics[width=95mm]{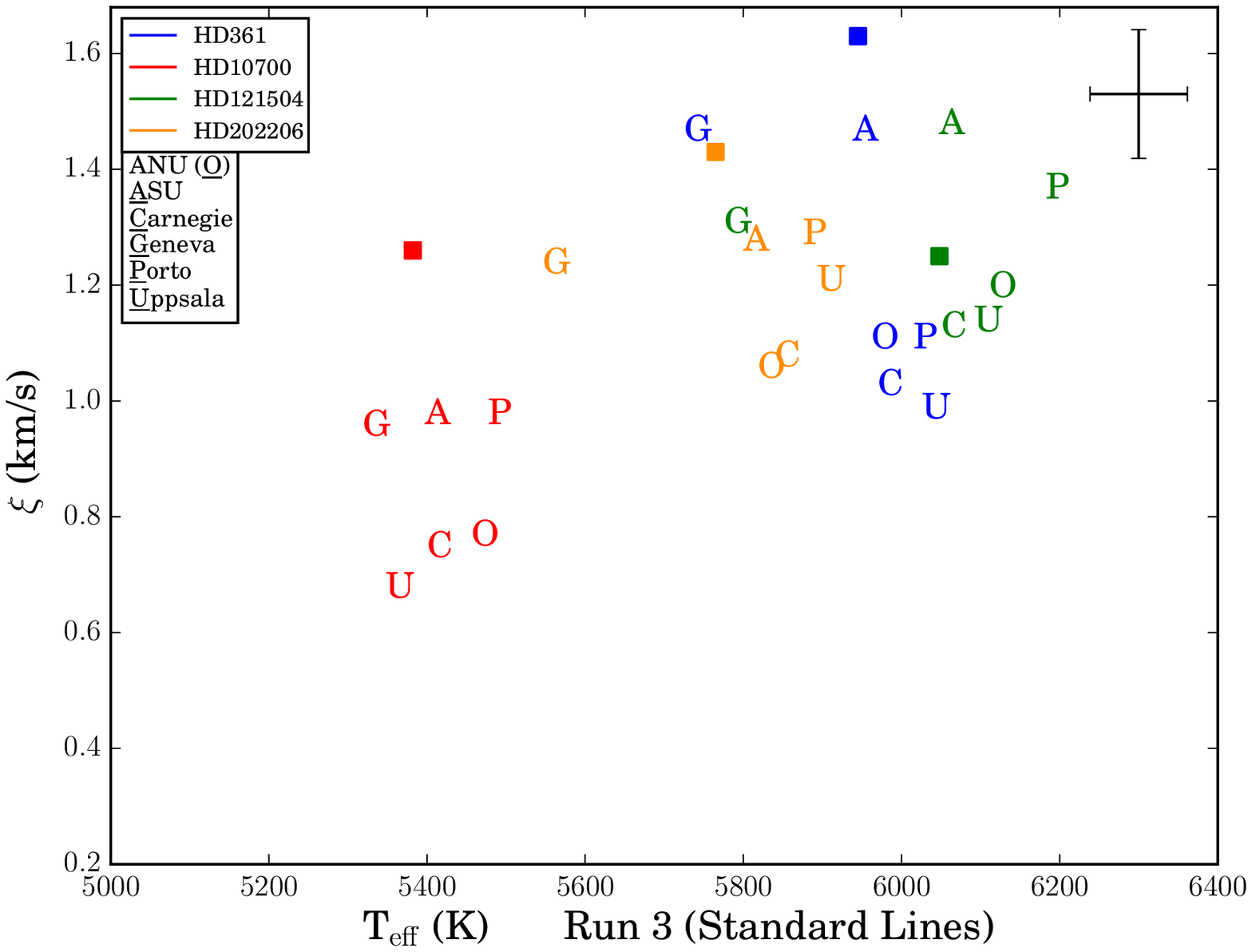} \\
\includegraphics[width=95mm]{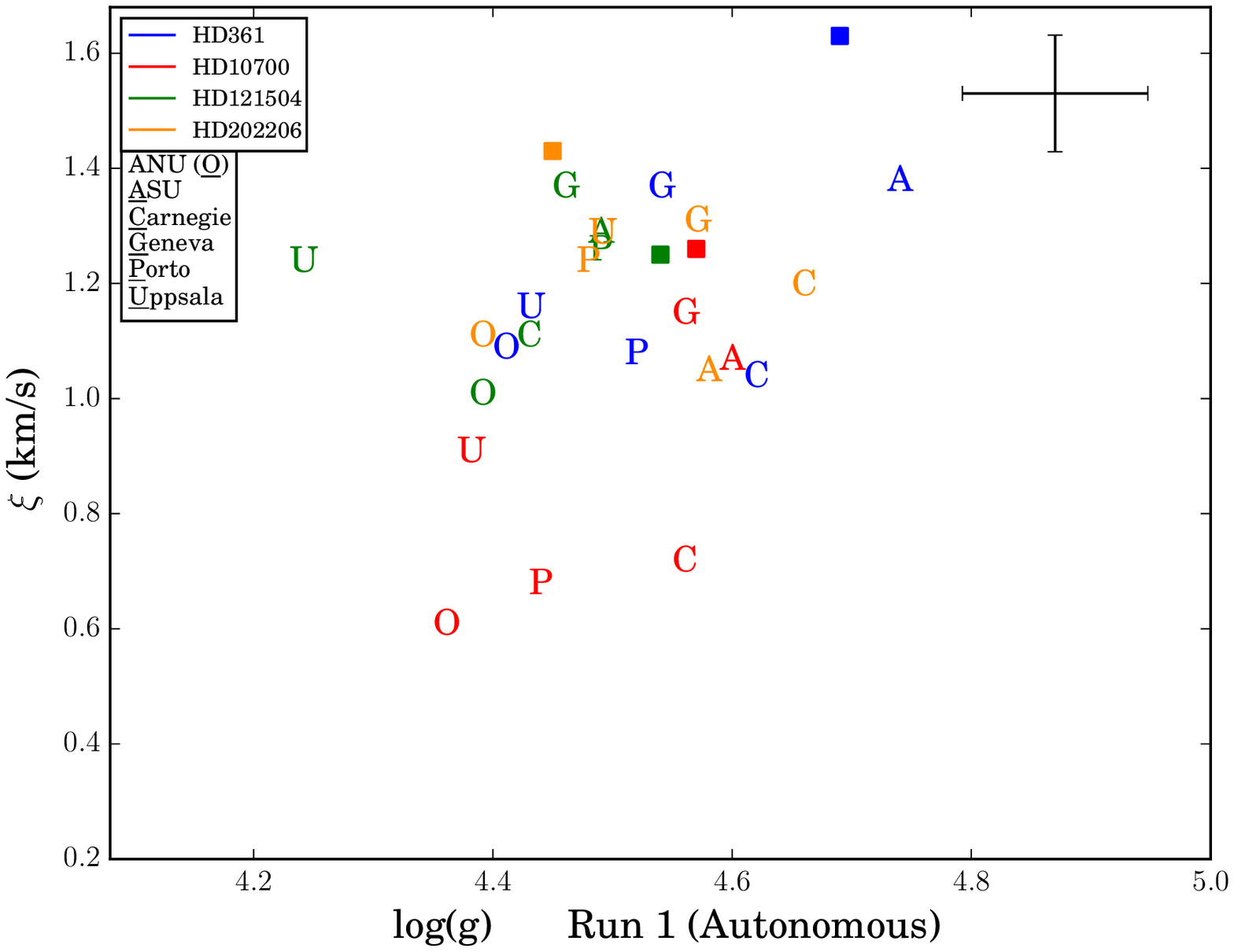} &   \includegraphics[width=95mm]{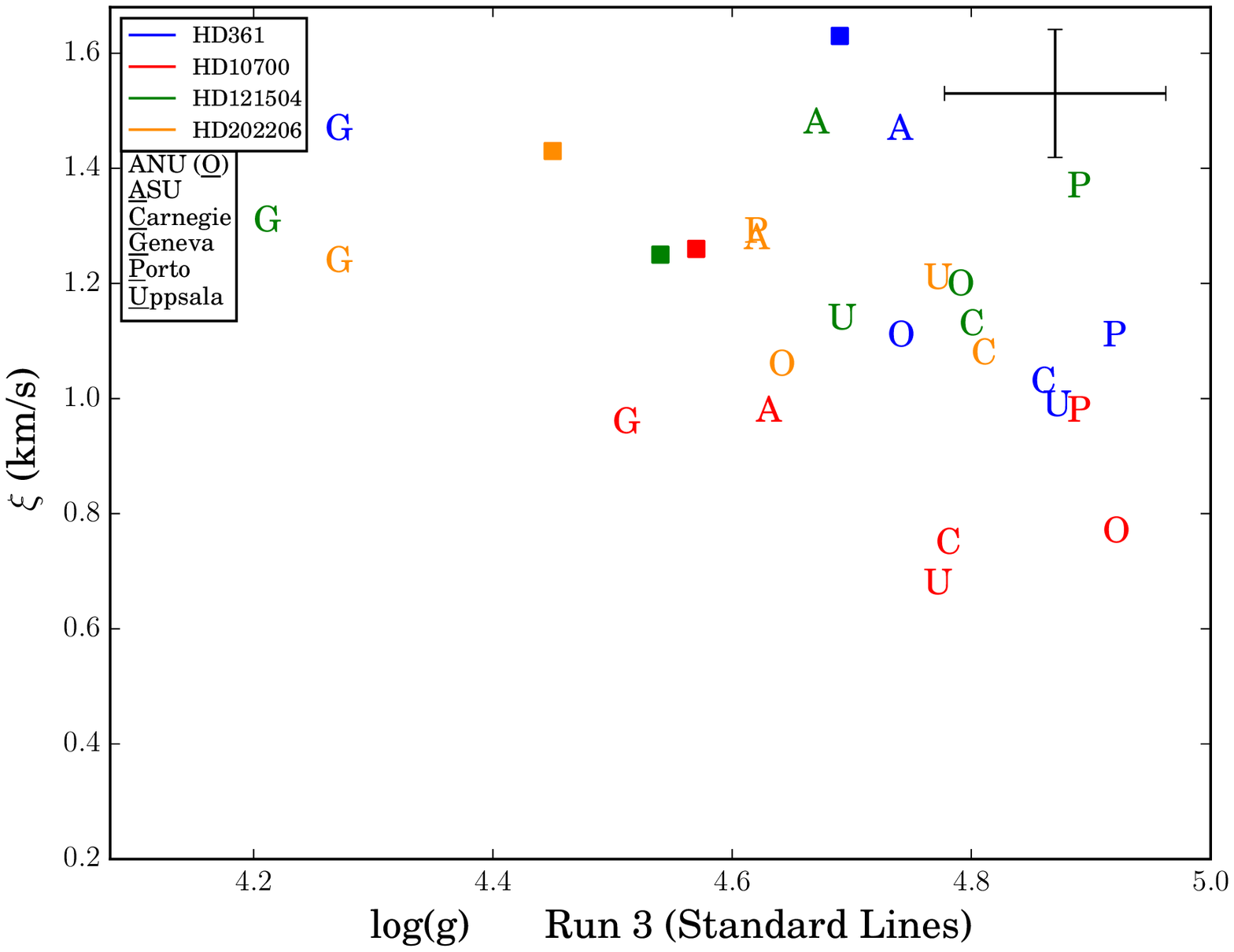} \\
\end{tabular}
\caption{Stellar parameters for the Autonomous Method Run 1 (left column) and Standard Lines Run 3 (right column). On the top row is \lg vs. \teff, middle row is $\xi$ vs. \teff, and the bottom row is $\xi$ vs. \lg. Colors are indicative of the star and each letter represents the group (see underlined letter in legend). There is a representative error bar in the upper right-hand corner for each plot. The standard parameters for each star, as given in Table \ref{standardparams}, are denoted as solid squares.}\label{params}
\end{figure*}

Unique stellar parameters were produced by each group for Run 1 (Autonomous) and Run 3 (Standard Line List), which can be found with their individual errors in Tables \ref{stellarparams1} and \ref{stellarparams3}, respectively. We have also plotted the \teff, \lg, and $\xi$ values with respect to each other in Figure \ref{params} (the A(Fe) comparisons will be discussed in \S \ref{s.elements}). A representative error bar is given in the top right-hand corner for each plot, calculated by taking the median error for each star (to reduce the effect of outliers) and then taking the mean of all four stars. We signify an `outlier' as a result which does not overlap any other data point to within either individual error or representative error, when the former was not reported. The standard stellar parameters, used during Run 2 (Standard Parameters) and Run 4 (Standard Line List and Parameters), are denoted as solid squares on all of the plots. 

We go into more detail in Appendix \ref{a.parameters} regarding the trends seen in the stellar parameters. However, in general we found that both \teff and \lg varied somewhat between the Autonomous Method in Run 1 and Standard Line List in Run 3. The total ranges in both parameters for all groups became worse in Run 3 and we found that removing obvious outliers decreased the range in all stars to values smaller than those in Run 1. In fact, the ranges for Run 3 sans outliers were comparable to the average-median errors for both \teff and \lg. We conclude that the standard line list had a polarizing effect on the methods analyzed in this study -- making some calculations more similar and others vastly different. This may be related to the fact that a number of groups determined extremely similar \lg values for all stars, regardless of spectral type, during the Autonomous Run 1. In contrast, the variation in $\xi$ was constant or decreased for all stars between Runs 1 and 3, although those differences in the ranges were within the error budget. In other words, $\xi$ was not noticeably affected by implementing a specific line list. We also found that the number of Fe lines, as compared to the Standard Line List (see Table \ref{models}), did not impact the stellar parameters whether comparing between groups or between Run 1 and 3. This may be because the number of Fe lines implemented by all groups was above a critical number needed to consistently and precisely determine the stellar parameters of our sample.

As a further test, we compared the stellar parameters (\teff, \lg, and $\xi$) determined by each group to the average for Runs 1 and 3. We found that everything but $\xi$ was clearly correlated, as indicated in Figure \ref{params}, which was likely due to parameter degeneracies. Group-to-group biases confused the parameter correlations, but were realized again once outliers were removed. With outliers removed, the adopted standard line list tended to make results tighter and less correlated, supporting the degeneracy scenario. Remaining correlations are unexplained, but the strong \teff-[Fe/H] correlation indicates that something rather fundamental is missing, possibly simply related to continuum placement or allowing [Fe/H] as a free parameter in Run 3.

\subsection{Element Analysis}
\label{s.elements}
As mentioned in \S \ref{s.homedescription}, we chose ten elements on which to focus our analysis: C, O, Na, Mg, Al, Si, Fe, Ni, Ba, and Eu. These elements cover a range of nucleosynthetic origins, from $\alpha$-type to neutron-capture elements, so that we may better understand the reproducibility and standardizations employed in our study. All elements were measured by the majority, if not all, of our groups for more consistent comparisons. The abundance results for Runs 1--4 from all groups are reported in Tables \ref{run1abs}--\ref{run4abs}. Note that, to avoid introducing more variability in the methods, the abundances are reported as absolute without any solar normalization. Therefore, 
\begin{equation}
\text{A(El)} = \log_{10} \text{N(El)/N(H)}+12,
\end{equation}  
where N(El)/N(H) is the atomic ratio of the element to hydrogen and normalized to $10^{12}$ hydrogen atoms on a logarithmic scale, such that $\log$ N(H) = A(H) = 12. Abundance results are also plotted in Figures \ref{elems1} and \ref{elems2}, using a similar legend as Figure \ref{params}. The y-axis ranges for all plots span the same distance (1.5 dex) for easier comparison between elements. While the individual group error estimations are given in Tables \ref{run1abs}--\ref{run4abs}, we have included a representative error bar in the top-right corner of all of the individual plots in Figures \ref{elems1} and \ref{elems2}. The representative error was calculated by first taking the median of all individual errors (to avoid outliers) determined for an element in a star per run. The median errors were then averaged across all stars and analyses, to calculate a unique representative error per element. Since we have provided all of the individual errors in Tables \ref{run1abs}--\ref{run4abs}, these representative errors are meant to guide the eye of the reader without becoming distracting. In the rest of the discussion, similar to \ref{s.parameters} and Appendix \ref{a.parameters}, we will describe an `outlier' as a result where a data point does not overlap to within individual error with any other data point (for example, the same star and run). If an individual error was not reported, we will use the representative error.

We have developed a metric to quantize how ``good" or similar the results are for a star per run. Assume that we have a total number, $T$, of groups who have provided $T$ data points for the abundance of a star for a particular run (most commonly, $T$ = 6). Qualitatively, we examine the distance between each individual data point with respect to the other data points, via permutation. In this way, we can calculate the average distance from each data point ($i$) with respect to the other data points ($j$). We then get the average distances for the $T$ individual data points, which we sum. Shorter distances between points are ideal, thus we use an inverse exponential. In an ideal case where all data points are the same value, the individual-metric for each data point is 1 and the total metric (summing of the individuals) yields a value of $T$ (for example, 6). The exponential has a varying constant per element, $\gamma$, such that if one standard deviation of the $T$ data points is below the representative error for that element, then the metric is weighted positively. If, however, there are large gaps or outliers between the data points, the metric is weighted negatively -- but with a limitation. We did not want an extreme data point or outlier to outweigh the other results, especially if the other data sets show good agreement. Thus after a point of ``extremeness", the bad point reaches a limit on how strongly it can affect the metric. To equally compare results when less than the $T$ number of groups have measured an element, we scale the metric by 6/$T$ -- so 6 is the ideal maximum and values close to 0 are the minimum. More quantitatively, the metric is calculated via:
\begin{equation}
\frac{6}{T} \sum\limits_{\substack{i=1 \\ i \ne j}}^{T} \bigg( \frac{1}{T} \sum\limits_{\substack{j=1 \\ i \ne j}}^{T} \frac{1}{10^{\,\gamma\, \left|x_i-x_j\right|}} \bigg)\,\, ,
\end{equation}  
where $x$ is the absolute abundance determination for a star. The results for each of the 10 elements discussed here can be found in Table \ref{metric}, where the second column lists each element's one standard deviation constant, $\gamma$.

When devising the metric, not only did we want a measurement of consistency between groups, but also the ability to compare the elements to each other. However, the use of different exponential constants for each element, $\gamma$, while mathematically rigorous, inhibits that functionality. Therefore, we have included normalized metric values (``Norm." column in Table \ref{metric}) where we have divided all metrics for a particular element (across all analyses and stars) by the maximum metric for that element. In this way, the reader does not have to reacclimate their sense of a ``good" metric number between elements and can easily identify which star in which run yielded the most consistent abundance results. For example, the carbon abundances for HD~361 in Run 4 produced the most uniform measurements between groups as compared to any of the other carbon determinations (see Fig. \ref {elems1}, top left). The metric and normalized metric will be utilized when discussing the individual element results in \S \ref{s.alpha}--\ref{s.neutron} and within Appendix \ref{a.alpha}--\ref{a.neutron}.

\subsubsection{$\alpha$-Elements: C, O, Mg, \& Si}
\label{s.alpha}
Examining Figure \ref{elems1}, and with respect to more detailed discussion in Appendix \ref{a.alpha}, the $\alpha$-elements have varying consistency throughout each run. A likely contribution may be due to the number of lines measured for many of these elements. Carbon, oxygen, and magnesium can range from one to three lines, with O and Mg often having only one reliable line. Silicon has many more lines, usually $>$8, which tends to give a more robust result. Analyzing Table \ref{metric}, the majority of stellar $\alpha$-element abundance determinations were found to be the most consistent between groups. For example, the highest metric value for all four elements was usually during Run 2 -- in 10 out of the 16 instances (four stars measured for four elements). Runs 1, 3, and 4 resulted in the best uniformity a total of 2 instances each. 

Looking at the maximum metric, however, may neglect the possibility that a method works particularly well in one case, but poorly for others. Therefore, we take the mean and the median of the normalized metric per run for the four elements, which should be equal for a uniform distribution, to gauge the variation. We find that in all cases the median and the mean of the normalized metric are similar to each other within a 0.02 tolerance and span a range of 0.13 and 0.12, respectively. In both cases, Run 2 has the highest mean/median normalized metric, followed by Run 1, Run 3, and Run 4, respectively. 

Therefore, standardizing the stellar parameters is ultimately favorable when calculating abundances for this set of elements and results in more consistent measurements between methods. However, it is unclear whether standardizing the line list was beneficial. In Appendix \ref{a.alpha} we go into more detail for the individual elements.

\begin{figure*}
\begin{tabular}{p{8.5cm}p{2.0cm}}
  \includegraphics[width=95mm]{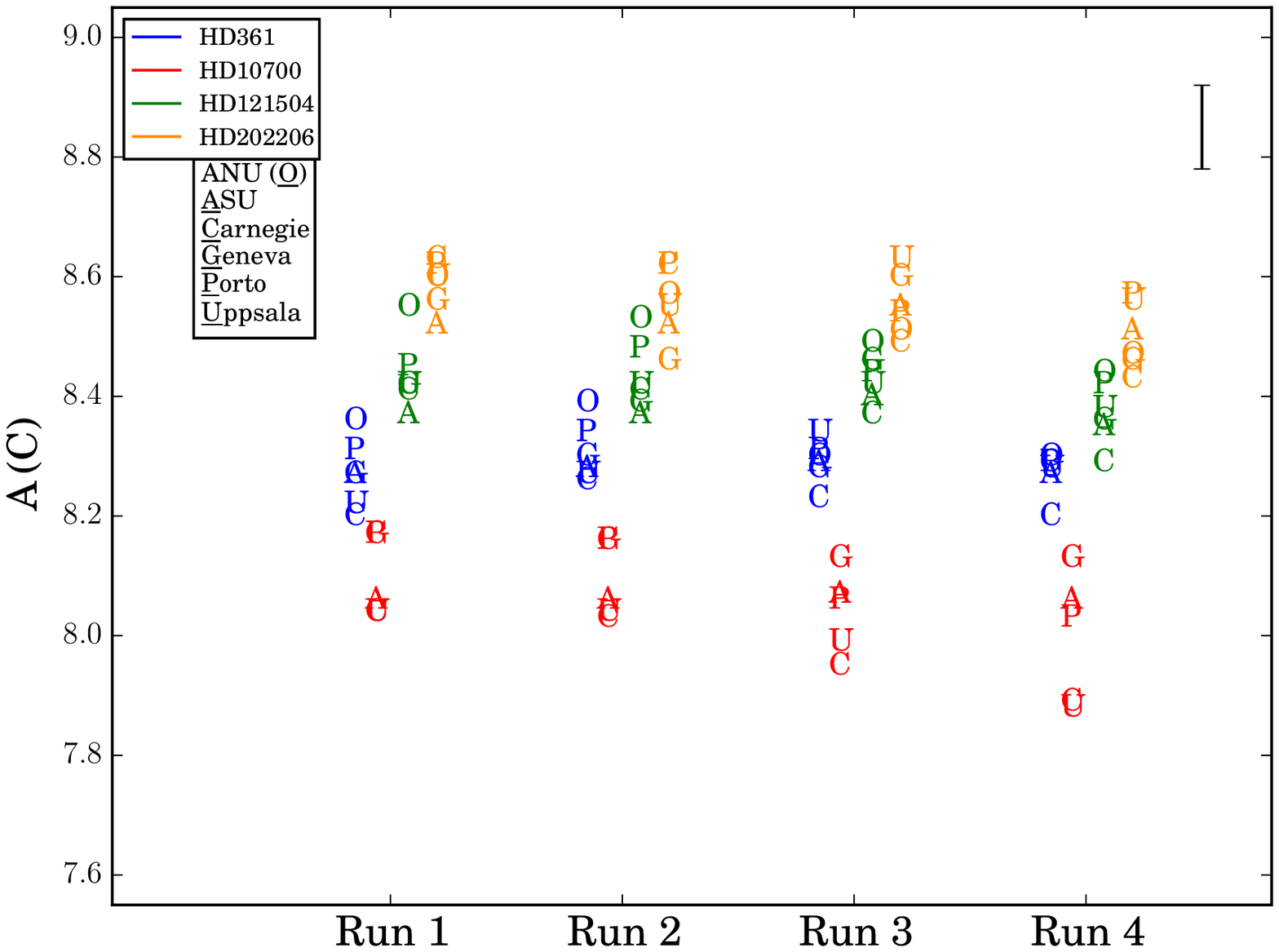} &   \includegraphics[width=95mm]{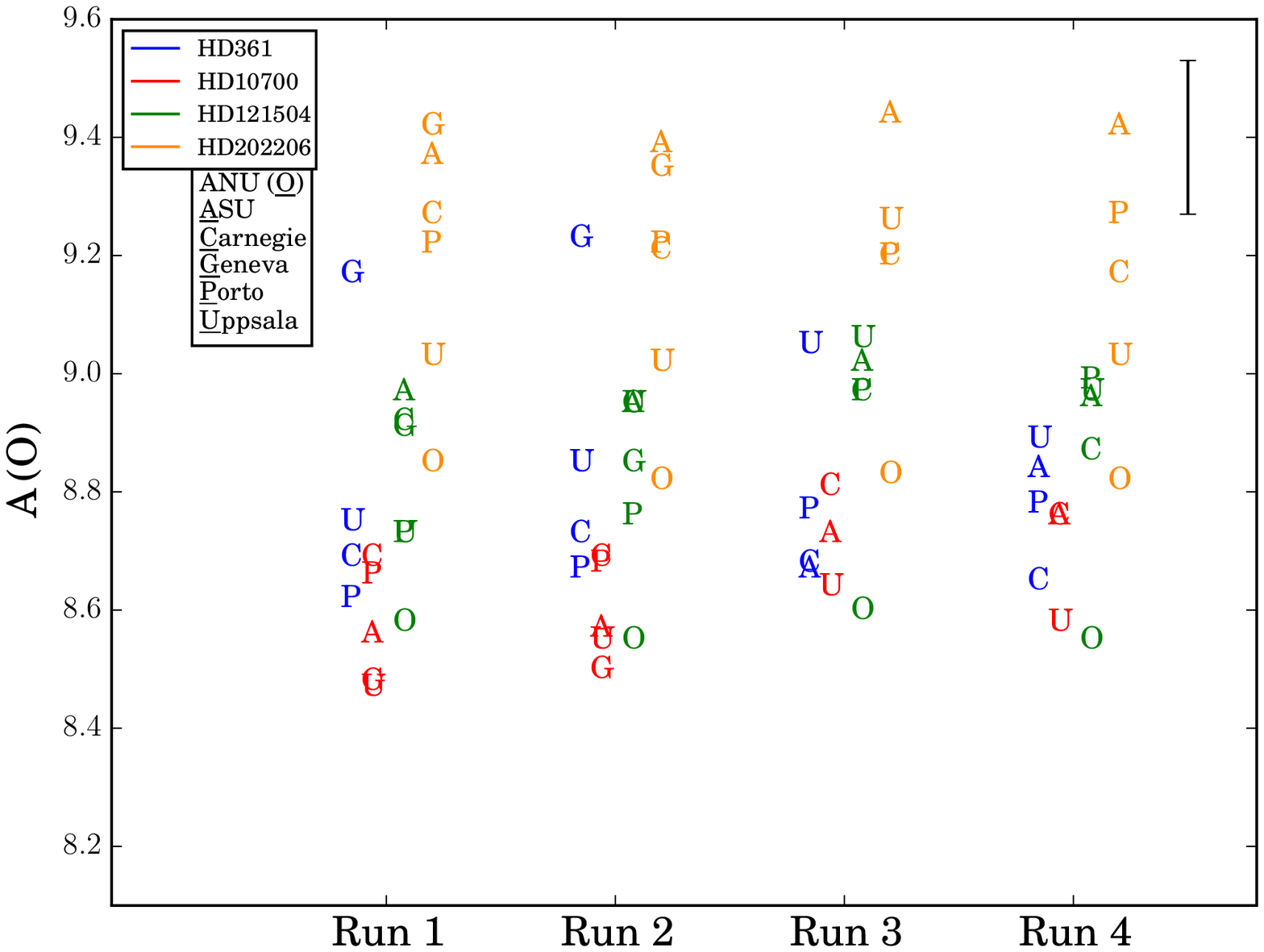} \\
  \includegraphics[width=95mm]{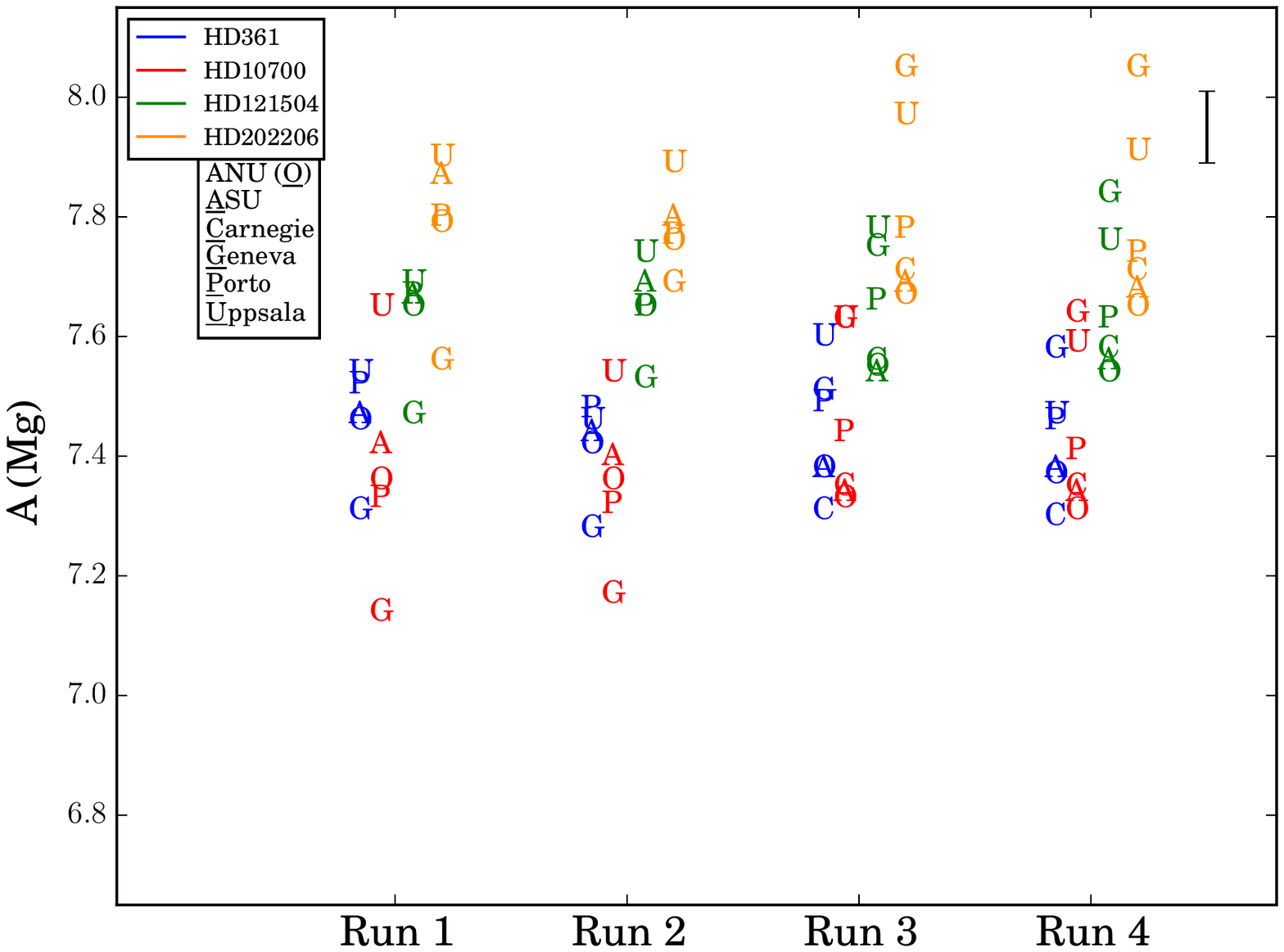} & \includegraphics[width=95mm]{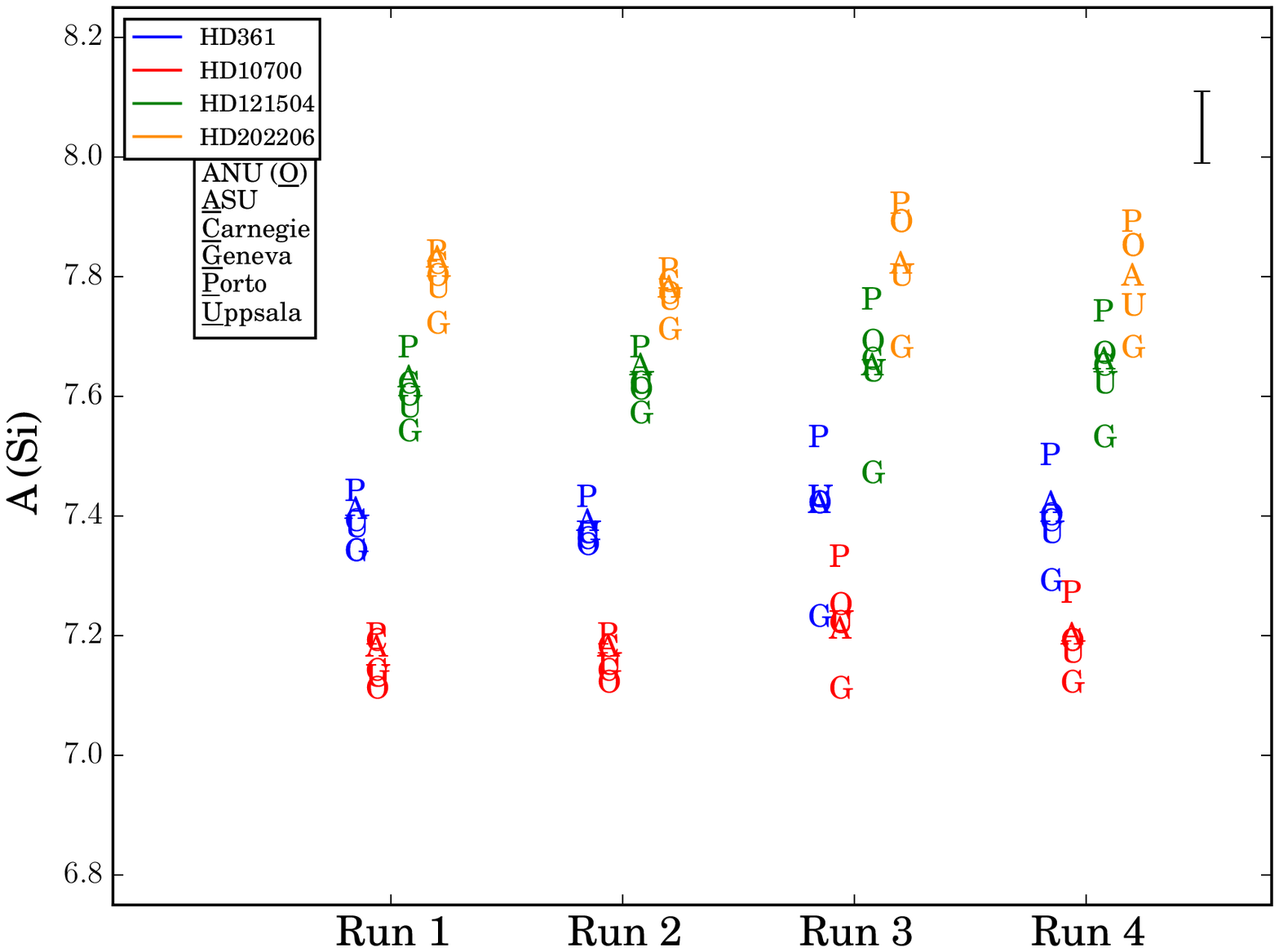} \\
  \includegraphics[width=95mm]{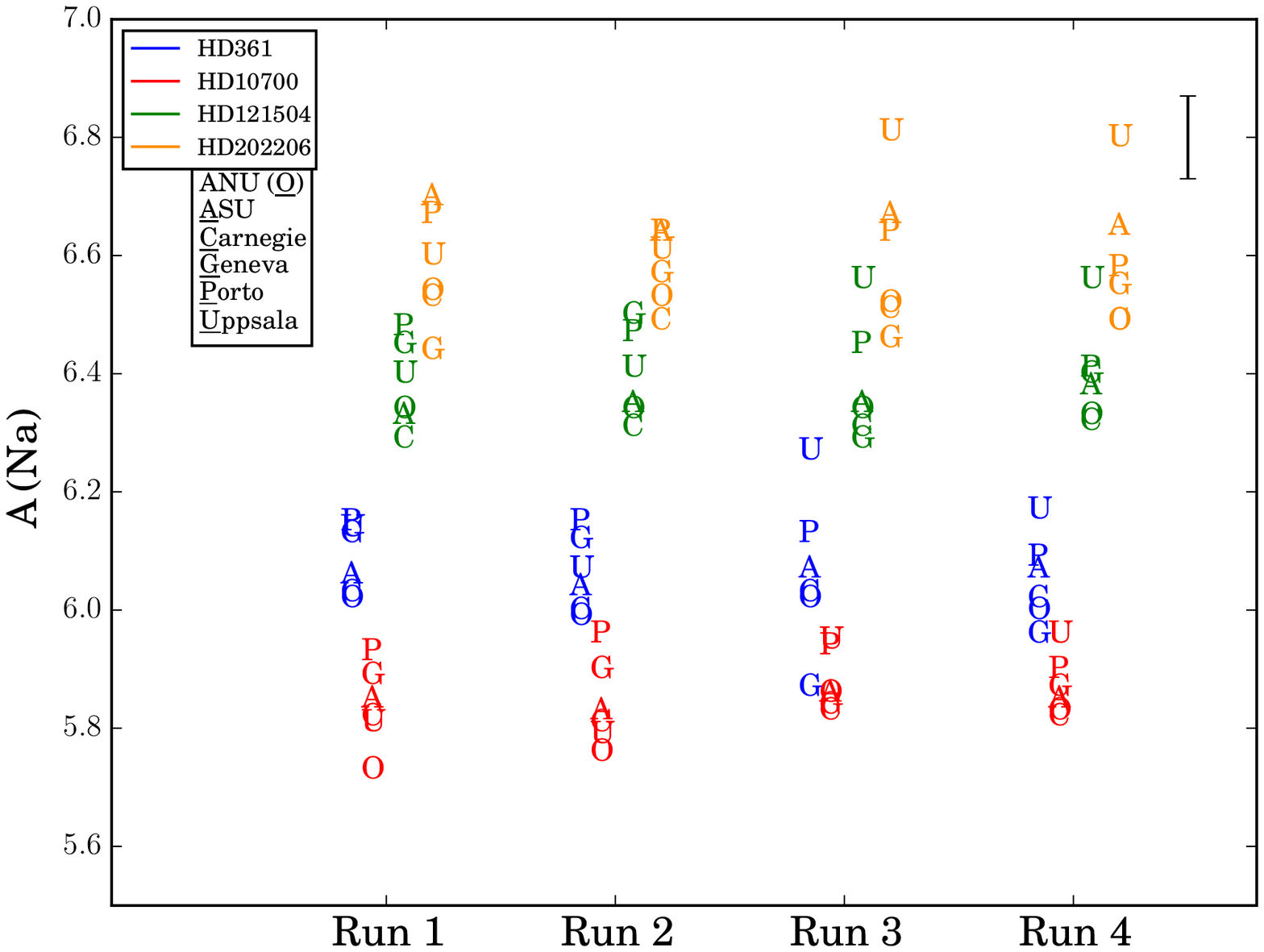} &  \includegraphics[width=95mm]{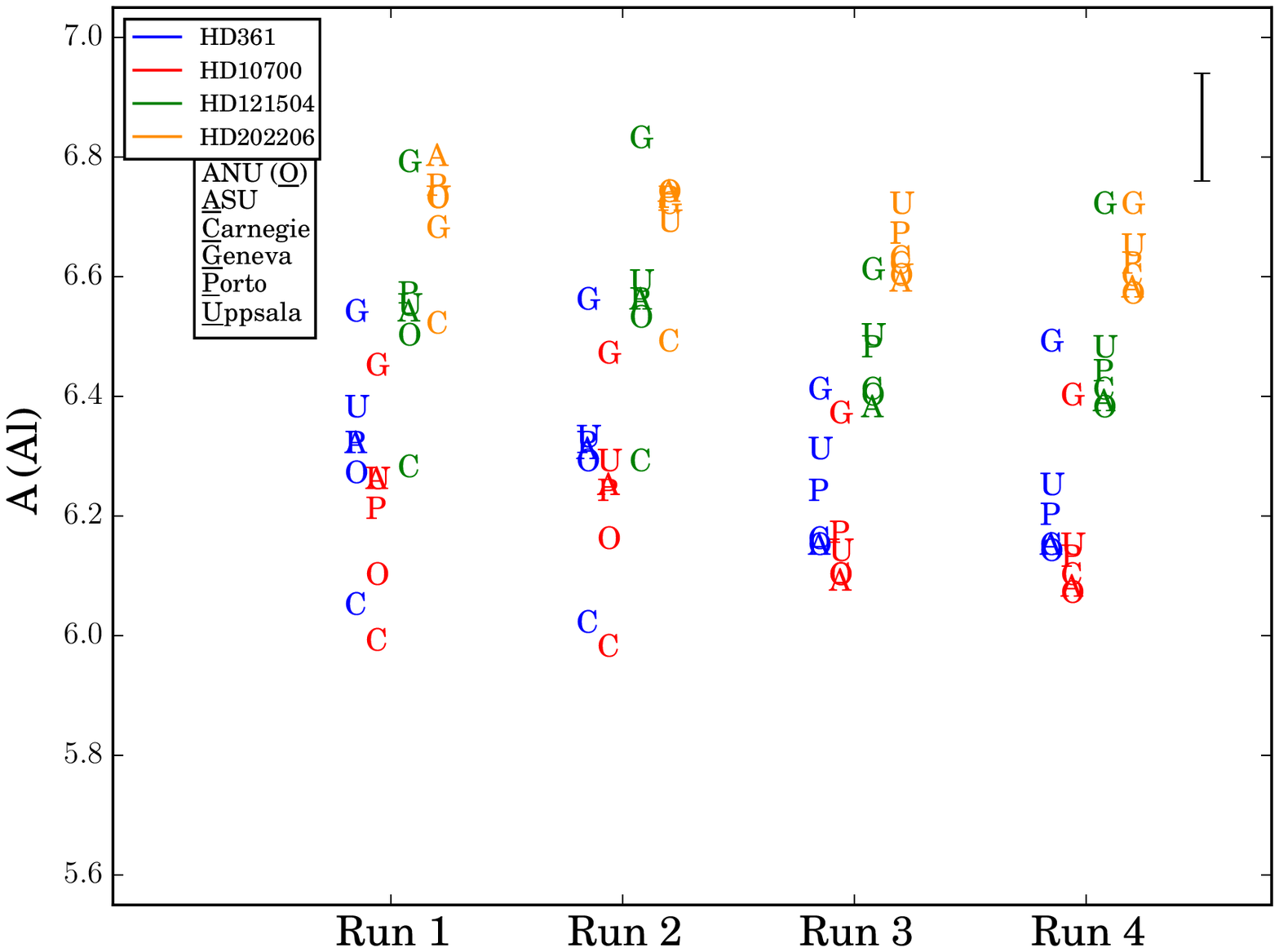} \\
\end{tabular}
\caption{Absolute abundances of A(C), A(O), A(Na), A(Mg), A(Al), and A(Si) as determined for all four measurements analyses (Run 1 is Autonomous, Run 2 is Standard Parameters, Run 3 is Standard Lines, and Run 4 is Standard Parameters and Lines). Colors are indicative of the star and each letter represents the group (see underlined letter in legend). A representative error is given in the top-right corner, taken as the average of the median individual errors per star.}\label{elems1}
\end{figure*}

\subsubsection{Odd-$Z$ Elements: Na \& Al} 
\label{s.oddz}
Per the last row in Figure \ref{elems2} and the discussion in Appendix \ref{a.oddz}, the odd-Z elements are improved when employing a standardized line list (Runs 3 and 4) in 6 out of eight cases. Looking at Table \ref{metric}, most of the maximum metric values occur during Runs 3 and 4. In addition, the mean and median normalized metrics for the two elements, similar to each other by $\le$ 0.03 and with a range of $\sim$ 0.15, are highest during Run 4. 

Both elements have two lines in our spectra (see Table \ref{standardlines}). Column 6 in Table \ref{standardlines} demonstrates that the EWs, specifically relevant to the CoG methods, for the first line of each element (6154.23\aaa and 6696.03\aaa, respectively) are similar to each other: 39.8 m\aaa and 38.1 m\aaa, respectively. However, the second lines (6160.75\aaa and 6698.67\aaa, respectively) are rather disparate: 58.4 m\aaa and 21.9 m\aaa, differing by a factor of $\sim$3. The weaker Al line may account for larger variation between groups.

\begin{figure*}
\begin{tabular}{p{8.5cm}p{2.0cm}}
  \includegraphics[width=95mm]{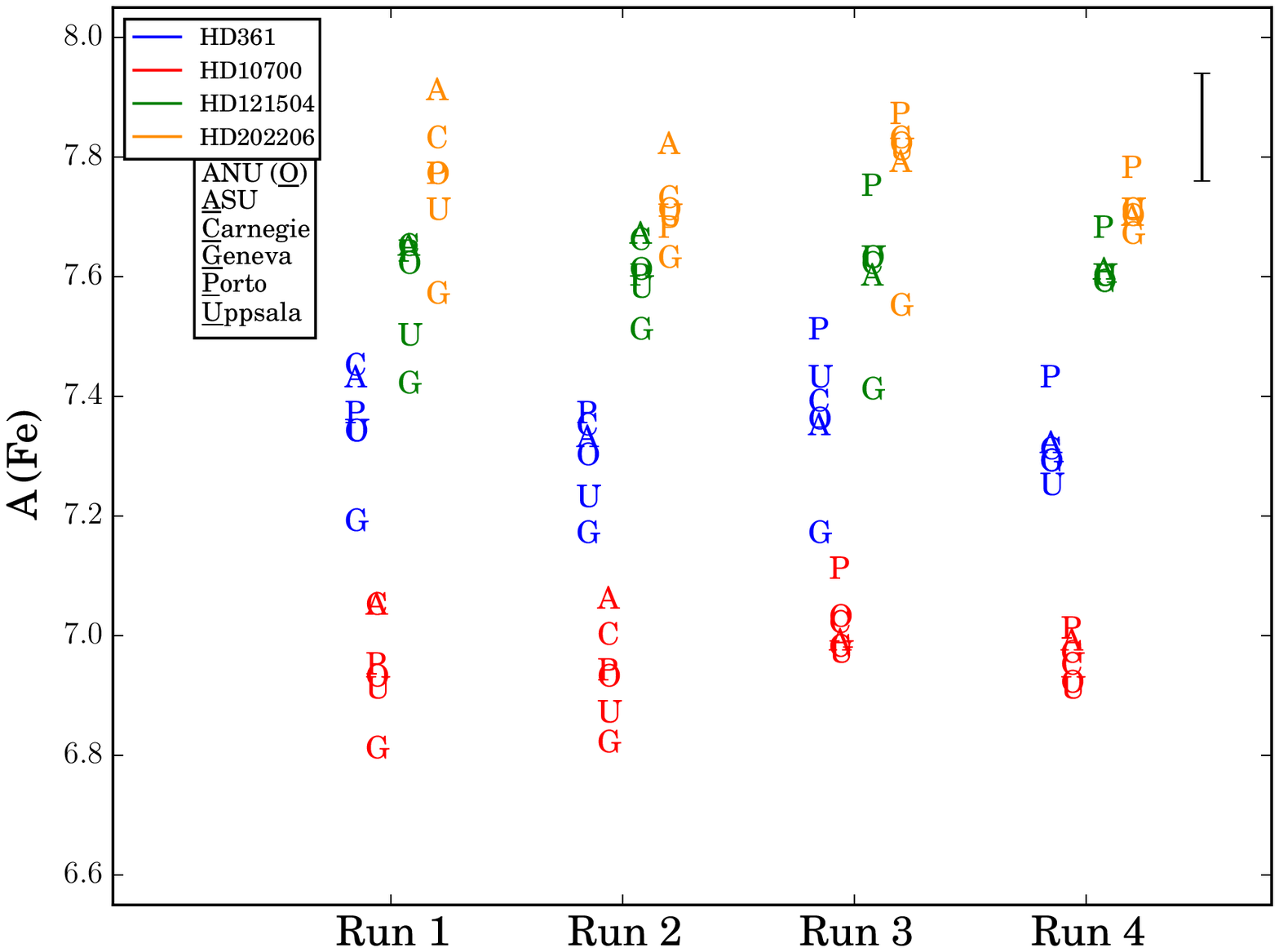} &   \includegraphics[width=95mm]{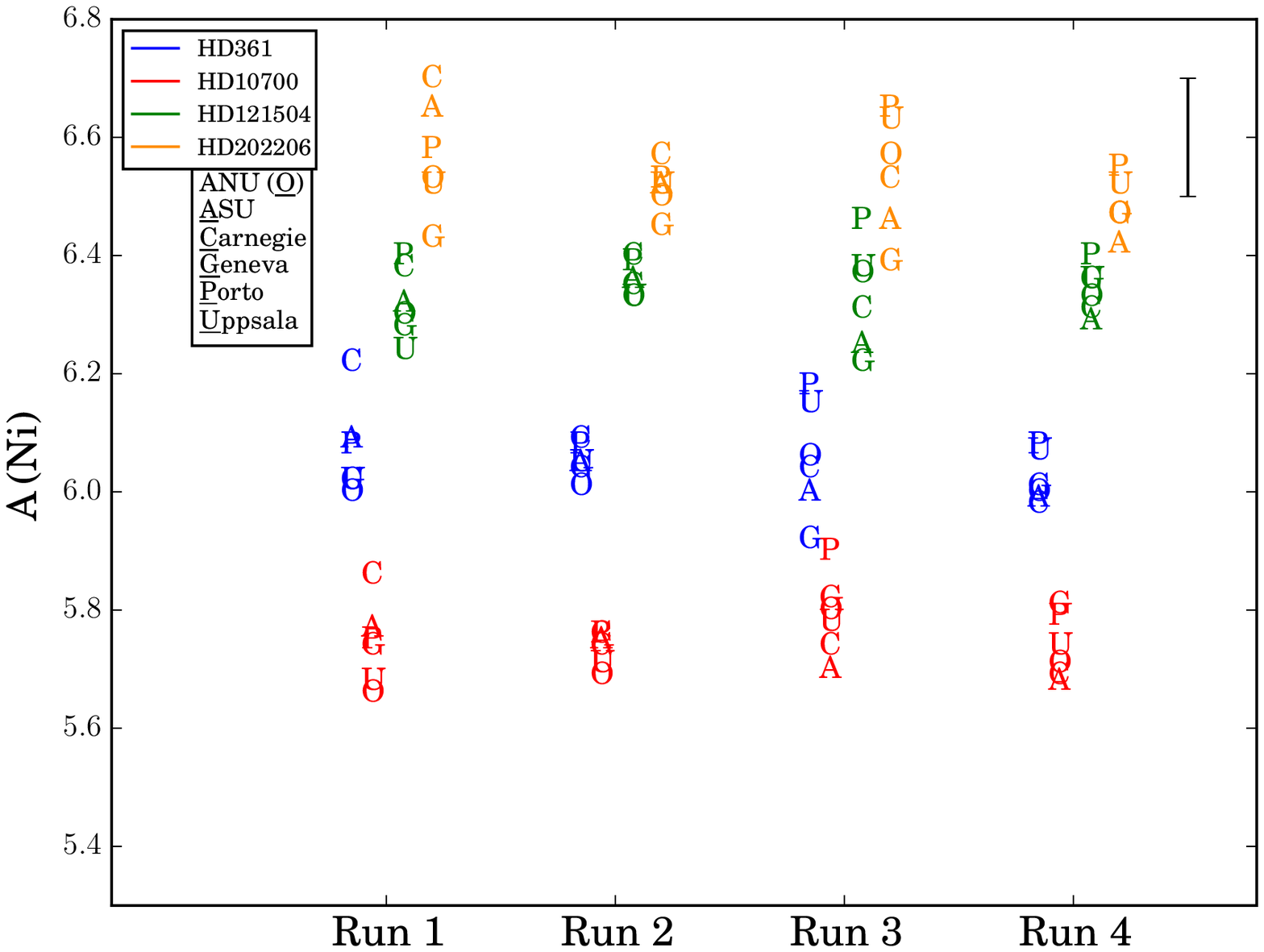} \\
 \includegraphics[width=95mm]{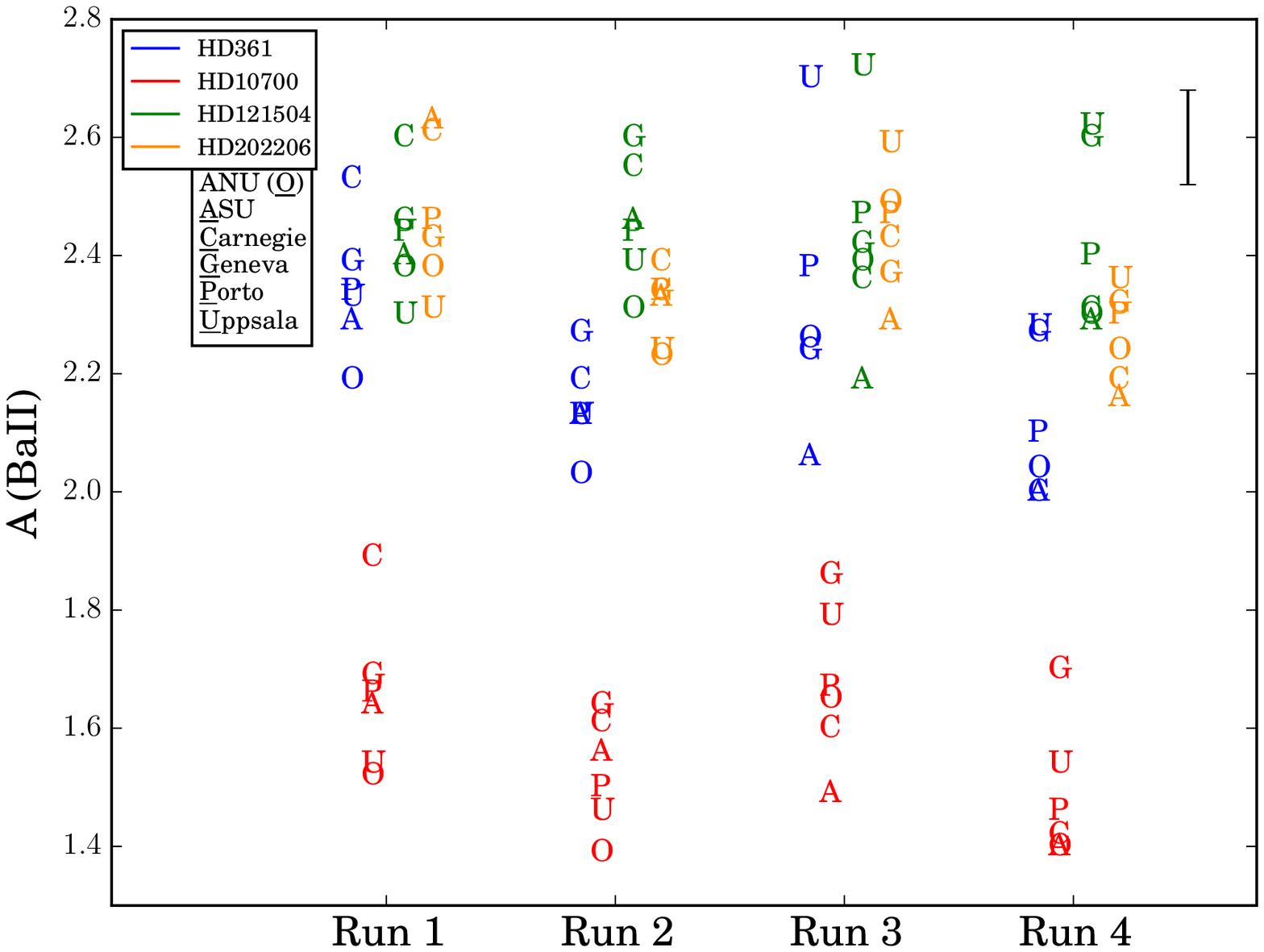} &   \includegraphics[width=95mm]{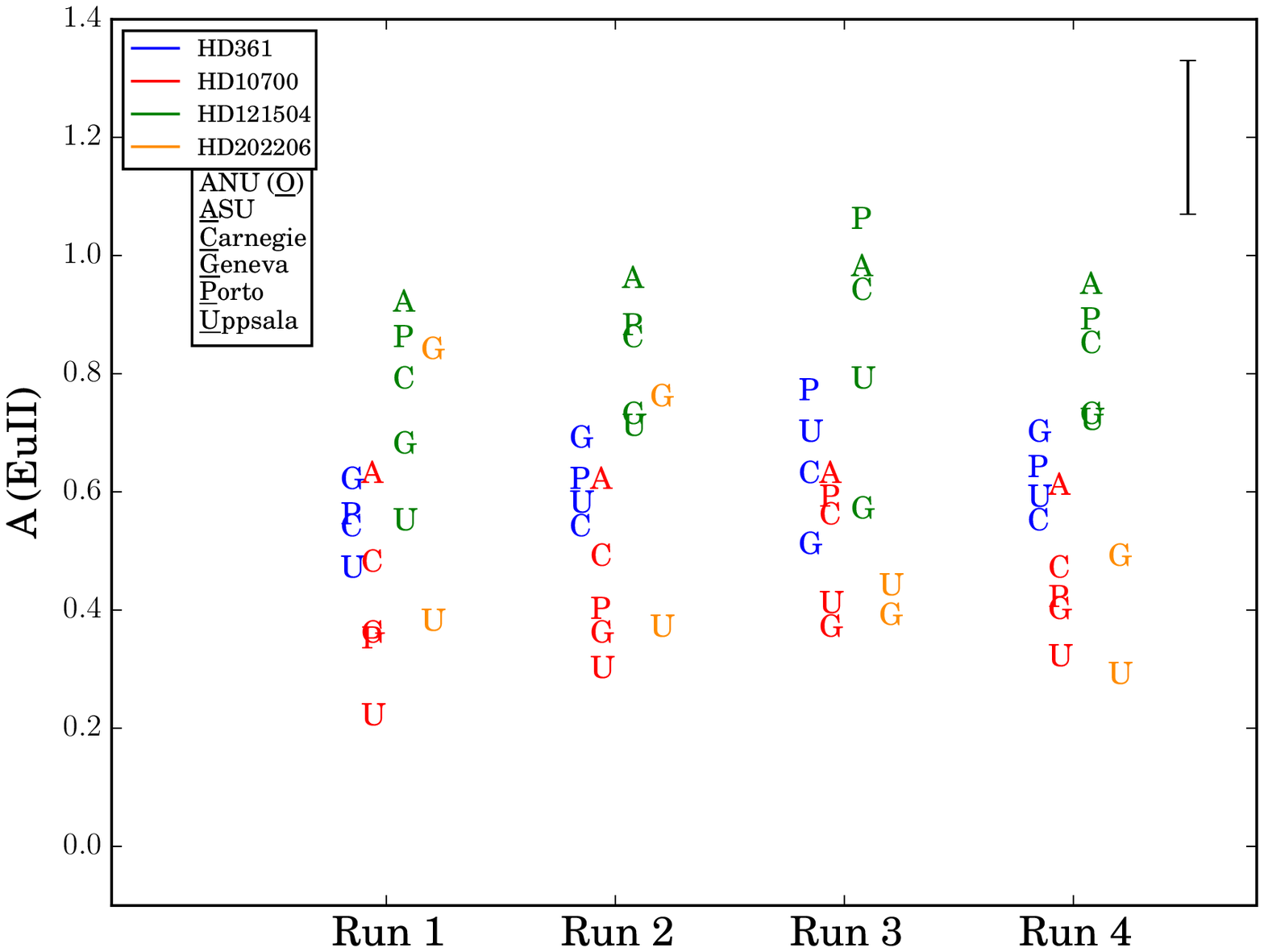} \\
\end{tabular}
\caption{\label{elems2} Similar to Figure \ref{elems1}, but for A(Fe), A(Ni), A(BaII), and A(EuII).}
\clearpage
\end{figure*}

\subsubsection{Iron-Peak Elements: Fe \& Ni}
\label{s.ironpeak}
For both A(Fe) and A(Ni) in the top-row Figure \ref{elems2}, we found that all of the groups measured the abundances rather consistently, to within error, for all the analyses. See Appendix \ref{a.ironpeak} for more details. Both elements benefit from the combination of standardized line lists and stellar parameters (Run 4), although Ni also improved when standardizing only the stellar parameters (Run 2). While no results were poor, A(Fe) measurements were most uniform during Run 4, with lower metrics for Runs 1--3. In comparison, the A(Ni) determinations were most consistent for Run 2, relatively good for Run 4, and substantially worse for Runs 1 and 3. The general A(Ni) similarity between Runs 2 and 4 is reflected when taking the mean and median of the normalized metric values. Namely, the mean and median, which were the same to within 0.01 for all runs, had the highest values for Run 2, followed by Run 4 (smaller by $\sim$0.07), then Runs 1 and 3 (smaller by $\sim$ 0.22). The total range was $\sim$0.3.

\subsubsection{Neutron-Capture Elements: Ba \& Eu}
\label{s.neutron}

For the neutron-capture elements shown in the last row of Figure \ref{elems2}, there was no clear indication which standardization, if any, was better when looking at the maximum stellar metrics. This issue was made worse by the difficulty that many groups had measuring the weak Eu II line. We discuss both elements in more details in Appendix \ref{a.neutron}. Unfortunately, analyzing the mean and median of the normalized metrics did not elucidate the situation, even when the Eu II values for HD~202206 were ignored. The variation in the mean and median determinations for each run made it clear that these distributions were not uniform: the mean values ranked the analyses from best to worst as Run 2, Run 4, Run 1, and Run 3 while the median indicated that the order should be Run 4, Runs 1 and 2 tied, and then finally Run 3. The only conclusion that can be drawn is that the standardized line list, on its own, is not conducive to making abundance measurements more uniform between groups.

\subsection{Line Analysis}
\label{s.elemlinebyline}
When trying to examine possible means of variation of elemental abundances, we initially hypothesized that the two biggest impacts on abundance determination would be the stellar parameters and the line list. We set up the four analyses to pinpoint any areas we believed introduced the greatest sensitivity. The differences between Run 1 and 2 and comparatively Run 3 and 4 should allow us to interpret the impact of stellar parameters. The differences between Run 1 and 3 should allow us to interpret the impact of the line lists chosen. Ideally, we imagined that there would be an increase in precision from Run 1 onwards; however this is not always the case.

After organizing the results we had a chance to examine the details of the individual line lists and the results that were calculated on a line-to-line basis. The line lists for each group for Runs 1 and 2 are included in Appendix \ref{a.tables} in Tables \ref{anulines}-\ref{uppsalalines}. The standard line list used in Runs 3 and 4 is in Table \ref{standardlines}. In this section we will look at a few selected lines to try and investigate the reasons for elemental variation, specific grouping of results, and improvements from one run to the next.

\subsubsection{Iron}
\label{s.ironlines}
With the multitude of iron lines used by each group, which ranged from 60 to 298 Fe I lines (Table \ref{models}), we decided to only compare the lines for one star: metal-rich HD~202206. Table \ref{felines} shows a few selected Fe I lines for this star. The atomic parameters input for each line are the excitation potential ($\chi_l$), the oscillator strength ($\log \,gf$), and the measured EW (m\aaa) when given. The final 4 columns in this table are the absolute iron abundance values determined for each of the four analyses. For Runs 3 and 4, the atomic parameters are those given in the standard line list, per Table \ref{standardlines}, which are not reproduced in Table \ref{felines} but implied via the two vertical lines. Throughout Table \ref{felines}, we have provided the average iron abundance for the 7 individual selected lines compared to the average over all iron lines, listed at the bottom.

Since the line list is the only thing standardized in Run 3, the individual line determinations can be used to understand the effect the atomic parameters have on individual lines. Additionally, it is possible to study the impact a single line may induce on the overall abundance measurement.  The EWs measured by the CoG groups are the same for Runs 1 and 2 (column 5 of Table \ref{felines}) and for Runs 3 and 4 (column 6 of Table \ref{standardlines}). This is not the case for the spectral fitting groups, Geneva and Uppsala, whose methods do not produce individual EWs. The variations in technique make group comparisons difficult when analyzing on a line-to-line basis.

We include a detailed discussion of the seven individual Fe lines in Table \ref{felines} in Appendix \ref{a.ironlines}. Overall, despite similar atomic parameters, there is a scatter in the abundance determinations for Runs 1 and 2 that is likely due to the variation in methods and models. The utilization of a standardized line list is apparent in Run 3, which typically sees larger line-by-line abundance measurements. The variation in measurements between the first two analyses and Run 3 can be greater than the $\pm$ 0.09 dex representative error for iron. This is evident in the averages and total-averages listed in the table. However, during Run 4, when both the stellar parameters and line list are homogenized, the abundance determinations are more consistent with the first two runs. As seen in Fig. \ref{elems2} (top-left) and in Table \ref{metric}, the stars are most similar between groups in Run 4. 

In general, the determinations between Runs 2 and 4 appear to be more consistent with each other to within error as compared to the Runs 1 and 3. This trend hints that stellar parameters are perhaps more influential in determining consistent abundances than standardizing the line list, at least for iron. The only remaining difference between these runs is the adopted [Fe/H] during the analysis, which was left as a free parameter and therefore varies between groups. The [Fe/H] content affects the temperature and electron density structure (important for stronger lines), the molecular equilibrium (important for C and O, as well as background H- opacity), and ionization equilibrium (important for neutral minority species, which may also depend on the H- opacity). Therefore, the [Fe/H] measurement itself may be an important missing ingredient for Runs 2 and 4.

\subsubsection{Elements Other than Iron}
\label{s.otherlines}
When examining the abundances of elements other than iron it is important to note that most elements have very few absorption lines. The exceptions are silicon and nickel, which both show small ranges and great correlation between groups, as seen in Figures \ref{elems1} (middle-right) and \ref{elems2} (top-right). Similar to iron, here we briefly discuss the line-by-line breakdown of a select group of elements for HD~202206. Table \ref{otherlines} shows the line-by-line analysis of carbon, magnesium, sodium, aluminum and barium, with a more detailed discussion in Appendix \ref{a.otherlines}.

For the carbon lines, we find that there is a $\sim$0.1 dex decrease in line-by-line abundance measurements when implementing the standardized line list. This indicates that the choice of carbon lines may outweigh the variation in stellar parameters or EW determinations. In general, while carbon abundance patterns emerge from the line-by-line comparison, they are not consistent between lines or analyses. The lack of obvious trend was also true for the Mg and Ba II lines. Analysis of some Si lines imply that the EW measurement may be more significant than the oscillator strength, although that was not clear for all lines. When measuring Al, even though the groups agree on the average abundance for each star, they are systematically offset. Unfortunately, it's not clear in which direction they are offset. The solution to the Al abundances may be that we need to agree on which $\log (gf)$ values are best, which would give consistent results with each other.

\section{Literature Abundances}
\label{s.litcomp}
We chose the stars analyzed in this Investigation for a number of reasons, one of those was to enable a comparison to literature values. First, we discuss the {\it Hypatia Catalog}, a compilation of stellar abundance literature sources and the update it has undergone since \citet{Hinkel14}. Second, we will directly compare the measurements from \S \ref{s.elements} to the overall {Hypatia 2.0} results as well as the individual data sources.

\subsection{Hypatia Catalog 2.0}
\label{s.hyp}
The {\it Hypatia Catalog}, originally introduced in \citet{Hinkel14}, is an amalgamation of stellar abundance data sets that have been renormalized to the same solar scale and homogenized to reduce systematic variations between techniques. The original catalog contained abundances for 50 elements within 3058 main-sequence (FGK-type) stars all within 150 pc of the Sun. Since then, {\it Hypatia} has been improved upon in multiple ways, which we are now calling {\it Hypatia 2.0}.

\begin{figure*}
\begin{center}
\centerline{\includegraphics[height=2.5in]{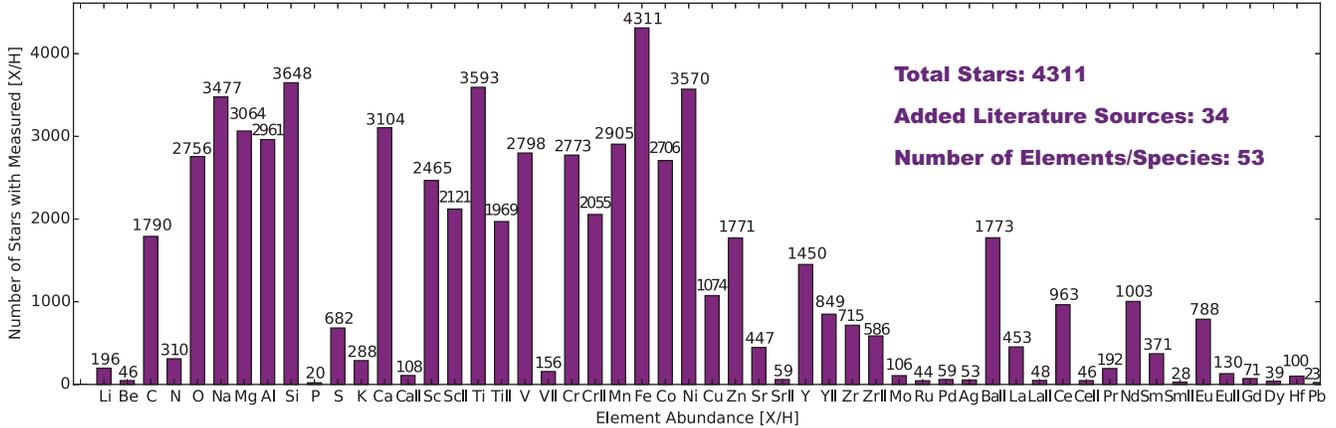}}
\end{center}
\caption{
Number of stars in the updated {\it Hypatia Catalog} 2.0 with measured abundances
for 53 different element species for stars all within 150 pc of the Sun. The catalog now includes an additional 1253 stars from 34 data sets, making the total 4311 stars.  
}\label{hist}
\end{figure*}

An additional 34 data sets have been added into {\it Hypatia} (see Table \ref{update}) which increased the stellar count by 1253 stars, making the total population 4311 main-sequence stars within 150 pc of the Sun. We have also included three new element and element-species, namely palladium (Pd), silver (Ag), and singly-ionized samarium (Sm II). A histogram showing the individual elements and the total number of stars in {\it Hypatia 2.0} for which they have been measured is shown in Fig. \ref{hist}. The {\it Hypatia Catalog} will continue to be upgraded such that new data sets are included as they are released and the back-end functionality expanded. In subsequent papers, we will increase the tenths value of the version number in the former case and the ones value in the latter case, or if the two were simultaneous. 

While the reduced abundance determinations, similar to those from \citet{Hinkel14}, will be made available on Vizier, we are excited to announce the creation of the Hypatia Catalog Database. The full multidimensional data found within the {\it Hypatia Catalog} will be available at \url{www.hypatiacatalog.com} as of Fall 2016. The online database will include catalog and abundance data manipulation, a variety of solar normalizations, planetary information (when available) from the Exoplanet Orbit Database \citep[\url{www.exoplanets.org,}][]{Wright:2011p3231}, as well as a graphical/statistical interface. The community be able to take full advantage of these compiled resources, either within the interface or by downloading the data through a terminal.

The variety in stellar abundance techniques is nothing if not impressive. Over multiple decades, a number of different instruments with various resolutions, S/N, models, and programs have been utilized in order to determine elemental abundances. We go into more detail regarding the differences between observations and data reduction techniques in Appendix \ref{a.field}. However, the problem arises when data sets are compared to one another. While data may be internally consistent, for example by measuring a unique solar abundance spectrum, it is unclear how those values will compare to data sets from other methods. As it stands, much of the abundance information that we get from different groups for the same stars often do not agree within error. 

A part of the analysis of the {\it Hypatia Catalog} involved understanding the {\it spread} in the data -- or the range of abundances determined by different groups but for the same element in the same star. 
Using {\it Hypatia 2.0}, we find there are 1244 stars that are so discrepant in their [Fe/H] measurements between groups, or with a spread above typical error, that it was impossible to homogenize them (see \citealt{Hinkel14}). Interestingly, the analysis of the spread was taken {\it after} all of the data sets were renormalized to the same solar scale! The average spread in [Fe/H] for all of {\it Hypatia 2.0} is 0.18 dex which is nearly twice the standard error associated with [Fe/H]. Not even iron, the most frequently measured element in the field, can be well agreed upon between groups. It is because of this that we believe it is time to better understand the abundance measurement techniques that are employed by different groups, such that we can understand these systematic offsets and variations. 

\subsection{Comparison to Literature}

The purpose of our Investigation is to test the effects of the various methodologies on the actual abundance measurements, in particular the determination of stellar parameters, the employed line list, the measurement of EWs, and the stellar atmospheres/abundance codes used by different groups. 
We cannot perform a controlled literature comparison to determine how other techniques or even other data sets/observations may vary, but we can compare the average values and spreads from our experiment to those of the literature as a whole. This enables us to estimate how much of the measurement variation could be accounted for by the particular methods tested here. A smaller spread in the Investigation results implies that some factor not present in any of our methods is affecting the values found by certain groups.

For the purpose of comparison and brevity, we will primarily focus on the averages and spreads of C, O, and Fe from the Investigation with respect to the those reported in the {\it Hypatia} catalog \citep{Hinkel14}. Discussion of the other seven elements can be found in Appendix \ref{a.litcomp}, which are also summarized at the end of this section. While the abundance measurements were defined both with and without outliers, as described in \S~\ref{s.elements}, comparisons here will be without outliers unless noted otherwise. In Table \ref{littable}, all known (at the time of this writing) literature sources are listed with an associated {\it Hypatia} value, which is the median of the A(X) values by the groups who measure that element per \citet{Hinkel14}.
Unfortunately, HD~121504 has no reported abundances in the literature for elements besides Fe and is not discussed. 
Carbon determinations behave similarly for the Investigation and the literature as a whole. For the unconstrained Run 1 comparison, the results of the Investigation show a larger range in HD~361 as compared to {\it Hypatia}. However, the {\it Hypatia} ranges are larger for HD~202206 and HD~10700. Notably, the {\it Hypatia} values for the iron-poor star yielded a spread of 0.78 dex and average of 8.17 dex while the Investigation's range was 0.13 dex. In the literature more than half of the spread is attributed to one group, \citet{Takeda:2005p1589, Takeda:2007p1531}. Removing this group results in average A(C) = 8.28 dex for the literature, a much more substantial discrepancy compared to the Investigation. \citet{Takeda:2005p1589, Takeda:2007p1531} use an EW measurement routine and abundance finding code that is different from any of the other groups who participated here. Results for Runs 2-4 are similar to literature averages for HD~361 and HD~202206. For HD~10700 the average abundance found in Run 4 is 8.01 dex, with a larger range of 0.25. Otherwise, ranges tend to be fairly consistent across analyses.

Oxygen values are much more discrepant with those found in the literature, with average A(O) differing by as much as 0.4 dex from the {\it Hypatia} values. For Runs 1 and 2, HD~10700 finds the best agreement, differing by only 0.08 dex in Run 1. HD~202206 sees the largest difference in these analyses, at A(O) = 9.24 dex vs. 8.82 dex. The difference for HD~361 is intermediate at 0.22-0.24 dex. There is a correlation between iron-content of the star and lack of agreement with literature values, though the sample is too small to be called robust. HD~10700 fares more poorly in Runs 3 and 4. From a difference smaller than the range, the disparity in A(O) values rises to 0.24 dex in Run 3. 

Imposition of the list of oxygen lines chosen for Runs 3 and 4 has an interesting effect. The fixed line list used only the 6156.8\aaa OI line while the 6300\aaa line is affected significantly by a blend with a NiI line \citep{Nissen14}. The triplet at 7774\aaa was outside the wavelength range of the spectra. Standardizing the oxygen lines has very little effect on the other two stars, leaving HD~202206 essentially unchanged and improving the agreement of HD~361 slightly. The spreads in these two stars are much larger than the {\it Hypatia} values for Runs 1 and 2. The spread for HD~10700 is the smallest in all analyses and smaller than the {\it Hypatia} value of 0.33 dex. Consistency improves for HD~361 in Runs 3 and 4, but remains poor ($\sim$0.6 dex) for HD~202206 in all cases.

Iron abundances are fairly consistent between analyses, with maximum variations of 0.07 to 0.08 dex for all three stars. The Investigation's A(Fe) values for all three stars are similar to the literature averages and, 
in all cases, the {\it Hypatia} values fell within the range of the Investigation spreads. HD~361 is an interesting case. All five surveys from the literature use the same telescope and instrument, the 3.6m at La Silla with the High Accuracy Radial velocity Planet Searcher (HARPS). The surveys also use the same set of Fe lines, model atmospheres (ATLAS9), and abundance finding code (MOOG). The EWs are determined using IRAF's {\sc SPLOT} routine or ARES. Four out of the five surveys \citep{Adibekyan12, Bertran15, delgado-mena_2010_aa, Neves:2009p1804} determined solar-renormalized values of A(Fe) = 7.35 dex while \citet{GonzalezHernandez:2010p7714} report A(Fe) = 7.391 dex. The only difference discernible in the methodology is the use of differential analysis by \citet{GonzalezHernandez:2010p7714}. The other groups all use the iron ratios from \citet{Santos:2004p2996}, where the \citet{Anders:1989p3165} solar abundance scale, which has since been superseded by much different solar abundances, was used for normalization, although A(Fe)$_{\odot}$ = 7.47. All groups agrees very well with this study's A(Fe). The same pattern appears in HD~202206, with the same four groups reporting lower A(Fe) abundances as compared to \citet{GonzalezHernandez:2010p7714}. Again, all five groups agree well with the abundances found by this study. Only three of five groups have abundances for HD~10700, not including \citet{GonzalezHernandez:2010p7714}, which have measurements similar to this study's values and the {\it Hypatia} value. 

\vspace{3mm}

For the most part, the abundances values found in the Investigation are consistent with the median of the literature values found in the {\it Hypatia Catalog}. At low metallicity, C diverges for all analyses but the spread is very large for both the Investigation and literature. The difference is not, therefore, as significant as it might first appear, emphasizing the care that must be taken in the comparisons. When analyzing the spreads, the Investigation captures the range of variation seen in the literature. We find that O is the only case with radically different abundance values and the difference between the Investigation and {\it Hypatia} increases with metallicity. Imposition of a specific line list significantly affects measured O abundances, improving the high iron-content star, worsening the low, and leaving the approximately solar-iron star mostly unaffected. Given the problematic nature of O lines, this is unsurprising. As discussed in Appendix \ref{a.litcomp}, Na, Al, Mg, and Si, are all very consistent. The spread in Al dropped to much lower than the {\it Hypatia} spread for Al in the two higher iron-content stars with a fixed line list, suggesting that Al is sensitive to the lines used. Magnesium has a similar amount of variation to the literature except for the low iron-content HD~10700. Comparison to literature allows us to deduce that Fe is sensitive to the technique, namely a differential line analysis or a bulk solar abundances scaling, which was a factor we do not test. The same may be true to a lesser extent for Na, Al, Si, and Ni, which merits a detailed follow-up. Barium appears to be particularly sensitive to stellar parameters but robust against the line list used.

\section{Discussion}
\label{s.disc}

It has been the goal of this research to better understand the abundance measuring techniques between groups and attempt to resolve the inherent issues. However, we wished to make it clear that it has not been our objective to try to determine which group determines the most ``accurate" stellar abundance, or who is closest to calculating the actual amount of an element within a stellar photosphere. While we agree that this is an important issue, we believe that it is necessary to understand what components of a technique give rise to the observed variations. For now our focus has to be on the precision of the techniques, not the accuracy of their results. This is the first step that needs to be taken such that similar issues don't arise again down the road. And if nothing else, once made to agree, we know if the stellar abundances do not reflect a true representation of the element within the star, these results can be scaled such that they are accurate without any loss of precision.

\subsection{Error Analysis}
\label{s.error}

A major facet of the Investigation was to include participants with a relatively wide variety of abundance determination methods. Not only did we want to see how the techniques compared to each other, but also the ways in which each of the methods were able to adapt to the restrictions placed on the abundance measurements. In this way, we were able to get a more in-depth understanding of the more commonly used techniques, including both their strengths and their limitations. The plots in Figs. \ref{elems1}--\ref{elems2} illustrated the overall abundance variation seen between the groups for all stars and analysis standardizations. However, by employing the metric of data similarity (Table \ref{metric}) and the respective error bars for each element, we glossed over the details of the individual uncertainties. 

In Appendix \ref{a.people}, each group described not only their technique for measuring stellar abundances but also the way in which their errors were calculated. Overall, the four groups who employed the CoG method (ANU, ASU, Carnegie, and Porto) had a similar uncertainty derivation: combining the errors calculated using a standard deviation of the abundances determined by a line-to-line basis and the effect of stellar parameter changes. In comparison, the spectral fitting groups had vastly different techniques for calculating error. For example, Geneva employed a weighted dispersion and Uppsala used a variety of integrated tools to contrast the data with the synthetic spectrum. All of the individually reported errors can be found in Tables \ref{stellarparams1}--\ref{run4abs}.

To more clearly evaluate the error determinations for the element abundances, we have compiled the error calculations in Table \ref{errors} in a variety of different manners. For example, the first set (rows 1-6) shows the average of each group's uncertainty determinations for all four stars over all four analyses with respect to the 10 elements (columns). The second set (rows 7-10) reports errors for each run, averaging together the error bars from all of the groups' measurements for all four stars. The last set (rows 11-14) gives errors for each star, taking the mean of all the uncertainties as reported by every group in every run. Finally, the last row shows the average from all 96 calculations, or the four stars across four analyses by six groups. These last values are similar if not the same as the representative error bars, calculated using both medians and means as discussed in \S \ref{s.results}, given in Figs. \ref{elems1}--\ref{elems2}. 

We see from the first set of error compilations that ASU has some of the overall lowest uncertainty reports for 9 of 10 elements. While this might imply that ASU's abundance determinations are more accurate, it may alternatively be the case that their uncertainties are simply underestimated. Porto's uncertainties tend to be relatively low, but not in all cases. There were few particularly high or low error determinations for ANU, Carnegie, and Uppsala, though Uppsala's tended to be above average.

Geneva report the highest errors in seven cases, where six of them were by a wide margin. In the cases of O and Eu II, Geneva's large errors arose because each element only had one line to measure. Therefore, the error did not come from the standard deviation, but directly from the singly fitted line. In these cases, the single-line error is estimated using the covariance matrix by the least square algorithm, and may be overestimated. In addition, the version of iSpec used in this study is not capable of scanning a small range of wavelength-space in order to determine the peak of the absorption line or do small radial velocity corrections to match the line with the synthetic spectra, as is possible in SME. Therefore, it is susceptible to imperfect wavelength calibration, which clearly affects the determination of individual abundances, as was the case for Ni, Eu II, and to a lesser degree Fe. We discuss these issues within the field in general in \S \ref{s.cali}. We also noted in \S \ref{s.parameters} that the Geneva measurements exhibited opposite trends in their stellar parameter determinations between Runs 1 and 3 as compared with the other groups. It is most likely that the mismatch in wavelength calibration, especially when implementing the standardized list, is responsible for the notably different calculations.  

Also discussed within \S \ref{s.parameters} were the size of the uncertainties for the stellar parameters calculated by the Uppsala group. These errors were not reflected in the abundances, as discussed in Appendix \ref{a.people} and per Table \ref{errors}. However, Tables \ref{stellarparams1}--\ref{stellarparams3} demonstrate that they are roughly twice those determined by the other groups. The uncertainties determined by the Uppsala group are consistent with other line-by-line abundance uncertainties and with the parameter uncertainties that were determined for non-solar type stars. Their stellar parameter uncertainties are essentially dominated by the line-to-line scatter which is caused by errors in $gf$ values as well as modeling shortcomings, missing blends, and continuum uncertainties. By adopting astrophysical $gf$ values, some of these errors would cancel and the precision, but not accuracy, would have improved. This would have resulted in error bars more closely aligned with the other groups. As is currently given, however, the values and their uncertainty are more representative of what is produced for the Gaia-ESO survey, which emphasize accuracy over.

In the second set of averages given in Table \ref{errors}, we find that for the majority of elements, the average stellar abundance error is relatively consistent when cycling from Run 1 through Run 4. There are, however, a few exceptions. For example, the errors for Mg, Al, Si, Ni, and Eu II get worse when implementing the standardized line list, although to varying degrees. However, the opposite was true for C and more predominantly O, where the average errors were halved in Runs 3 and 4 as compared to Runs 1 and 2. 

The results for Eu II, O, and to some degree Mg may be related to the method in which errors are calculated for elements with one absorption line by each group. The tables for Runs 3 and 4 (Tables \ref{run3abs} and \ref{run4abs}) illustrate how Geneva was not able to calculate the A(O) abundance for any star while ANU and Porto had null results for HD~361 and HD~10700 during Run 4. Also ANU, ANU, and Carnegie had sporadic, if any, results for A(Eu II). Similarly, Porto did not report errors, only abundances, for A(Mg) and A(Eu II). It is clear that many of the groups had difficulty with respect to these single-line elements, since omissions did not occur for any other elements (save that ANU was unable to measure carbon in HD~10700 for any run). However, the uncertainties given in Tables \ref{run3abs}, \ref{run4abs}, and \ref{errors} do not always reflect the lower confidence associated with fewer lines, especially when the lines are weak. The case for the 6156.8 \aaa oxygen line, which had a EW of 4.1 m\aaa, is an excellent example. We therefore make note that special precautions should be made, both when measuring and determining the uncertainties, for elements with only a single absorption line. This is an issue that should be addressed within the community and implemented with more transparency 
and reproducibility
within stellar abundance techniques. 

Finally, for the last section in Table \ref{errors}, with respect to the average errors associated with the individual stars, the most metal-poor star, HD~10700, typically had the largest error bars. This makes sense given the smaller and weaker lines within the spectrum. The other relatively metal-poor star, HD~361, also had slightly larger associated errors. The converse was not always true for HD~202206, which we expected to achieve the lowest uncertainties given that it is metal-rich. However, the average error for A(Fe) was largest for HD~202206.

\subsection{Wavelength Calibration}
\label{s.cali}

As mentioned previously in the paper, we found a significant mismatch in the provided stellar spectra that resulted in a misalignment between spectral features and wavelength numbers below 5050 \aaa. Prior to sending the data to the participating groups, the spectra was tested using the ASU methodology and the results were comparable to other data and literature sources. The ASU team has also looked into the typical wavelength calibration from other data/telescopes and found the provided spectra to be similarly, if not better, calibrated than other (published) sources. In addition, per our analysis of the abundances produced here with respect to the literature sources and their uncertainties (\S \ref{s.litcomp} and Appendix \ref{a.litcomp}), we find that all results are rather standard for the field. 

However, while the data provided here was ``typical," the wavelength calibration was not perfect. In general, the EW/CoG methods will find the appropriate line, due to either automatic procedures that have a margin of error in which they can search for the correct spectral feature or because many groups tend to measure the EWs manually. While SME has an option to do small radial velocity corrections, that margin is limited. The version of iSpec used in this study does not implement that kind of correction, so it was more affected by the imperfect wavelength calibration than the rest of techniques.

Rather than see spectral fitting techniques, which have a number of benefits as compared to the CoG techniques, at a disadvantage, it makes sense that proper wavelength calibration should become more of a priority within the field. Offset spectra could affect the shape of the absorption lines, especially when taking into account convective shifts (see Appendix \ref{a.field}). In fact, the overall impact that wavelength calibration may have on spectroscopic analysis is not clearly understood and could result in some of the systematic scatter that has been noted here and throughout the field. Therefore, although not specifically tested here, we believe that more precise wavelength calibration could benefit all methodologies for determining stellar abundances.

\subsection{Spectral Fitting vs. Curve-of-Growth Techniques}

Over the course of our analysis, we compared different levels of standardization throughout the analyses, the absolute abundance results between elements, variations seen in stars of assorted spectral types, and calculations from separate abundance techniques. With respect to the stellar parameters, namely Figure \ref{params} and Tables \ref{stellarparams1}--\ref{stellarparams3}, we did not find any correlations between the groups who used spectral fitting (Geneva and Uppsala) and those who used the CoG technique (ANU, ASU, Carnegie, and Porto). In fact, in terms of stellar parameters, the two spectral fitting groups tended to be at the opposite end of the ranges both in terms of values calculated and size of errors reported. No systematic difference was found between groups using the MARCS and ATLAS9 atmosphere models. While EW measurements did affect the results for the CoG techniques, there was no systematic difference between different EW measuring tools.

There were a few instances of similarity for the spectral fitting groups when determining the stellar abundances. For example, both Geneva and Uppsala were often at the extreme ends of the spread, either the highest or lowest, when measuring A(Mg) in general. In addition, both groups had the largest determinations for most A(Al) calculations compared with the other groups, while they had lowest results in general for A(Fe) and A(Eu II). However, we note that (1) the individual and representative error bars associated with these elements often overlap with the CoG groups, (2) all measurements for Eu II are inconsistent (or unmeasurable) given the weakness of the line, and (3) there are counterexamples prevalent in each of the above elements. Therefore, we are left to conclude that correlations between the spectral fitting and CoG techniques are more likely coincidental and are not strongly related to the differing methods. 

\subsection{Varying Line Lists}

As discussed in \S \ref{s.elemlinebyline}, during both analyses where the groups could choose which lines they measured, a variety of different line lists were used, especially with respect to iron (see Tables \ref{anulines}-\ref{uppsalalines}). However, iron is one of the most important factors when determining stellar abundances, since it influences stellar parameter calculations and the measurements of other elements.  With so few overlapping lines between the groups in this study, it was hard to pinpoint areas of variation. 

Many line lists used atomic parameters from laboratory or observational sources, such as listed VALD \citep{Kupka00}. However, we found that the atomic parameters, particularly the value of the oscillator strengths, were correlated with abundance discrepancies. The issue with atomic parameters is well known \citep[e.g.][]{Thevenin1999, Doyle13} and often corrected by via a differential analysis technique. This involves using a benchmark star, typically a solar spectrum, to redetermine the $\log(gf)$ values. In this way, an inconsistent atomic parameter can be corrected by reconciling an over- or under-estimated abundance calculation in the solar spectrum, which is then removed from the final result. Differential analysis is a common practice, as seen in \citet[][Table 3]{Hinkel14} and Table \ref{update}, but can be difficult to compare with results that do not normalize to the same solar abundance scale. 

\citet{Sousa14} tackles the issue of a homogenous line lists specifically with respect to stellar parameter calculations. They found that a consistent line list would improve the accuracy of parameter results, but would require a selection of the ``best" iron lines and agreement within the community. With a variation in the number of lines, the particular lines, and the  atomic parameters for Runs 1 and 2, we dealt with many free variables in the calculation of stellar parameters. As shown in the last column Table \ref{models}, the number of Fe I lines varied from 60--298 between the six groups. Yet, in general, the ranges in stellar parameters became worse when implementing the standardized line list, as compared to the autonomous run and shown in Tables \ref{stellarparams1}--\ref{stellarparams3}. However, as discussed in \S \ref{s.elements} and shown in Tables \ref{run1abs}--\ref{metric}, the abundances improved when standardizing the stellar parameters and, to a lesser extent, the line list. Therefore, it would appear that perhaps a line list could be chosen such that the stellar parameters are consistent and precise, which would therefore allow the elemental abundances to be more consistent between groups, such as described in \citet{Smiljanic14}. \citet{Pagano16} test the amount of variation in stellar parameters and final abundance determinations as a function of the number of Fe lines in the linelist using spectra from the same survey used in this work. They find that above $\sim$70 lines parameters and abundances vary by less than the error, with little improvement at higher numbers of lines. For smaller linelists the determinations could change substantially with number of lines and with lines chosen.

\subsection{Consistency}

In this study, we have attempted to illuminate some of the issues plaguing the stellar abundance community and determine a solution to rectify those differences. We identify quantities that give rise to significant variation and recommend directions for further research. We do not provide specific recommendations of lines, techniques, or parameters to use. Determinations of best practices require a great deal of further investigation.

In a very broad sense, we saw that standardizing either the stellar parameters, the line list, or both typically, but not universally, minimized the discrepancies between groups. Whether a fixed line list or fixed stellar parameters were more beneficial changed on an element-by-element basis. By directly comparing the values for measured EWs on a line-by-line basis, it was found that some discrepancies could be attributed to differences in the EW measurements. However, some lines resulted in discrepant abundances even when the EWs were similar or identical. In such cases the atomic constants, in particular oscillator strength, were often found to be the culprits. Therefore, it is our recommendation that a standard group of benchmark stars (for example \citealt{Jofre14, Blanco14a, Heiter15a, Jofre15}) be established to calibrate abundance finding techniques in order to produce consistency in stellar abundances regardless of measurement technique. Any group could adjust their procedure to reproduce the benchmark EWs (if measured), stellar parameters, and abundances. The community may then directly compare the group's abundances using the benchmark technique, which should be consistent across all groups, to the group's preferred method if they believe it to be more accurate.  It would also benefit the community to agree upon the best elemental absorption lines and atomic parameters, which would also yield more reliable results. In this way, we may systematically reduce the variation between methods. 

While we believe our recommendation is reasonable and consistent with our results, we recognize that this will not solve all of the inherent issues. For example, it was our hypothesis at the beginning of this endeavor that stellar parameters and line lists would account for most of the discrepancies, and increasing standardization of these components would improve the agreement of the results. Namely, if identical stellar parameters and line lists were used, the abundances would be very similar and the results in Run 4 would be best. However, we saw throughout the paper that this was not always the case and particular methods or elements were adversely affected. The overall implication is that there are other aspects of the abundance measurement techniques that need to be addressed or standardized, since the remaining scatter is largely unexplained.

In addition, elements with only one absorption line, blends, non-LTE effects, HFS, et cetera, require special treatment in order to attain even a hint of uniformity, for example O, Mg, and Eu II. We strongly urge the community to better understand their abundance techniques and the ways in which they may break down, such that these special cases may be properly handled. We also seek to attain error determinations that truly reflect the uncertainties within not only the elements but also the stellar parameters. The error bars may be calculated with respect to benchmark stars, which would ideally provide a basis on which to evaluate abundance variations between groups on a large scale, for example the work conducted by \citet{Smiljanic14}. 

To date, the results of multiple analyses of the same targets do not produce the same results. Therefore, we call for transparency in presented results to unravel the systematic offsets that are clearly present in our current stellar abundance data. 
In other words, it would be beneficial if groups published their full line lists, the EW measurements for their lines, atomic parameters, stellar parameters, error bars, and methods of measurement and error calculation in detail. We also believe that it is important to report absolute abundances unnormalized to solar values wherever possible. While this clearly results in longer papers, as demonstrated here and in the Appendices, we believe that it is an important step towards resolving the issues prevalent in the field. 

{\bf However, transparency of techniques alone is not sufficient if the community does not come together to increase reproducibility. }
In this vein, we believe that more rigorous comparisons between data sets should be implemented, specifically those which involve statistics, and an avoidance of purely graphical comparisons. 
We also encourage members of the field to openly discuss offsets and variations found between techniques in order to discover their fundamental root. Since there is a current lack of ``ground-truthing" within astronomy, methodologies are not so much ``wrong" as incomparable, which has lead to the current lack of consensus. Once abundance results are accurate and agree between techniques, only then can we focus on determining the true, precise stellar chemical make-up. We are optimistic that stellar abundances can become both precise and accurate, however, it is clear that it requires an effort that spans the entire community.

\section{Summary}
\label{s.summary}
As a result of the preliminary discussion and tests at ASU's Stellar Stoichiometry workshop, we have formulated an investigation to understand the causes of variations between stellar abundance measurement techniques. A total of six groups were asked to determine the abundances within four stars, exhibiting a range in iron-content, over a series of multiple tests. Each group was asked to calculate the abundances for each star a total of four times, where Run 1 utilized their own autonomous method, Run 2 employed standardized stellar parameters, Run 3 implemented a standardized line list, and Run 4 had both standardized stellar parameters and element absorption lines. The results for the stellar parameters and the 10 elements, namely C, O, Na, Mg, Al, Si, Fe, Ni, Ba II, and Eu II, can be found in Tables \ref{stellarparams1}--\ref{run4abs}.  

The effect of the standardized line list on the stellar parameters appeared to have a polarizing effect, as summarized in Table \ref{sum}. In some cases, using the same lines allowed measurements to become more consistent between groups. On the other hand, a number of the stellar parameter values became more disparate during Run 3. These variations were not due to whether the methodology employed spectral fitting or CoG. However, the cause may be related to the inherent degeneracy between the stellar parameters or the number of Fe lines employed by each of the groups. 

It was our expectation at the beginning of the Investigation that Run 4 would yield the most comparable results between groups. While the spread in the abundance determinations did decrease during this run as compared to Run 1, the results were not as consistent as we had expected. We implemented a metric of data similarity (Table \ref{metric}) in order to quantify the similarity/dispersion of the abundance determinations. In general, we found that Run 2 had consistently better results between the elements and stars, followed by Run 4. As seen in Table \ref{sum}, out of the 40 possibilities (10 elements within 4 stars), using the standardized parameters (only) achieved the most similar measurements between groups a total of 16 times. Using both the standardized parameters and lines resulted in the most comparable abundances in 12 instances. Given that both Runs 1 and 3 had similar success rates for best abundance similarity for a star between analyses, the implication is that the standardized line list may have adversely affected the abundance techniques for particular elements. This conclusion is mirrored in our line-by-line abundance analysis with respect to iron. Unfortunately, the small number of non-Fe lines measured by all groups for all four analyses made comparisons and broader conclusions difficult. Given that Run 3 generated much larger spreads in stellar parameters than Run 1, it may be that a standardized line list could improve consistency, but the choice of Fe lines used must be optimized. On the whole, elements that had a strong dependence on stellar parameters, as measured by an improvement between Run 1 to Run 2, did not improve or get worse during Run 3.

We saw similar spreads in the data when comparing the values determined here to those found within the literature. Optimistically, this shows that the setup of the Investigation accurately reflected the techniques and measurements found within the community. In addition, it emphasized the point that particular elements, such as C and O, require special attention to determine precise abundances. For elements like Fe and Al, and to a lesser extent Na, Al, Si, and Ni, the overall abundance determinations may be sensitive to either the lines, the technique, or the stellar parameters used in their calculation. 

The Investigation that was conducted here is complimentary to studies by \citet{Smiljanic14, Jofre14, Jofre15}, all of which are relevant to ongoing and upcoming large stellar surveys, such as the Gaia-ESO survey, the Transiting Exoplanet Survey Satellite (TESS), the Characterizing Exoplanets Satellite (CHEOPS), and the James Webb Space Telescope (JWST). It is important that stars be precisely and consistently measured such that they, and possible orbiting exoplanets, may be well characterized. 

It was our hope that either standardized stellar parameters, line list, or both would ameliorate the disparate abundance measurements between different techniques. Instead, we have found a deeper understanding of the problems inherent to the stellar abundance field. The Investigation has shown that, in order for measurements to be copacetic, 1) the details of the methods and their input parameters need to be directly compared, 2) particular elements require special care, 3) line lists and/or stellar parameters should be standardized, and importantly, 4) all research should be presented with the utmost transparency to ensure reproducibility of the results. Given the huge amount of determination, flexibility, support, and feedback offered by all of the participating groups, we are hopeful that the community can work together to rectify these issues and work towards not only precise but accurate stellar abundances.

\section*{Acknowledgements}
The authors would like to thank Paul Butler for providing the original stellar spectra in addition to Eric Mamajek for his help determining accurate stellar types for our sample. 
They would also like to thank the anonymous referee for their support and guidance, which has greatly improved the manuscript.
N.R.H. would like to thank CHW3. The ASU team (N.R.H., P.A.Y., M.D.P., S.J.D. and A.D.A.) acknowledge that the results reported herein benefitted from collaborations and/or information exchange within NASA's Nexus for Exoplanet System Science (NExSS) research coordination network sponsored by NASA's Science Mission Directorate. E.D.M and V.A. acknowledge the support from the Funda\c{c}\~ao para a Ci\^encia e Tecnologia (FCT, Portugal) in the form of the grants SFRH/BPD/76606/2011 and SFRH/BPD/70574/2010, respectively.  J.K.C. acknowledges partial support from an appointment to the NASA Postdoctoral Program at the Goddard Space Flight Center, administered by Universities Space Research Association through a contract with NASA. T.N., A.K., and U.H. acknowledge support by the Swedish National Space Board (SNSB). N.C.S. and S.G.S. acknowledge the support from FCT through Investigador FCT contracts of reference IF/00169/2012 and IF/00028/2014, respectively, and POPH/FSE (EC) by FEDER funding through the program ``Programa Operacional de Factores de Competitividade". The Porto group also acknowledges the support from FCT in the form of grant reference PTDC/FIS-AST/7073/2014 (POCI-01-0145-FEDER-007672) and project PTDC/FIS-AST/1526/2014. This research has made use of the SIMBAD database and VizieR catalogue access tools operated at CDS, Strasbourg, France as well as the Exoplanet Orbit Database at exoplanets.org.

\clearpage
\begin{deluxetable}{lcccccc}
  \tablecaption{\label{standardparams} Stellar Properties and Standard Parameters}
  \tablehead{
    \colhead{Star} &
    \colhead{Spectral Type} &
     \colhead{RA} &
     \colhead{Dec} &
    \colhead{\teff (K)} &
    \colhead{\lg} &
    \colhead{$\xi$ (km/s)} 
  }
 \startdata
 HD~361	 & G1V  &  00h 08m 16s &  -14$^{\circ}$ 49' 28'' & $5945  $ & $  4.69$ & $  1.63 $ \\  
HD~10700 & G8V &  13h 44m 04s &  15$^{\circ}$ 56' 14''&$5382$ & $  4.57$ & $  1.26 $ \\
HD~121504 & G2V &  13h 57m 17s &  -56$^{\circ}$ 02' 24''&$6048   $ & $4.54$ & $  1.25 $ \\
HD~202206 & G6V &  21h 14m 58s &  -20$^{\circ}$ 47' 21'' &$5765  $ & $  4.45$ & $  1.43 $ 
  \enddata
\end{deluxetable}

\newpage

\begin{deluxetable}{lccccc}
  \tablecaption{\label{models} Abundance Models for Investigation Participants}
  \tablehead{
    \colhead{Name} &
    \colhead{Calculations Performed By:} &
    \colhead{Stellar Atmosphere} &
    \colhead{EW} &
    \colhead{CoG or Spectral Fitting} &
    \colhead{FeI/FeII lines}
  }
 \startdata
ANU & Fan Liu & ATLAS9 & ARES & CoG (MOOG 2014) & 102/13\\
ASU & Michael Pagano & ATLAS9 & ARES/IRAF {\sc SPLOT} & CoG (MOOG 2014) & 69/14\\
Carnegie & Joleen Carlberg & MARCS & ARES/IRAF {\sc SPLOT} & CoG (MOOG 2014) & 74/13\\ 
Geneva & Sergi Blanco-Cuaresma & MARCS & SPECTRUM & SF (SPECTRUM) & 298/19\\
Porto & Vardan Adibekyan, & ATLAS9 & ARES/IRAF {\sc SPLOT} & CoG (MOOG 2014) & 210/23\\
 & Elisa Delgado Mena, Sergio Sousa  & & & & \\
Uppsala & Thomas Nordlander & MARCS & SME & SF (SME) & 60/12\\ 
  \enddata
\end{deluxetable}

\begin{deluxetable}{lcccc|c}
  \tablecaption{\label{rew} REW vs. $\chi_l$ (from Figure \ref{rew-ep}) Slopes}
  \tablehead{
    \colhead{Group} &
    \colhead{HD~361} &
    \colhead{HD~10700} &
    \colhead{HD~121504} &
    \colhead{HD~202206} &
    \colhead{mean}
  }
 \startdata
 \vspace{3mm}
ANU &   -0.043 & -0.089 & -0.032 & -0.027 &  -0.048\\ 
\vspace{3mm}
ASU &  0.002 & -0.054 & 0.039 & -0.002 & -0.012 \\ 
 \vspace{3mm}
Carnegie &  -0.013 & -0.059 & -0.007 & 0.004 & -0.019 \\ 
Porto &  -0.066 & -0.083 & -0.058 & -0.055 & -0.065 \\ 
 \enddata
 \end{deluxetable}

\clearpage
\LongTables
\begin{deluxetable}{llllll}
  \tablecaption{\label{stellarparams1} Stellar Parameters (Run 1)}
  \tablehead{
    \colhead{Star} &
    \colhead{Group} &
    \colhead{\teff} &
    \colhead{\lg} &
    \colhead{$\xi$} &
    \colhead{[Fe/H]} \\
    \colhead{} &
    \colhead{} &
    \colhead{(K)} &
    \colhead{} &
    \colhead{(km/s)} &
    \colhead{(dex)} 
  }
 \startdata
 HD~361 &ANU      & 5847 $\pm$ 47 & 4.40 $\pm$ 0.07 & 1.08 $\pm$ 0.07 & -0.12 $\pm$ 0.05 \\
& ASU      & 5970 $\pm$ 47 & 4.73             & 1.37 $\pm$ 0.18 & -0.02 $\pm$ 0.03 \\
&Carnegie  & 5980 $\pm$ 86 & 4.61 $\pm$ 0.06 & 1.03 $\pm$ 0.13 & 0.00 $\pm$ 0.01 \\
& Geneva  & 5864 $\pm$ 12 & 4.53 $\pm$ 0.02 & 1.36 $\pm$ 0.02  & -0.26 $\pm$ 0.13 \\
 & Porto   & 5898 $\pm$ 31 & 4.51 $\pm$ 0.03 & 1.07 $\pm$ 0.04 & -0.09 $\pm$ 0.02 \\
\vspace{0.4mm}
& Uppsala  & 5875 $\pm$ 126  & 4.42  $\pm$ 0.28  & 1.15 $\pm$ 0.39  & -0.11 $\pm$ 0.07  \\ 
& {\bf Avg.} & {\bf 5906} & {\bf 4.53} & {\bf 1.18 } & {\bf -0.10} \\
\vspace{3mm}
& {\bf Range} & {\bf 133} & {\bf 0.33} & {\bf 0.34} & {\bf 0.26} \\
\enddata
\end{deluxetable}

\clearpage
\LongTables
\begin{deluxetable}{llllll}
  \tablecaption{\label{stellarparams3} Stellar Parameters (Run 3)}
  \tablehead{
    \colhead{Star} &
    \colhead{Group} &
    \colhead{\teff} &
    \colhead{\lg} &
    \colhead{$\xi$} &
    \colhead{[Fe/H]} \\
    \colhead{} &
    \colhead{} &
    \colhead{(K)} &
    \colhead{} &
    \colhead{(km/s)} &
    \colhead{(dex)} 
  }
 \startdata
HD~361 & ANU & 5962 $\pm$ 50 & 4.73 $\pm$ 0.08 & 1.10 $\pm$ 0.08 & -0.10 $\pm$ 0.05 \\
 & ASU & 5938 $\pm$ 81 & 4.73 & 1.46 $\pm$ 0.21 & -0.10 $\pm$ 0.06 \\
 & Carnegie & 5970 $\pm$ 70 & 4.85 $\pm$ 0.08 & 1.02 $\pm$ 0.16 & -0.06 $\pm$ 0.10   \\
 & Geneva & 5725 $\pm$ 29 & 4.26 $\pm$ 0.05 & 1.46 $\pm$ 0.05  & -0.28 $\pm$ 0.16 \\
 & Porto & 6015 $\pm$ 62 & 4.91 $\pm$ 0.07 & 1.10 $\pm$ 0.07 & \,0.05 $\pm$ 0.04 \\
  \vspace{0.4mm}
  & Uppsala & 6026 $\pm$ 126 & 4.86 $\pm$ 0.28 & 0.98 $\pm$ 0.39 & -0.02 $\pm$ 0.1 \\
& {\bf Avg.} & {\bf 5939} & {\bf 4.72} & {\bf 1.19} & {\bf -0.09} \\
 \vspace{3mm}
& {\bf Range} & {\bf 301} & {\bf 0.65} & {\bf 0.48} & {\bf 0.33} \\
\enddata
\end{deluxetable}

\clearpage
\begin{landscape}
\begin{deluxetable}{|llllllllllll|}\tiny
  \tablecaption{\label{run1abs} Absolute Abundances by Investigation Participants for the Stellar Sample (Run 1)}
  \tablehead{
    \colhead{Star} &
    \colhead{Group} &
    \colhead{A(C)} &
    \colhead{A(O)} &
    \colhead{A(Na)} &
    \colhead{A(Mg)} &
    \colhead{A(Al)} &
    \colhead{A(Si)} &
    \colhead{A(Fe)} &
    \colhead{A(Ni)} &
    \colhead{A(BaII)} &
    \colhead{A(EuII)}
  }
 \startdata
HD~361&ANU&  8.37 $\pm$ 0.09 &        --       & 6.03 $\pm$ 0.06 & 7.47 $\pm$ 0.05 & 6.28 $\pm$ 0.06 & 7.35 $\pm$ 0.05 & 7.35 $\pm$ 0.05 & 6.01 $\pm$ 0.05 & 2.20 $\pm$ 0.06 &-- \\
&ASU&  8.28 $\pm$ 0.04 &        --       & 6.07 $\pm$ 0.03 & 7.48 $\pm$ 0.03 & 6.33 $\pm$ 0.02 & 7.42 $\pm$ 0.02 & 7.44 $\pm$ 0.03 & 6.10 $\pm$ 0.04 & 2.30 $\pm$ 0.09 &-- \\
&Carnegie&  8.21 $\pm$ 0.10 & 8.70 $\pm$ 0.09 & 6.04 $\pm$ 0.04 &        --       & 6.06 $\pm$ 0.04 & 7.40 $\pm$ 0.04 & 7.46 $\pm$ 0.10 & 6.23 $\pm$ 0.18 & 2.54 $\pm$ 0.08 & 0.55 $\pm$ 0.04   \\
&Geneva&  8.28 $\pm$ 0.04 & 9.18 $\pm$ 0.34 & 6.14 $\pm$ 0.23 & 7.32 $\pm$ 0.08 & 6.55 $\pm$ 0.09 & 7.35 $\pm$ 0.08 & 7.20 $\pm$ 0.13 & 6.03 $\pm$ 0.16 & 2.40 $\pm$ 0.16 & 0.63 $\pm$ 0.28   \\
&Porto&  8.32 $\pm$ 0.03 & 8.63 $\pm$ 0.10 & 6.16 $\pm$ 0.01 & 7.53 $\pm$ 0.05 & 6.33 $\pm$ 0.01 & 7.45 $\pm$ 0.02 & 7.38 $\pm$ 0.02 & 6.09 $\pm$ 0.03 & 2.35 $\pm$ 0.03 & 0.57              \\
\vspace{3mm}
&Uppsala&  8.23 $\pm$ 0.08 & 8.76 $\pm$ 0.09 & 6.15 $\pm$ 0.07 & 7.55 $\pm$ 0.07 & 6.39 $\pm$ 0.10 & 7.39 $\pm$ 0.04 & 7.35 $\pm$ 0.07 & 6.03 $\pm$ 0.14 & 2.34 $\pm$ 0.08 & 0.48 $\pm$ 0.10   \\
 & {\bf Avg.} & {\bf 8.28 } & {\bf 8.82 } & {\bf 6.10 } & {\bf 7.47 } & {\bf 6.32 } & {\bf 7.39 } & {\bf 7.36 } & {\bf 6.08 } & {\bf 2.36 } & {\bf 0.56}\\
  \vspace{3mm}
& {\bf Range} & {\bf 0.16 } & {\bf 0.55 } & {\bf 0.13 } & {\bf 0.23 } & {\bf 0.49 } & {\bf 0.10 } & {\bf 0.26 } & {\bf 0.22 } & {\bf 0.34 } & {\bf 0.15}\\
\enddata
\end{deluxetable}
\clearpage
\end{landscape}

\clearpage
\begin{landscape}
\begin{deluxetable}{|llllllllllll|}\tiny
  \tablecaption{\label{run2abs} Absolute Abundances by Investigation Participants for the Stellar Sample (Run 2)}
  \tablehead{
    \colhead{Star} &
    \colhead{Group} &
    \colhead{A(C)} &
    \colhead{A(O)} &
    \colhead{A(Na)} &
    \colhead{A(Mg)} &
    \colhead{A(Al)} &
    \colhead{A(Si)} &
    \colhead{A(Fe)} &
    \colhead{A(Ni)} &
    \colhead{A(BaII)} &
    \colhead{A(EuII)}
  }
 \startdata
HD~361&ANU&  8.40 $\pm$ 0.08 &        --       & 6.00 $\pm$ 0.06 & 7.43 $\pm$ 0.06 & 6.30 $\pm$ 0.06 & 7.36 $\pm$ 0.05 & 7.31 $\pm$ 0.05 & 6.02 $\pm$ 0.05 & 2.04 $\pm$ 0.05 &-- \\
&ASU&  8.29 $\pm$ 0.03 &        --       & 6.05 $\pm$ 0.02 & 7.45 $\pm$ 0.01 & 6.32 $\pm$ 0.01 & 7.40 $\pm$ 0.02 & 7.34 $\pm$ 0.11 & 6.06 $\pm$ 0.02 & 2.14 $\pm$ 0.03 &-- \\
&Carnegie&  8.27 $\pm$ 0.08 & 8.74 $\pm$ 0.04 & 6.01 $\pm$ 0.03 &        --       & 6.03 $\pm$ 0.03 & 7.38 $\pm$ 0.04 & 7.36 $\pm$ 0.12 & 6.10 $\pm$ 0.10 & 2.20 $\pm$ 0.11 & 0.55 $\pm$ 0.03  \\
&Geneva&  8.31 $\pm$ 0.04 & 9.24 $\pm$ 0.37 & 6.13 $\pm$ 0.19 & 7.29 $\pm$ 0.08 & 6.57 $\pm$ 0.08 & 7.37 $\pm$ 0.07 & 7.18 $\pm$ 0.15 & 6.05 $\pm$ 0.16 & 2.28 $\pm$ 0.14 & 0.70 $\pm$ 0.27   \\
&Porto&  8.35 $\pm$ 0.03 & 8.68 $\pm$ 0.10 & 6.16 $\pm$ 0.01 & 7.49 $\pm$ 0.09 & 6.33 $\pm$ 0.01 & 7.44 $\pm$ 0.03 & 7.38 $\pm$ 0.10 & 6.09 $\pm$ 0.04 & 2.14 $\pm$ 0.03 & 0.63             \\
\vspace{3mm}
&Uppsala&  8.28 $\pm$ 0.06 & 8.86 $\pm$ 0.08 & 6.08 $\pm$ 0.08 & 7.47 $\pm$ 0.06 & 6.34 $\pm$ 0.10 & 7.38 $\pm$ 0.03 & 7.24 $\pm$ 0.13 & 6.06 $\pm$ 0.15 & 2.14 $\pm$ 0.12 & 0.59 $\pm$ 0.10   \\
 & {\bf Avg.} & {\bf 8.32 } & {\bf 8.88 } & {\bf 6.07 } & {\bf 7.43 } & {\bf 6.32 } & {\bf 7.39 } & {\bf 7.30 } & {\bf 6.06 } & {\bf 2.16 } & {\bf 0.62} \\ 
  \vspace{3mm}
 & {\bf Range} & {\bf 0.13 } & {\bf 0.56 } & {\bf 0.16 } & {\bf 0.20 } & {\bf 0.54 } & {\bf 0.08 } & {\bf 0.20 } & {\bf 0.08 } & {\bf 0.24 } & {\bf 0.15} \\
\enddata
\end{deluxetable}
\clearpage
\end{landscape}

\clearpage
\begin{landscape}
\begin{deluxetable}{|llllllllllll|}\tiny
  \tablecaption{\label{run3abs} Absolute Abundances by Investigation Participants for the Stellar Sample (Run 3)}
  \tablehead{
    \colhead{Star} &
    \colhead{Group} &
    \colhead{A(C)} &
    \colhead{A(O)} &
    \colhead{A(Na)} &
    \colhead{A(Mg)} &
    \colhead{A(Al)} &
    \colhead{A(Si)} &
    \colhead{A(Fe)} &
    \colhead{A(Ni)} &
    \colhead{A(BaII)} &
    \colhead{A(EuII)}
  }
 \startdata
HD~361&ANU&  8.31 $\pm$ 0.08 &        --       & 6.03 $\pm$ 0.06 & 7.39 $\pm$ 0.07 & 6.16 $\pm$ 0.14 & 7.43 $\pm$ 0.06 & 7.37 $\pm$ 0.05 & 6.07 $\pm$ 0.05 & 2.27 $\pm$ 0.06 &-- \\
&ASU&  8.30 $\pm$ 0.06 & 8.68 $\pm$ 0.05 & 6.08 $\pm$ 0.05 & 7.39 $\pm$ 0.03 & 6.16 $\pm$ 0.11 & 7.43 $\pm$ 0.03 & 7.36 $\pm$ 0.06 & 6.01 $\pm$ 0.05 & 2.07 $\pm$ 0.08 &-- \\
&Carnegie&  8.24 $\pm$ 0.07 & 8.69 $\pm$ 0.06 & 6.04 $\pm$ 0.05 & 7.32 $\pm$ 0.08 & 6.17 $\pm$ 0.20 & 7.43 $\pm$ 0.10 & 7.40 $\pm$ 0.10 & 6.05 $\pm$ 0.06 & 2.25 $\pm$ 0.10 & 0.64 $\pm$ 0.04   \\
&Geneva&  8.29 $\pm$ 0.01 &        --       & 5.88 $\pm$ 0.07 & 7.52 $\pm$ 0.06 & 6.42 $\pm$ 0.07 & 7.24 $\pm$ 0.08 & 7.18 $\pm$ 0.16 & 5.93 $\pm$ 0.33 & 2.25 $\pm$ 0.18 & 0.52 $\pm$ 0.27   \\
&Porto&  8.32 $\pm$ 0.04 & 8.78 $\pm$ 0.06 & 6.14 $\pm$ 0.03 & 7.50            & 6.25 $\pm$ 0.13 & 7.54 $\pm$ 0.06 & 7.52 $\pm$ 0.04 & 6.19 $\pm$ 0.06 & 2.39 $\pm$ 0.05 & 0.78              \\
\vspace{3mm}
&Uppsala&  8.35 $\pm$ 0.00 & 9.06 $\pm$ 0.10 & 6.28 $\pm$ 0.16 & 7.61 $\pm$ 0.15 & 6.32 $\pm$ 0.14 & 7.44 $\pm$ 0.08 & 7.44 $\pm$ 0.10 & 6.16 $\pm$ 0.13 & 2.71 $\pm$ 0.06 & 0.71 $\pm$ 0.10   \\
 &{\bf Avg.} & {\bf 8.30} & {\bf 8.80} & {\bf 6.08} & {\bf 7.46} & {\bf 6.25} & {\bf 7.42} & {\bf 7.38} & {\bf 6.07} & {\bf 2.32} & {\bf 0.66} \\
  \vspace{3mm}
 &{\bf Range} & {\bf 0.11} & {\bf 0.38} & {\bf 0.40} & {\bf 0.29} & {\bf 0.26} & {\bf 0.30} & {\bf 0.34} & {\bf 0.26} & {\bf 0.64} & {\bf 0.26} \\
\enddata
\end{deluxetable}
\clearpage
\end{landscape}

\clearpage
\begin{landscape}
\begin{deluxetable}{|llllllllllll|}\tiny
  \tablecaption{\label{run4abs} Absolute Abundances by Investigation Participants for the Stellar Sample (Run 4)}
  \tablehead{
    \colhead{Star} &
    \colhead{Group} &
    \colhead{A(C)} &
    \colhead{A(O)} &
    \colhead{A(Na)} &
    \colhead{A(Mg)} &
    \colhead{A(Al)} &
    \colhead{A(Si)} &
    \colhead{A(Fe)} &
    \colhead{A(Ni)} &
    \colhead{A(BaII)} &
    \colhead{A(EuII)}
  }
 \startdata
HD~361&ANU&  8.31 $\pm$ 0.08 &        --       & 6.01 $\pm$ 0.06 & 7.38 $\pm$ 0.07 & 6.15 $\pm$ 0.14 & 7.41 $\pm$ 0.06 & 7.30 $\pm$ 0.05 & 6.01 $\pm$ 0.05 & 2.05 $\pm$ 0.06 &-- \\
&ASU&  8.28 $\pm$ 0.03 & 8.85 $\pm$ 0.07 & 6.08 $\pm$ 0.02 & 7.39 $\pm$ 0.05 & 6.16 $\pm$ 0.10 & 7.43 $\pm$ 0.03 & 7.33 $\pm$ 0.02 & 6.00 $\pm$ 0.01 & 2.01 $\pm$ 0.01 &-- \\
&Carnegie&  8.21 $\pm$ 0.06 & 8.66 $\pm$ 0.05 & 6.03 $\pm$ 0.04 & 7.31 $\pm$ 0.07 & 6.16 $\pm$ 0.20 & 7.40 $\pm$ 0.09 & 7.32 $\pm$ 0.13 & 5.99 $\pm$ 0.06 & 2.01 $\pm$ 0.08 & 0.56 $\pm$ 0.03  \\
&Geneva&  8.30 $\pm$ 0.02 &        --       & 5.97 $\pm$ 0.06 & 7.59 $\pm$ 0.06 & 6.50 $\pm$ 0.07 & 7.30 $\pm$ 0.09 & 7.30 $\pm$ 0.16 & 6.02 $\pm$ 0.32 & 2.28 $\pm$ 0.14 & 0.71 $\pm$ 0.28   \\
&Porto&  8.30 $\pm$ 0.03 & 8.79 $\pm$ 0.06 & 6.10 $\pm$ 0.03 & 7.47            & 6.21 $\pm$ 0.13 & 7.51 $\pm$ 0.05 & 7.44 $\pm$ 0.15 & 6.09 $\pm$ 0.05 & 2.11 $\pm$ 0.03 & 0.65             \\
\vspace{3mm}
&Uppsala&  8.29 $\pm$ 0.06 & 8.90 $\pm$ 0.08 & 6.18 $\pm$ 0.11 & 7.48 $\pm$ 0.13 & 6.26 $\pm$ 0.12 & 7.38 $\pm$ 0.04 & 7.26 $\pm$ 0.14 & 6.08 $\pm$ 0.10 & 2.29 $\pm$ 0.09 & 0.60 $\pm$ 0.10   \\
&{\bf Avg.} & {\bf 8.28} & {\bf 8.80} & {\bf 6.06} & {\bf 7.44} & {\bf 6.24} & {\bf 7.41} & {\bf 7.33} & {\bf 6.03} & {\bf 2.13} & {\bf 0.63} \\
  \vspace{3mm}
&{\bf Range} & {\bf 0.10} & {\bf 0.24} & {\bf 0.21} & {\bf 0.28} & {\bf 0.35} & {\bf 0.21} & {\bf 0.18} & {\bf 0.10} & {\bf 0.28} & {\bf 0.15} \\
\enddata
\end{deluxetable}
\clearpage
\end{landscape}

\begin{deluxetable*}{ccl|rlrlrlrl}
  \tablecolumns{10}
  \tablewidth{0pc}
  \tablecaption{\label{metric} Metric of Data Similarity}
  \tablehead{
    \colhead{} &
    \colhead{} &
    \colhead{} &
    \multicolumn{2}{c}{Run 1} &
    \multicolumn{2}{c}{Run 2} &
    \multicolumn{2}{c}{Run 3} &
    \multicolumn{2}{c}{Run 4} \\
    \colhead{Element} &
    \colhead{$\gamma$} &
    \colhead{Star} &
    \colhead{Metric}  &
    \colhead{Norm.} &
    \colhead{Metric} &
    \colhead{Norm.} &
    \colhead{Metric}  &
    \colhead{Norm.} &
    \colhead{Metric} &
    \colhead{Norm.} \\
   }  
  \startdata
C & 7.1 & HD~361 & 2.30 & {\it 0.62} & 2.74 & {\it 0.74 }& 3.25 & {\it 0.88 }& 3.71 & {\it 1.00} \\ 
 & & HD~10700 & 2.13 & {\it 0.57} & 2.05 & {\it 0.55 }& 1.56 & {\it 0.42 }& 1.16 & {\it 0.31} \\ 
 & & HD~121504 & 2.68 & {\it 0.72} & 2.35 & {\it 0.63} & 2.77 & {\it 0.75} & 2.45 & {\it 0.66} \\ 
 \vspace{3mm}
 & & HD~202206 & 3.07 & {\it 0.83} & 2.24 & {\it 0.60 }& 2.43 & {\it 0.65 }& 2.35 & {\it 0.63} \\  
 O & 3.1 & HD~361 & 1.17 & {\it 0.44 }&  0.98 & {\it 0.37 }& 1.48 & {\it 0.56 }& 1.76 & {\it 0.67} \\ 
 & & HD~10700 & 2.34 & {\it 0.89} & 2.64 & {\it 1.00 }& 1.39 & {\it 0.53 }& 1.55 & {\it 0.59} \\ 
 & & HD~121504 & 2.26 & {\it 0.86} & 2.51 & {\it 0.95} & 2.22 & {\it 0.84} & 2.15 & {\it 0.81} \\ 
 \vspace{3mm}
 & & HD~202206 & 1.54 & {\it 0.58 }& 1.63 & {\it 0.62 }& 1.57 & {\it 0.59 }& {\it 0.99} & {\it 0.38} \\
\enddata
\end{deluxetable*}

\clearpage
\LongTables
\begin{deluxetable}{llllllllllll}
  \tablecaption{\label{littable} Absolute Abundances from Hypatia Sources for the Stellar Sample}  \tablehead{
    \colhead{Star} &
    \colhead{Group} &
    \colhead{A(C)} &
    \colhead{A(O)} &
    \colhead{A(Na)} &
    \colhead{A(Mg)} &
    \colhead{A(Al)} &
    \colhead{A(Si)} &
    \colhead{A(Fe)} &
    \colhead{A(Ni)} &
    \colhead{A(BaII)} &
    \colhead{A(EuII)}
  }
 \startdata
HD~361 & \citet{Adibekyan12} & -- & -- & 6.175 & 7.467 & 6.31 & 7.437 & 7.35 & 6.095  & -- & -- \\ 
& \citet{Bertran15} & -- & 8.67 & -- & -- & -- & -- & 7.35 & --  & -- & -- \\ 
& \citet{delgado-mena_2010_aa} & 8.36 & 8.49 & -- & 7.42 & -- & 7.43 & 7.35 & 6.09  & -- & -- \\ 
& \citet{GonzalezHernandez:2010p7714} & 8.26 & 8.58 & 6.085 & 7.47 & 6.28 & 7.387 & 7.391 & 6.057  & 2.345 & 0.48 \\ 
\vspace{3mm}
& \citet{Neves:2009p1804} & -- & -- & 6.15 & 7.45 & 6.31 & 7.43 & 7.35 & 6.09  & -- & -- \\ 
& {\bf Hypatia} & 8.31 & 8.58 & 6.15 & 7.46 & 6.31 & 7.43 & 7.35 & 6.09  & 2.345 & 0.48 \\
\vspace{3mm}
& {\it Range} & {\it 0.1} & {\it 0.18} & {\it 0.09} & {\it 0.05} & {\it 0.04} & {\it 0.05} & {\it 0.16} & {\it 0.04}  & {\it 0.0} & {\it 0.0} \\
HD~10700 &  \citet{Adibekyan12} & -- & -- & 5.885 & 7.297 & 6.20 & 7.186 & 6.95 & 5.737  & -- & -- \\
\enddata
 \tablenotetext{*}{Errors were not included for the literature due to the varying ways in which they are calculated, for example, on a star-by-star basis or as a representative error per element. In addition, by converting error determinations to an absolute scale may not accurately represent the true error in the abundance calculation.}
  \tablenotetext{**}{Abundances have been normalized to the same solar scale, namely \citet{Lodders:2009p3091}, per the analysis within the Hypatia Catalog \citep{Hinkel14}.}
\end{deluxetable}

\clearpage
\LongTables
\begin{deluxetable}{lllllll||ll}
  \tablecaption{\label{felines} Sample of Line-by-line Comparisons for Iron in HD~202206 (Runs 1--4)}
  \tablehead{
    \colhead{Wavelength} &
    \colhead{Group} &
    \colhead{$\chi_l$} &
    \colhead{$\log (gf)$} &
    \colhead{EW} &
    \colhead{A(Fe)$_{Run 1}$} &
    \colhead{A(Fe)$_{Run 2}$} &
    \colhead{A(Fe)$_{Run 3}$} &
    \colhead{A(Fe)$_{Run 4}$} \\
    \colhead{($\text{\AA}$)} &
    \colhead{} &
    \colhead{(eV)} &
    \colhead{} &
    \colhead{(m$\text{\AA}$)} &
    \colhead{(dex)} &
     \colhead{(dex)} &
      \colhead{(dex)} &
     \colhead{(dex)} \\
  }
5522.45 & ANU & 4.21 & -1.45 & 60.9 & 7.76 & 7.72 & 7.93 & 7.83 \\ 
 & ASU & 4.21 & -1.55 & 59.7 & 8.00 & 7.84 & 7.97 & 7.84 \\ 
 & Carnegie & 4.21 & -1.56 & 58.4 & 7.88 & 7.84 & 7.87 & 7.78 \\ 
 & Geneva & 4.21 & -1.45 & -- & 7.64  & 7.66 & 7.57 & 7.66 \\ 
 & Porto & 4.21 & -1.42 & 61.0 & 7.77 & 7.71 & 7.97 & 7.88 \\ 
   \vspace{2mm}
 &  &  &  & {\bf Avg.} & 7.81 & 7.75 & 7.86 & 7.80 \\ 
5778.45 & ANU & 2.59 & -3.44 & 36.2 & 7.69 & 7.70 & 7.80 & 7.71 \\ 
 & ASU & 2.59 & -3.48 & 35.7 & 7.87 & 7.75 & 7.88 & 7.75 \\
\enddata
  \tablenotetext{*}{Geneva and Uppsala do not report abundances from all or some individual lines and Uppsala doesn't report some equivalent widths. The $\chi_l$, $\log (gf)$, and EW values are only applicable to Runs 1 and 2, as indicated by the vertical lines, since Runs 3 and 4 use a standardized line list.}
\end{deluxetable}

\clearpage
\begin{deluxetable}{llllllll||ll}
  \tablecaption{\label{otherlines} Sample of Line-by-line Comparisons for Additional Elements in HD~202206 (Runs 1--4)}
  \tablehead{
    \colhead{Element} &
    \colhead{Wavelength} &
    \colhead{Group} &
    \colhead{$\chi_l$} &
    \colhead{$\log (gf)$} &
    \colhead{EW} &
    \colhead{A(X)$_{Run 1}$} &
    \colhead{A(X)$_{Run 2}$} &
    \colhead{A(X)$_{Run 3}$} &
    \colhead{A(X)$_{Run 4}$} \\
    \colhead{} &
    \colhead{($\text{\AA}$)} &
    \colhead{} &
    \colhead{(eV)} &
    \colhead{} &
    \colhead{(m$\text{\AA}$)} &
    \colhead{(dex)} &
     \colhead{(dex)} &
      \colhead{(dex)} &
     \colhead{(dex)} \\
  }
   \vspace{1mm}
C & 5052.15 & ANU & 7.68 & -1.3 & 43.8 & 8.57 & 8.55 & 8.56 & 8.53 \\ 
 &  & ASU & 7.68 & -1.3 & 42.2 & 8.63 & 8.63 & 8.59 & 8.60 \\ 
 &  & Carnegie & -- & --  & -- & --  &--  & 8.46 & 8.42 \\ 
 &  & Geneva & 7.68 & -1.3 & -- & 8.59  & 8.48 & 8.62 & 8.48 \\ 
 &  & Porto & 7.68 & -1.303 & 44.4 & 8.66 & 8.66 & 8.59 & 8.61 \\ 
 &  & Uppsala & 7.69 & -1.3 & --  & 8.57 & 8.54 & 8.48 & 8.47 \\ 
   \vspace{2mm}
 &  &  &  &  & {\bf Avg.} & 8.60 & 8.57 & 8.55 & 8.51 \\ 
 & 5380.32 & ANU & 7.68 & -1.6 & 25.0 & 8.51 & 8.50 & 8.53 & 8.50 \\
\enddata
   \tablenotetext{*}{Groups that did not measure particular lines are not listed. Additionally, averages are not given in the cases where only one group measured a line. The $\chi_l$, $\log (gf)$, and EW values are only applicable to Runs 1 and 2, as indicated by the vertical lines, since Runs 3 and 4 use a standardized line list.}
\end{deluxetable}

\begin{deluxetable}{l|cccccccccc}
  \tablecolumns{11}
  \tablewidth{0pc}
  \tablecaption{\label{errors} Compilation of Abundance Errors}
  \tablehead{
    \colhead{Averages} &
    \colhead{A(C)} &
    \colhead{A(O)} &
    \colhead{A(Na)} &
    \colhead{A(Mg)} &
    \colhead{A(Al)} &
    \colhead{A(Si)} &
    \colhead{A(Fe)} &
    \colhead{A(Ni)} &
    \colhead{A(BaII)} &
    \colhead{A(EuII)} \\
   }  
  \startdata
ANU & 0.09 & 0.07 & 0.07 & 0.06 & 0.09 & 0.06 & 0.05 & 0.05 & 0.06 & -- \\ 
ASU & 0.04 & 0.05 & 0.04 & 0.04 & 0.06 & 0.03 & 0.05 & 0.03 & 0.05 & 0.02 \\ 
Carnegie & 0.13 & 0.07 & 0.06 & 0.09 & 0.12 & 0.06 & 0.11 & 0.10 & 0.09 & 0.04 \\ 
Geneva & 0.06 & 0.35 & 0.11 & 0.07 & 0.08 & 0.08 & 0.15 & 0.24 & 0.16 & 0.30 \\ 
Porto & 0.06 & 0.13 & 0.04 & 0.04 & 0.07 & 0.05 & 0.07 & 0.06 & 0.04 & -- \\ 
   \vspace{2mm}
Uppsala & 0.04 & 0.10 & 0.08 & 0.07 & 0.11 & 0.07 & 0.10 & 0.13 & 0.08 & 0.15 \\ 
Run 1 & 0.08 & 0.16 & 0.07 & 0.06 & 0.05 & 0.05 & 0.08 & 0.10 & 0.08 & 0.14 \\ 
Run 2 & 0.07 & 0.14 & 0.06 & 0.05 & 0.05 & 0.05 & 0.10 & 0.08 & 0.08 & 0.13 \\
\enddata
  \tablenotetext{*}{To determine these numbers, the group/run/star listed in the left-column was kept constant and the errors were averaged across the other two factors. For example, for `Run 1 Avg,' the errors were averaged across all four analyses and all six groups but only for Run 1 determinations. See text for a further explanation.}
\end{deluxetable}

\begin{deluxetable*}{ccccc}
  \tablecolumns{10}
  \tablewidth{0pc}
  \tablecaption{\label{sum} Summary of Optimized Results}
  \tablehead{
    \colhead{Element} &
    \colhead{HD~361} &
    \colhead{HD~10700}  &
    \colhead{HD~121504} &
    \colhead{HD~202206} \\
   }  
     \startdata
C & Std. Params \& Lines  & Autonomous & Std. Lines & Autonomous \\
O &  Std. Params \& Lines  & Std. Params & Std. Params & Std. Params \\
Na & Autonomous & Std. Lines & Std. Params \& Lines  & Std. Params \\
Mg & Std. Params & Std. Lines & Autonomous & Std. Params \\
Al & Std. Params \& Lines  & Std. Lines & Std. Params \& Lines  & Std. Lines \\
Si & Std. Params & Std. Params & Std. Params & Std Params \\
Fe & Std. Params \& Lines  & Std. Params \& Lines  & Std. Params \& Lines  & Std. Params \& Lines  \\
Ni & Std. Params & Std. Params & Std. Params & Std. Params \\
Ba II & Std. Params & Std. Params \& Lines  & Autonomous & Std. Params \\
Eu II & Autonomous & Std. Params \& Lines  & Std. Params \& Lines  & Std. Lines \\
\enddata
 \tablenotetext{*}{Here we show a summary table of the results quantified in Table \ref{metric}, where the run with the most similar element abundances within a star (or the highest metric per row of Table \ref{metric}) is listed.}
\end{deluxetable*}

\clearpage

\appendix 
In an effort to follow our own recommendations regarding method transparency, we have provided a variety of detailed explanations and tables in the Appendix. 
\vspace{3mm}

\section{Appendix A: Techniques in the Field}\label{a.field}
Because the stellar abundance field has been evolving for decades, there is a great deal of diversity in terms of the techniques used to quantify the proportions of elements within a star's photosphere. For example, a wide variety of telescopes and spectrographs have been employed in order to acquire stellar spectra: everything from the High Resolution Echelle Spectrometer (HIRES) on the relatively large 10 m Keck telescope \citep{Boesgaard11, Gratton:2003p1182} to the CORALIE echelle spectrograph on the comparatively smaller 1.2 m telescopes at the Euler Swiss telescope \citep{Bodaghee:2003p4448, Ecuvillon:2004p2198, Gilli:2006p2191, Mortier13}. 

There is a spectral resolution threshold, typically considered to be $\Delta \lambda / \lambda$ $\approx$ 50\,000,  which must be attained or exceeded in order to produce respectable error bars for most elements. However, there is quite a variation not only between groups who use single data sources, but also when multiple data sources are employed by one group. For example, \citet{Ramirez:2007p1819} incorporated data from four telescopes/spectrographs in order to study iron and oxygen in FGK stars; the resolution of this data varied between $\Delta \lambda / \lambda$ $\approx$ 45\,000 -- 120\,000. Similarly, \citet{Battistini15} used data from 6 different telescopes/spectrographs where their resolution has a range of $\Delta \lambda / \lambda$ $\approx$  42\,000--120\,000. In general, $\sim$35\% of groups use spectra from multiple (two or more) instruments that are then homogenized to form one data set (statistics are from the {\it Hypatia Catalog} per \citealt{Hinkel14}, see \S \ref{s.hyp}). The average spectral resolution for all groups, using single or multiple telescopes, is $R$ $\sim$ 65\,000 while the median is $R$ $\sim$ 60\,000; the maximum is $R$ $\sim$ 120\,000 and the minimum is $R$ $\sim$ 20\,000.
 
While spectral resolution is important for measuring precise stellar abundances, so too is the signal-to-noise (S/N) of the data. In the literature, there are variations from 20 $<$ S/N $<$ 2000, with the average S/N midpoint (typically defined as a range or a lower-bound) is $\sim$ 280 with a median of 200. Groups that use multiple instruments also have a range of S/N. The most extreme ranges were given by \citet{Neves:2009p1804} and \citet{Adi15} who both reported S/N $=$ 70--2000 and \citet{Bertran15} with S/N $=$ 40--2000.

There are two stellar atmosphere models that are very common throughout the literature, namely some flavor of ATLAS \citep{Castelli03} or MARCS \citep{Gustafsson08}. In terms of abundance measuring methods, there are two standard techniques: curve-of-growth (CoG) and spectral fitting. Some of the more common approaches to the CoG method, although certainly not all, utilize IRAF {\sc SPLOT} and ARES \citep{Sousa07} to measure equivalent widths (EWs).  However, there are a few groups who use more in-house procedures, such as ABON-TEST8 (via P. Magain) and DECH20 \citep{Galazutdinov92}, or their own analyses. The CoG groups predominantly use MOOG \citep{Sneden73}, WIDTH9 \citep{Kurucz1993}, or Uppsala EQWIDTH for the abundance measurements. In general, 23\% of groups use spectral fitting techniques in order to determine stellar abundances, such as \citet{RecioBlanco06, Koleva09}.

There are a number of smaller, less obvious variations that occur while calculating stellar abundances. For example, when implementing a CoG technique the spectral continuum fit may vary between the global and local scales, or when one has zoomed-in to measure a particular line. Continuum placement strongly affects the EW measurement, 
and may vary depending on the person who is measuring the line. Continuum placement presents a new level of internal uncertainty when more than one person makes these calculations for the same stellar sample. Spectral fitting techniques are limited by the flexibility of their routines to accurately determine the central wavelength of (the correct) absorption features, a process which is often done by-hand for CoG. In addition, the placement of the continuum is also an issue for spectral fitting since it strongly affects the element abundance measurement. The number of lines used to determine the abundance of an element plays an integral role in the total abundance measurement, yet varies widely between groups even for iron (see the penultimate column of Table \ref{update} as well as the discussion in \S \ref{s.elemlinebyline}.

Another aspect to consider when determining stellar abundances is that only the equivalent widths for unblended lines can be estimated by direct integration. For blended spectra, particularly when using synthesis approaches, the \textit{shape} of the line must be taken into account. Physically, line shapes are the result of intrinsic effects of local broadening and velocity fields in the stellar photosphere, rotation, and the extrinsic instrumental profile. If one of these broadening effects is much larger than the others, e.g. in the case of low resolution or very high rotational velocity, the others can be neglected. As no intrinsic line shape is available in CoG methods, a Gaussian profile is usually assumed.
In solar-type stars, vertical (radial) and horizontal (tangential) large-scale motions in convective cells on the observed stellar disk result in superimposed line profiles of varying Doppler shift, which result in a broadened but not strengthened line profile \citep[see][Fig. 3]{Asplund05}. To-date, this is the method typically employed by spectral fitting techniques.

Additionally, convective velocities are predicted to vary smoothly with stellar parameters. For example, larger velocities are for higher effective temperatures or lower surface gravity, which also varies with metallicity \citep[Fig. 17 in][]{Magic13}. In one-dimensional radiative transfer, these three-dimensional hydrodynamical effects are usually represented by radial-tangential macroscopic broadening \citep{Gray75}, where the radial and tangential velocity distributions are typically assumed to be equal. With sufficient data quality, the broadening parameter can be determined simultaneously with rotational velocity. The results in the literature are generally in line with theoretical predictions with very little scatter \citep[e.g. Figs. 4-7 of][]{Fuhrmann04}. However, the velocity fields vary with depth \citep[Fig. 17 in][]{Magic13} such that weak lines form in deep layers where velocities are high and stronger lines form over a range of depths, including where velocities are lower. This pattern is indeed found in the solar spectrum, where e.g. \citet{Gray77} determined that the macroturbulent velocity v$_{\text{mac}}$ = 3.8 and 3.1 km/s for weak and saturated lines, respectively. A single value of macroturbulent velocity does thus not generally apply to any line of any strength, but should be measured on individual lines if the spectrum quality allows it. Unfortunately, there do not appear to be any studies which test the direct effect of  v$_{\text{mac}}$ errors on measured abundances.

Finally, once the total absolute abundances have been determined in a star, they are commonly defined as a ratio with respect to the same element abundance within the Sun. This ratio-of-ratios determines whether a star is `rich' or `poor' in that element with respect to the Sun. To date, there are over 40 solar abundance scales, the most popular of which are \citet{Anders:1989p3165}, \citet{Grevesse:1998p3102}, and \citet{Asplund:2009p3251}. Note: the total number of solar abundance scales does not include those techniques that use differential analysis to renormalize to the Sun on a line-by-line basis. The majority of solar abundance scales stem from individual determinations usually made from solar spectra reflected off a solar system body during the observation run. Even direct solar observations analyzed for the express purpose of determining solar abundances can vary substantially, for example due to the use of 3D vs. 1D solar atmosphere models \citep{Asplund:2009p3251, Lodders:2009p3091, Caffau11}. Unfortunately, the choice of solar abundance scale may significantly change the overall abundance determination. For example, the range in solar carbon abundance, A(C), where A(C) = log(atom number) of carbon normalized to A(H)=12 (described more in \S \ref{s.elements}), is 0.28 dex while A(O) varies by 0.33 dex across all scales. The range of A(Si) abundances found within the Sun is 0.28 dex and even A(Fe) may change by 0.26 dex. While it is understandable to measure the solar scale under the same conditions as the other data in order to ensure internal consistency, care must be taken when comparing abundance measurements that use different scales.

\section{Appendix B: Investigation Participants}\label{a.people}
\vspace{1mm}

{\it Australian National University (ANU):} The ANU team started their analysis by automatically measuring the EWs of the spectral lines using the ARES code. The line list employed in their analysis was adopted mainly from \citet{Asplund:2009p3251} and augmented with additional unblended lines from \citet{Bensby:2005p526, Neves:2009p1804}. They performed a 1D, local thermodynamic equilibrium (LTE) abundance analysis using the 2014 version of MOOG \citep{Sneden73,Sobeck11} with the ODFNEW grid of Kurucz ATLAS9 model atmospheres \citep{Castelli03}. In order to determine the stellar parameters (effective temperature, surface gravity, microturbulence and metallicity), they force the excitation/ionization balance by minimizing the slopes in A(Fe I) versus excitation potential and reduced the EWs as well as the difference between A(Fe I) and A(Fe II), simultaneously for all of the lines. They also required the derived average [Fe/H] to be consistent with the adopted model atmospheric value. They adopted the final results by iterating the whole processes until the balance was exactly achieved. Lines whose abundances departed from the average by $> 2.5\sigma$ were clipped during the analysis. See Table \ref{anulines} for the ANU line list. They emphasize that their final results do not rely on the initial guess of the stellar parameters. Having established the stellar parameters for the sample stars, they derived chemical abundances for 9 elements: C, O, Na, Mg, Al, Si, Fe, Ni, and Ba. Note that Eu was not measured by the ANU group. 

The uncertainties in the stellar parameters were derived with the method described by \citet{Epstein10} and \citet{Bensby14}, which accounts for the co-variances between changes in the stellar parameters and the chemical abundances. The errors in the abundances were calculated following the manner of \citet{Epstein10}: the standard errors of the mean abundances are added in quadrature to the errors introduced by the uncertainties in the atmospheric parameters. This method is more robust than the simple line-to-line differences and allows for an error calculation when there is only one absorption line.

\vspace{1mm}

{\it Arizona State University (ASU):} The ASU analysis was done using the CoG method of stellar abundance determination, relying on the 2014 version of the {\sc MOOG} radiative transfer code. A previously created script was used to run through all of the steps, then edited to meet the requirements and line list for this study \citep{Pagano15, Pagano16}. The stellar atmospheres were calculated through {\sc ATLAS9} grid model atmospheres \citep{Castelli03} via the {\sc MSPAWN72} script. These model atmospheres were updated when calculating the stellar parameters for the analyses where they were free variables. 

The EWs were measured with the {\sc ARES} v2 code, such that the output was entered into {\sc MOOG} with the created model atmosphere. Iron lines that were three standard deviations or greater from the average were automatically removed from further calculations. The stellar parameters were calculated using an iterative script to achieve simultaneous excitation, ionization, and iron abundance balances. The temperature was varied until any trend with Fe I and excitation potential was removed to a Pearson correlation of less than 0.06. The microturbulent velocity was varied to remove any correlation between the Fe I line abundances and the reduced EW of each line. Lastly, the surface gravity and metallicity were varied until the overall abundances of Fe I and Fe II agreed to within 0.005 dex. The iterative process constantly returned to the previous restrictions until all balances were met. The errors on the stellar parameters were calculated after the balances were achieved, derived from varying the stellar parameters until a 1$\sigma$ correlation was achieved. 

After the stellar parameters were calculated, or the parameters were set for Runs 2 and 4, the individual elemental abundances were derived by putting the EW and line information through the {\sc abfind} function in {\sc MOOG}. See Table \ref{asulines} for the ASU line list. The final abundance for each element was the average of the abundance calculated for each absorption line. Errors for individual elements were determined by adding in quadrature the variations associated with the parameter errors and the differences between abundances for separate absorption lines. When each abundance for a given element was calculated, the abundance was immediately redone using the same parameters plus and minus their errors, varying only one parameter at a time. The differences between the calculated abundance and the two plus/minus values were averaged to get error based on the parameters. This was done for temperature and microturblent velocity, with the error for gravity set at 0.06 dex. Additional error was calculated from the standard deviation between the different element abundance lines divided by the total number of lines used and added in quadrature.

\vspace{1mm}

{\it Carnegie Institution of Washington/Department of Terrestrial Magnetism (Carnegie):} Runs 1 and 2 use the autonomous line list of 74 \ion{Fe}{1} and 13 \ion{Fe}{2} lines compiled by \cite{Carlberg12} for use in red giant stars. This list was tested on the atlas spectra of the Sun and Arcturus \citep{Hinkle00} and successfully reproduces the stellar parameters of both. A first pass at measuring the EWs of all of the lines automatically was made with ARES.  ARES is run in interactive mode to ensure proper treatment of the continuum. Lines for which the ARES fit was questionable were measured by hand in IRAF using the {\sc {\sc SPLOT}} procedure; these are lines that generally require deblending. Gaussian profiles are used for the fitting, and multiple profiles are fit for blended profiles. The high-resolution solar atlas of \cite{Hinkle00} was used to identify lines in the blend. The stellar parameters were measured by requiring excitation balance of \ion{Fe}{1} lines and ionization balance of \ion{Fe}{1} and \ion{Fe}{2} lines.  Abundances were computed with the 2014 version of MOOG using the {\sc ABFIND} driver and MARCS plane-parallel, standard composition atmosphere models \citep{Gustafsson08}. An IDL-wrapper code interacts with MOOG and searches through stellar parameter space following the algorithm described in \cite{Carlberg12}. A solution occurs when the input model Fe abundance and output \ion{Fe}{1} and \ion{Fe}{2} abundances agree better than 0.02~dex.  To avoid potential biases, this run was done blinded to the stellar parameters. No {\sc Simbad}\footnote[2]{\url{http://simbad.u-strasbg.fr}} searches to obtain spectral types or looking at the parameters provided for the later analyses was allowed beforehand. The initial guess of the stellar parameters for all stars was $T_{\rm eff}=5700$~K, $\log g=4.0$~dex, [Fe/H]=0.0, and $\xi = 1.0$~kms. Abundances of nine additional elemental species were measured using the line list in \cite{Smith01}, given in Table \ref{carnegielines}.  No Mg lines appear in this list and thus were not measured in the autonomous run. Due to the small number of lines, all EWs for these additional elements were measured by hand in IRAF. Abundances were derived using the \textit{abfind} driver in MOOG.

The uncertainties in $T_{\rm eff}$ and microturbulence are calculated as in \cite{Neuforge97} using the 1$\sigma$ uncertainties on the slopes of the line abundances versus excitation potential and reduced EWs, respectively. The $T_{\rm eff}$ uncertainty also includes the contribution from the microturbulence uncertainty. The uncertainty in A(Fe) and \lg are given by the standard deviations in the  \ion{Fe}{1} and \ion{Fe}{2} lines, respectively. These errors dominate uncertainties arising from sensitivities to the other stellar parameters. The uncertainties in the non-Fe elemental abundances are the quadrature sum of the standard deviation of the line-by-line abundance measurements (when more than line was measured) plus the error from varying the stellar parameters within their quoted uncertainties.

\vspace{1mm}

{\it Observatoire de Gen\`eve (Geneva):} The iSpec\footnote[3]{\url{http://www.blancocuaresma.com/s/}} program is a spectroscopic framework that implements routines for the determination of chemical abundances by using the spectral fitting technique \citep{Blanco14b}. Given a group of spectral regions, iSpec computes synthetic spectra on-the-fly and minimizes the differences with the observed spectra following a Levenberg--Marquardt (least square) algorithm. The spectroscopic analyses were performed using iSpec \citep{Blanco14b} with MARCS model atmospheres \citep{Gustafsson08}, Gaia-ESO Survey line list \citep{Heiter15a} and SPECTRUM \citep{Gray94} as the radiative transfer code. 

Before analyzing the spectra, they were homogenized for a robust analysis, following the same procedure detailed in \cite{Blanco15} that globally consists on the following stages:

\begin{itemize}[noitemsep]
   \item Cleaning: Wavelengths below 5050\aaa were ignored due to a issues with the wavelength calibration.
    \item Radial velocity determination and correction: They cross-matched each spectrum with a solar template built from co-adding several NARVAL observations \cite{Aurieere03} and calibrated with HARPS solar spectrum \cite{Mayor03}. The derived velocity corrections were smaller than 2.5 km/s.
    \item Resampling: Spectra was modified to have a homogeneous sampling with a wavelength step of 0.001 nm using a Bessel interpolator.
    \item Normalization: A re-normalization was applied to correct observed global trends. Continuum regions were automatically found by applying median/maximum filters and $\sim$150 splines were fitted.
    \item Atmospheric parameter determination: They derived effective temperature, surface gravity, metallicity (i.e. [M/H]), micro and macroturbulence using a selection of absorption lines and the H-$\alpha$ and magnesium triplet wings. Errors were determined from the covariance matrix reported by the least square algorithm, which is strongly influenced by the spectral flux errors (in our case they considered all spectra to have an average signal-to-noise ratio of 200).
    \item AP corrections: The atmospheric parameter determination was assessed with the Gaia-ESO FGK benchmark stars \citep{Jofre14, Blanco14a, Heiter15a, Jofre15} and it was determined that a surface gravity correction of 0.15 dex is required, which was applied to the atmospheric parameter derived in the previous step.
    \item Individual abundances: A line-by-line abundance determination was performed where lines with bad solutions were discarded, see Table \ref{genevalines}. The final abundance per element is calculated from a weighted average, while the errors correspond to the weighted dispersion.
\end{itemize}

\vspace{1mm}

{\it Instituto de Astrof\'isica e Ci\^encias do Espa\c{c}o, Universidade do Porto (Porto):}  We derived the stellar atmospheric parameters, metallicity and chemical abundances of individual elements using 
our standard spectroscopic analysis technique, as described in \citet[e.g.][]{delgado-mena_2010_aa, Adibekyan12, Sousa14}. To summarize, the EWs of the spectral lines were first automatically measured using the {\sc ARES} v1 \footnote[4]{The last version of {\sc ARES} code (ARES v2) can be downloaded at \url{http://www.astro.up.pt/$\sim$sousasag/ares} - \citep{Sousa15}.} code for a line list of Fe lines, both atomic and ionized. The spectroscopic parameters are derived by imposing both excitation and ionization balance assuming LTE where the iron abundance is used as a proxy for the metallicity. In this process, the 2014 MOOG radiative transfer code was used together with the grid of {\sc ATLAS9} plane-parallel model of atmospheres.

To derive the abundance of Na, Mg, Al, Si, and Ni, the EWs of the spectral lines were measured using the ARES v2 \citep{Sousa15}. Then a standard CoG approach was employed to derive the abundances of iron (as a
proxy for the metallicity) and other elements assuming LTE. The initial line-list of these elements were taken from \citet[]{Adibekyan12}, but several lines (two Si lines at $\lambda$5701.11\aaa, $\lambda$6244.48\aaa, and two Ni lines at $\lambda$5081.11\aaa, $\lambda$6767.78\aaa) were excluded because of large [X/Fe] star-to-star scatter at solar metallicities \citep{Adi15b}. The final abundance of elements when several spectral lines were available were calculated as a weighted mean of all the abundances, where as a weight the distance from the median abundance was considered. As demonstrated in \citet{Adi15b}, this method can be effectively used without removing suspected outlier lines.

The EWs for C, O, Ba, and Eu were measured manually using the task {\sc {\sc SPLOT}} in IRAF. The abundances were derived in the same way as for the rest of elements. For carbon, atomic optical lines from \citet{delgado-mena_2010_aa} was used together with the line at $\lambda$6587.62\aaa, and atomic data from VALD3\footnote[5]{\url{http://vald.astro.univie.ac.at/$\sim$vald3/php/vald.php}} database. For oxygen, the lines and atomic data from \citet{Bertran15} were used while for Ba data from \citet{prochaska_2000_aa} was employed. For Eu, the line at $\lambda$6587.15\aaa was analyzed with atomic data from VALD3. See Table \ref{portolines} for the Porto line list for non-Fe elements. When considering hyper-fine splitting for both Ba and Eu, the same lines as the Geneva and Uppsala groups were used. The 'abfind' driver of MOOG for normal abundances and 'blends' driver for the hyper-fine splitting treatment. The final abundance of these elements is calculated as the mean of all the individual abundances (the maximum number of lines per element is three) and the error is determined by the standard deviation of the abundances given by different lines.

\vspace{1mm}

{\it Uppsala University (Uppsala):} The spectroscopic analyses were performed using the {\sc SME} package \citep{Valenti96,Valenti:2005p1491}. The SME pipeline performs on-the-fly spectrum synthesis, explicitly taking into account blends in both lines and (pseudo-)continuum. Stellar parameters are optimized simultaneously using a Levenberg-Marquardt $\chi^2$ minimization technique applied to pixels defined in continuum and line masks. The optimization simply continues until $\chi^2$ is no longer improving, and the gradient search is based on double-sided partial derivatives. The Gaia-ESO Survey line list \citep{Heiter15a} was adopted. The line synthesis takes into account radiative, Stark, and van der Waals broadening -- the latter using either BPO self-broadening \citep{Cowley02}, Kurucz's calculations \citep{Kurucz93}, or the classical Uns{\"o}ld approximation \citep{Unsold55}. Hyperfine structure (HFS) lines were adopted when available. The latest MARCS model atmospheres \citep{Gustafsson08} were used, which are interpolated by {\sc SME} to the stellar parameters of the requested model. Instrumental broadening is taken into account by convolving the synthetic spectrum with a Gaussian kernel, at an assumed resolution of $R$ = 65\,000. Additionally, a (fixed) rotational broadening and radial-tangential macroturbulence was applied.
The list of lines to analyze was taken from the LUMBA Gaia-ESO survey analysis, and every line was inspected for every star. In regions blueward of 5050\aaa, an essentially linear wavelength drift reaching 50 km/s was identified and corrected. Unfortunately, the drift varies slightly across the spectral orders, such that overlapping regions suffer line doubling. Thus, most lines blueward of 5050\aaa were rejected, where elements were only retained when a few reliable lines were available.
Line masks are determined automatically, discarding pixels where more than 1\% of absorption is attributed to blends. Continuum normalization likewise automatically selects the pixels with least amount of line absorption.

We determine stellar parameters using the LUMBA Gaia-ESO survey stellar parameter pipeline. Normally, this pipeline determines the stellar parameters of dwarf stars using Balmer lines for \teff, and neutral and singly ionized iron lines of varying strength for [Fe/H] and \lg. The wings of the Balmer lines turned out to be unreliable in these spectra, and so the pipeline was executed in its iron line mode, normally used only for giant stars. The starting guess is \teff = 5700 K, \lg = 4.5, [Fe/H] = 0.0 and $\xi$ = 1.0 km/s, while $v\sin(i)$ is kept fixed at 1 \kms, and the macroturbulence is fitted to the spectrum. 

Stellar parameters are determined from a global optimal match of the spectrum taking into account not only line strengths but also their shapes. This means that trends in iron abundance with excitation energy, line strength, or ionization stage are not explicitly balanced. However, additional spectral information carried in the line profiles is taken into account, e.g., the shape of the wings of strong or saturated lines. 

Parameter uncertainties are based on a newly implemented technique based on the fit residuals, partial derivatives, and data uncertainties. A cumulative distribution was constructed using all data pixels for a given free parameter {\it p}. This distribution describes the fraction of spectral pixels that require a change of $\delta p$ or less to achieve a ``perfect" fit. The change is estimated from the residuals and the partial derivative of the synthetic spectrum. Such distributions typically have very wide wings due to pixels insensitive to the selected parameters or pixels impossible to fit due to erroneous observations or atomic/molecular data, but the central part with the highest gradient is not too far from a Gaussian and can be used to estimate realistic uncertainties of the free parameters. While this method still ignores the cross-talk between different parameters it is a significant improvement over the uncertainties based on the covariance matrix (also evaluated by {\sc SME}). The uncertainties derived from the diagonal of the covariance matrix are effectively a transformation of the observational errors onto the model parameter space using partial derivatives averaged over all data pixels. For high S/N observations the contribution of observational errors is negligible in comparison with model limitations and the resulting uncertainty estimates are unrealistically small. The mathematical background of the new approach can be found in the upcoming paper by Piskunov et al. (2016, in preparation) while practical tests and comparisons with other methods are presented in \citet{Ryabchikova15}. Here, the uncertainties in \teff, \lg and [Fe/H] are in the range 100--130 K, 0.2--0.3 dex and 0.11--0.15 dex, respectively. 

Abundances are determined using the LUMBA Gaia-ESO pipeline, line by line. Only lines that did not converge were rejected. See Table \ref{uppsalalines} for the Uppsala line list. To lessen the influence of extreme outliers, unweighted medians as averages were adopted, and standard deviations were estimated using the median absolute deviation error statistic. When only a single line has been analyzed, the estimated uncertainty in that measurement is used instead.

\section{Appendix C: Standardized Parameters and Line Lists}\label{a.standard}

During Runs 2 and 4, standardized stellar parameters, namely effective temperature, surface gravity, and microturbulence, were employed by all groups. These are listed in Table \ref{standardparams}. We chose to implement a standard set of stellar parameters bearing in mind techniques in astronomy which are able to determine high precision measurements. For example, asteroseismology \citep[e.g.][]{Huber12} can measure highly accurate stellar surface gravity which may be applied on a larger scale. While stellar properties must be calculated to measure the element abundances, it follows that stellar abundance techniques should be able to take advantage of high precision parameters offered by other methods while still remaining internally consistent. To this end, the standard parameters were determined using a combination (average) of appropriate literature sources that measured the respective stars (see \S \ref{s.litcomp} and Appendix \ref{a.litcomp}). 

The standardized line list is given in Table \ref{standardlines}, used in Runs 3 and 4, and was compiled directly from the respective sources. We focused on predominantly strong lines, although weaker and blended lines were included in order to broaden the scope of our analysis. In addition, the investigated lines span a wide range of ionization potentials, lower excitation energies and line strengths, as well as sampling both the minority and majority state of neutral or ionized species. While the line list included additional elements, we concentrated on C, O, Na, Mg, Al, Si, Fe, Ni, Ba, and Eu, which were consistently measured by most of the groups and included elements with varied nucleosynthetic origins. Namely, C, O, Mg, and Si are all $\alpha$-elements with varied origins; Na and Al are odd-$Z$ elements; Fe and Ni are iron-peak elements; and Ba and Eu are both neutron-capture elements via the s-process (primarily) and r-process, respectively. We go into a more detailed analysis of the individual lines in \S \ref{s.elemlinebyline}. 

Due to the nature of the spectral fitting techniques used by the Geneva and Uppsala groups, namely iSpec and {\sc SME} respectively, they required a complete line list including background features blending into targeted lines as well as the continuum. They isolated the lines that were used by the other groups in Table \ref{standardlines} and added the appropriate broadening parameters per \citet{Heiter15b}. In other words, the standardized $\log (gf)$ values were maintained across all groups to be as consistent as possible. In addition, Geneva and Uppsala added lines that were part of the Gaia-ESO line list \citep{Heiter15b}, including HFS lines for both Ba and Eu, while molecules were ignored. High ionization stages were also removed from the Gaia-ESO line list, retaining only neutrals and singly ionized species. The full line list used by the spectral fitting groups can be found in Table \ref{sfstandardlines}. 

The individual line lists used by the respective groups during Runs 1 and 2 are given in Tables \ref{anulines}-\ref{uppsalalines}, in alphabetical order by group name. The groups who used the CoG method, namely ANU, ASU, Carnegie, and Porto, all had the option to include the Ba and Eu HFS lines given in Table \ref{sfstandardlines}, however only Porto opted to use them. We will discuss the comparison of groups who used HFS lines for Ba and Eu in \S \ref{s.elements}.

\section{Appendix D: Investigation Results}\label{a.results}
In this appendix, we go into more detail regarding the trends that were found within the data output of the Investigation. While the key results were summarized in the corresponding sections of the main paper, namely those sections with the same name, we wanted to include a more thorough discussion for those interested.

\subsection{Appendix D.1: Stellar Parameters}\label{a.parameters}

This section continues discussion from \S \ref{s.parameters} with respect to Figure \ref{params}. 
To begin, we consider how parameter balance is achieved, specifically with respect to the CoG method. Stellar effective temperature is determined by adjusting the temperature solution until the correlation between the abundances and excitation potential are identically zero. To determine temperature errors, the effective temperature of the solution was varied above and below the optimal solution until the square of the sample correlation coefficient, $r^2$, was equal to the 1$\sigma$ variance of the sample. A typical value of $|r|$ that agreed with the 1$\sigma$ variance was around 0.05-0.06, as mentioned in the ASU section of Appendix \ref{a.people}. Therefore, there is no additional error term from stopping the regression before r = 0, since it did not stop at an arbitrary limit.

The top-left figure in Figure \ref{params} shows \lg versus \teff for Run 1. While the groups' determinations do not completely overlap to within average-median error ($\pm$52 K and $\pm$0.08, respectively), they do produce results for each star that follow a consistent internal stellar trend such that as \teff is increased for each star, \lg also increases in a monotonic fashion. In general and within the average-median error for each star, the ANU and Uppsala groups calculated the lowest \teff and \lg values for the stellar sample. While the measurements for both parameters in HD~121504 by Uppsala appear rather low, both are within individual error of other data points for the sample. Although, it should be noted that the Uppsala error estimates are at least double the average-median errors for \teff, \lg, and $\xi$, which will be discussed more in \S \ref{s.error}. Both Carnegie and the ASU teams determined consistently higher values for \teff and \lg for all four stars during Run 1. Of note, the ANU and Porto groups calculated nearly identical \lg values, to within individual error, for all stars regardless of temperature or spectral type. All methods measured HD~10700 to have the lowest \teff, while the other three stars do not show any general trends that aren't blurred by the error determinations (see Table \ref{stellarparams1}).

Compared to Run 1, the \lg and \teff parameters in Figure \ref{params} (top-right) were both consistently increased when using a standard line list (Run 3), where the errors are $\pm$62 K and $\pm$0.09, respectively. The exception to this trend was seen in the Geneva group, which saw a decreases in both parameters for HD~361, HD~121504, and HD~202206; the \lg and \teff determinations for HD~10700 were the same between Runs 1 and 3 to within individual error. Despite the overall increase in the parameter values, the general internal stellar trend between the groups for each star was maintained, as compared to Run 1. We find that only the Carnegie group maintains consistent \lg values between stars, similar to Run 1. Geneva determines significantly lower \teff and \lg values for all stars, which may be reconciled with the other groups via their large error bars, which we discuss more thoroughly in \S \ref{s.error}. On the other hand, if we were to temporarily ignore the `outlier' stellar parameters from Geneva in both \teff and \lg for all but HD~10700, since the results from ASU were similar to within error for the metal-poorest star, we would find a significant improvement in the range for both parameters in all three stars. Namely, the updated ranges for HD~361, HD~121504, and HD~202206 (without the respective Geneva outliers) in \teff would be 88 K, 134 K, and 93 K while \lg would be 0.18, 0.22, and 0.19, which are smaller than the ranges in Run 1. 

In the middle-left panel of Figure \ref{params} is a plot showing $\xi$ with respect to \teff for Run 1, where the average-median errors are $\pm$ 0.10 km/s and $\pm$ 52 K, respectively. Similar to the graphs in the top panel of the figure, HD~361, HD~121504, and HD~202206 are clustered together, while HD~10700 is at the lower end of both parameters. The Geneva group tends to have consistently higher $\xi$ values for all stars while the ANU group has consistently lower, although all are within the error of each other. In addition, beyond a certain temperature, the ANU, Carnegie, Geneva, and Uppsala measurements have $\xi$ values for HD~361, HD~121504, and HD~202206 that are all similar per group to within individual error. 

The $\xi$ versus \teff plot in Figure \ref{params} (middle-right) for Run 3 has average-median errors of $\pm$ 0.11 km/s and $\pm$ 62 K, respectively. The dispersion between the groups for the ``clustered" HD~361, HD~121504, and HD~202206 measurements is relatively similar to Run 1, per Table \ref{stellarparams3}. Again, Geneva has consistently lower \teff as compared to the other groups, but their $\xi$ values are all the same to within error, which we discuss in \S \ref{s.error}.  Also, ANU, ASU, Carnegie, and Uppsala all measure similar $\xi$ values for the three higher-metallicity stars, showing that their methods do not have a sensitive relationship between $\xi$ and \teff. 

Given that we have already discussed the trends between stars and groups for both $\xi$ and \lg, we will now examine the relationship between these two parameters with respect to Run 1 (Figure \ref{params}, bottom-left) and Run 3 (bottom-right). Unlike any of the other plots in Figure \ref{params}, there is less segregation between stars for $\xi$ with respect to \lg, namely that HD~10700 is no longer occupying an entirely separate parameter space. Instead, the HD~10700 measurements are relatively lower in $\xi$, but similar with \lg. A noticeable variation between Runs 1 and 3 is the shift in \lg while $\xi$ remains roughly the same. The outlying measurements follow a similar pattern, such that they are similar to within error for Run 1 but vary dramatically for Run 3.

\subsection{Appendix D.2: C, O, Mg, \& Si}\label{a.alpha}

Here we go into more detail from \S \ref{s.alpha}. Figure \ref{elems1} (top-left) shows the absolute carbon abundances, A(C), in dex for our stellar sample along the y-axis with respect to all four analyses. 
In general, we find that A(C) measurements were relatively consistent between groups to within representative error, with an average range of 0.15 dex. There was no clear indication that standardizing the stellar parameters or line list consistently improved the A(C) measurements within all stars. However, standardization of some kind did generally improve the group's results, since all stars but HD~202206 had optimized results in Runs 2--4 (see Table \ref{metric}).
The measurements for HD~202206 have a relatively consistent range across all four analyses, varying from 0.11--0.16 dex, per Tables \ref{run1abs}--\ref{run4abs}. Yet the metric values in Table \ref{metric} show that the autonomous Run 1 yielded more uniform results. HD~121504 also had a small variation in abundance ranges between analyses, varying from 0.12--0.18 dex, except in this case Run 3 had the highest metric. 
For HD~361, the combination of standardized parameters and line list produced the best results in Run 4.
Finally, HD~10700 had bimodal measurement groupings for Runs 1 and 2.
Despite this, Run 2 exhibited a slightly smaller range (0.13 dex) and larger metric than the other analyses.

Oxygen was not easily accessible in these spectra, since the wavelength cutoff was after the near-UV OH lines and before the O I triplet at 7700~\aaa. While the forbidden [O I] lines at 6300.3\aaa and 6360\aaa were considered for the standardized line list, the blending from the Ni I line and CN lines, respectively, would have created significant issues \citep[e.g][and references therein]{Ecuvillon06, Caffau13, Nissen14}. Rather than test how each groups' method responded to heavily blended lines, we opted to use the 6156.8\aaa line. This line does not suffer from blending and is stronger than the 6158.2\aaa line, which was thoroughly discussed in \citet{Bertran15}. We found that most groups chose not to use the [O I] lines when implementing their own line lists (see Tables \ref{anulines}-\ref{uppsalalines}). The lack of discussion in the literature regarding the 6156.8\aaa oxygen line also made its use more interesting, especially when considering a metal-poor star such as HD~10700.

Figure \ref{elems1} (top-right) shows the A(O) values for all groups, analyses, and stars. Measuring the oxygen values with the given spectra was difficult. In fact, not all groups were able to measure A(O) in all of the stars (Tables \ref{run1abs}--\ref{run4abs}), especially for the two more metal-poor stars, HD~361 and HD~10700. Despite the large representative error, $\pm$0.13 dex which was the largest of the 10 elements, the A(O) measurements have a number of outliers. Overall though, standardizing stellar parameters gave the best results for A(O), where Run 2 produced the best results in three stars (HD~10700, HD~121504, and HD~202206). Run 4 was best for HD~361 which benefited from a lack of outliers. Interestingly, the groups were most inhomogeneous when measuring the metal-rich star, HD~202206.

With regard to the ANU oxygen measurements for HD~121504 and HD~202206, they believe the cause for their deviation from the other groups has to do with their inability to differentially adjust the oxygen determinations to match measured solar values (see Appendix \ref{a.people}). Since the original data set from which the four target stars were selected did not include a solar spectrum, ANU could not adjust to solar values when using the standard line list, which affected their oxygen measurements very strongly. This deviation serves as an excellent consideration when implementing standardized line lists in the future.

Figure \ref{elems1} (middle-left) shows A(Mg) for all analyses with a representative error of $\pm$ 0.06 dex. No clear pattern emerged for Mg. For example, the standard stellar parameters gives the best results for two stars (HD~361 and HD~202206), the standardized line list for one (HD~10700), and the autonomous analysis for one (HD~121504). 
Given the prevalence of outliers, is hard to ignore them and gauge the range, since that means that a significant fraction of the data is ignored. Fortunately, the metric weighs the outliers less and instead emphasizes the total grouping of the remaining data. The case of two outliers puts more onus on the total spread between groups' measurements and therefore isn't as strongly rated on the metric scale. 
For HD~121504 Run 1 yielded the most consistent results between groups. In this case, ignoring Geneva, the four groups have a range of 0.04 dex. Both HD~361 and HD~202206 had the highest metric value per Table \ref{metric} during Run 2 while HD~10700 was best during Run 3.

Figure \ref{elems1} (middle-right) shows A(Si) abundances. Of all the elements shown in Figures \ref{elems1} and \ref{elems2}, silicon has the most consistent measurements, overall lowest ranges, and similar metric values for all stars across all analyses. The ranges had an average of 0.16 dex, where the maximum was 0.29 dex and the minimum was 0.07 dex. The representative error is $\pm$ 0.06 dex, the lowest among the elements (along with magnesium). This shows that Si is overall a robustly useful element. The similarity of A(Si) abundances between the groups is not surprising given the number of silicon lines typically used for measurements and overall agreement currently seen between literature sources \citep{Hinkel14}. For the Investigation, we found that the standardized stellar parameters produced the most uniform measurements and the smallest ranges between stars: 0.08--0.11 dex. Notably, the A(Si) measurements were not improved by the standardized line list in Runs 3 and 4. The ranges and metrics were considerably worse for both of these analyses, for example, the average ranges for these two runs was 0.23 dex. .

\subsection{Appendix D.3: Na \& Al}\label{a.oddz}
We continue the discussion from \S \ref{s.oddz}. Figure \ref{elems1} (bottom-left) shows A(Na) for our stellar sample across the four analyses. Similar to the A(Mg) abundances, we find that there was no method that reliably refined the measurement homogeneity for all stars. Per Table \ref{metric}, HD~361 has the best results in Run 1, HD~10700 in Run 3, HD~121504 in Run 4, and HD~202206 in Run 2. Both the plot and Table \ref{run4abs} reveal that there is an outlier for both HD~121504 and HD~202206 during Run 4, namely Uppsala. However, despite the outlier, HD~121504 had the highest metric in Run 4 due to the very tight clustering of the other five groups. HD~361 has an outlier during Run 3, although the large spread in group measurements dominated the metric calculation in this case. The representative error for A(Na) is $\pm$ 0.07 dex. 

The A(Al) results are shown in Figure \ref{elems1}, bottom-right, where the representative error is $\pm$ 0.09 dex. In the majority of cases, a standardized line list -- even one employing a weaker line -- is beneficial for determining similar abundance measurements with different techniques. HD~10700 and HD~202206 are the most homogeneous in Run 3 while HD~361 and HD~121504 are better in Run 4, as is shown in Table \ref{metric}. 

Both Runs 1 and 2 resulted in substantially different A(Al) determinations between groups for all stars. In many cases, Carnegie was a lower outlier, for example HD~361 and HD~121504 in Run 1; HD~361, HD~121504, and HD~202206 in Run 2. However, Geneva was oftentimes an upper outlier, namely for HD~10700 and HD~121504 in Run 1; HD~361 and HD~121504 in Run 2. The other four groups were typically clustered in the middle having measured more similar A(Al) abundances. While the removal of these outliers drastically reduced the ranges, it is at the loss of one third of data. 
Beyond a certain extremity, the metric is no longer affected by the presence of an outlier.  However two outliers dramatically reduces the quality and robustness of the metric.

\subsection{Appendix D.4: Fe \& Ni}\label{a.ironpeak}
Now we go into more detail from \S \ref{s.ironpeak}. Iron is the most commonly measured elemental abundance since there are hundreds of iron lines in the optical wavelength range. A large number of iron lines were employed by all of the groups in this study (see column 6 in Table \ref{models}), such that no group measured less than 60 Fe lines. We note that 70 Fe lines were used in the standard line list, see Table \ref{standardlines}. Iron is used as a proxy to indicate the overall metallicity within a star (to the extent that people will often say `metallicity' when in fact they are discussing iron-content), which affects stellar lifetime, color, structure, and a multitude of other properties. Nickel is also an iron-peak element and shares a number of characteristics with iron, for example, similar atomic structures and ionization potentials as well as a large number of optical lines. In Figure \ref{elems2}, top-left and -right, we show the A(Fe) and A(Ni) abundance results, respectively.

It clear from both Figure \ref{elems2} (top-left) and Tables \ref{run1abs}--\ref{metric} that the A(Fe) measurements for all stars are the most comparable during Run 4. The ranges for Run 4, 0.11--0.18 dex, are significantly smaller compared to the other runs, which vary from 0.14--0.34 dex with an average of 0.25 dex. 
This clear trend indicates that both standard stellar parameters and line list are beneficial for determining similar A(Fe) abundance measurements between methods. There are only two outliers for the A(Fe) determinations, occurring during Run 3, which indicates that the standardized line list may have polarizing effects on the participants' methods. Note, the representative error bar is $\pm$ 0.09 dex.

Despite similar nucleosynthetic origins and atomic structure between iron and nickel, the A(Ni) abundance measurements in Figure \ref{elems2} top-right show better agreement across all analyses as compared to the A(Fe) determinations. The average A(Ni) range from all four analyses is 0.16 dex, while the average A(Fe) range was 0.21 dex. The groups have the strongest overlap and highest metric for the A(Ni) measurements during Run 2, where the ranges vary from 0.07--0.12 dex. It is worth noting that, while Ni lines are the most predominant in our wavelength band compared to any other non-Fe element, the number of lines are only a fraction of the total Fe lines. Namely, ANU measured 48 Ni lines, ASU had 20 Ni lines, Carnegie used 15 Ni lines, Geneva had 32 Ni Lines, Porto measured 43 Ni lines, and Uppsala had 14 Ni lines. A total of 15 Ni lines were used in the Standard Line List, , as compared to 70 Fe I and 14 Fe II lines, where only four had an EW below 30 m\aaa. 

\subsection{Appendix D.5: Ba \& Eu}\label{a.neutron}

We continue the analysis from \S \ref{s.neutron}. The A(Ba II) metric values in Table \ref{metric} for the different analyses vary depending on the star: HD~121504 is the most uniform during Run 1, HD~361 peaks during Run 4, and both HD~361 and HD~202206 are the most consistent during Run 2. The A(Ba II) absolute abundance measurements in Fig. \ref{elems2} (bottom-left) have some of the largest group spreads for all of the stars across the four analyses (see Tables \ref{run1abs}--\ref{run4abs}). Namely, the average range is 0.33 dex, where the maximum is 0.64 dex and the minimum is 0.16 dex. However, this may be biased by the outlier for HD~10700 in Run 1 as well as the outliers in HD~361 and HD~121504 in Run 3. In addition, we note the similar stellar abundance clustering in Runs 1 and 3, which shows the average A(Ba II) measurement linearly increasing from HD~361 to HD~121504 and then to HD~202206. The pattern changes when compared to Runs 2 and 4, which shows the abundances for HD~361 again at a lower value, then HD~121504 is maximum, finally HD~202206 falls between the other two stars. The conclusion is that the implementation of the standardized stellar parameters has a marked effect on the overall abundances for this element.  

As discussed in Appendix \ref{a.people}, three groups utilized the HFS lines for Ba II within the standardized line list (as given in Table \ref{sfstandardlines}). Namely, the two spectral fitting groups, Geneva and Uppsala, used HFS lines in addition to one CoG group (Porto). In general, the three groups measure higher abundances values for all three stars during Run 3, though most are within error.

There are three very strong BaII lines in our stellar spectra, see Table \ref{standardlines}. In comparison, there is only one weak Eu II line, which made the measurement of this element so difficult that ANU was not able to determine any measurements. Additionally, ASU, Carnegie, and Porto did not report values for all stars (see in Figure \ref{elems2} bottom-right). In the case with so few measurements, in addition to the relatively large $\pm$ 0.13 dex representative error, it makes determining outliers questionable. 

For A(EuII), Table \ref{metric} shows that HD~361 is most uniform during Run 1 while HD~10700 and HD~121504 were best during Run 4. Since HD~202206 was only measured by two groups, we are unable to accurately determine the best run. We discuss the errors associated for an element with a single absorption line in \S \ref{s.error}. Geneva, Porto, and Uppsala utilized the HFS lines for Eu II in a similar way as for Ba II. As is shown in Fig. \ref{elems2} (bottom-right), both Geneva and Uppsala determine smaller A(Eu II) abundances during Run 3 for HD~10700 and HD~121504, while Porto had some of the larger values (when measured). On the other hand, Run 4 did not have any discernible pattern between these three groups. It is interesting to note that only the spectral fitting groups, Geneva and Uppsala, were able to measure A(Eu II) in HD~202206 for any of the four analyses.

\subsection{Appendix D.6: Iron Line Analysis}\label{a.ironlines}

This appendix gives a detailed discussion of the iron lines found in Table \ref{felines}, which were summarized in \S \ref{s.ironlines}. Of all the iron lines used for HD~202206, only 5 lines were used by all 6 groups: 5855.08 \aaa, 5856.08 \aaa, 6027.05 \aaa, 6151.62 \aaa, and 6156.36\aaa. A total of 16 lines were commonly measured by 5 groups and 27 lines were measured by 4 groups. These numbers are only a small percentage of the total discrete lines used when adding all of the groups' lines for HD~202206. Along with the varying atomic parameters for the same lines, the inconsistent lines used between groups will be discussed in \S \ref{s.disc}.

For the 5522.45\aaa iron line, the three CoG groups that used both MOOG and ATLAS9, namely ANU, ASU, and Porto, have the same excitation potential but different oscillator strengths. Additionally, their measured EWs are within 1.3 m\aaa of each other. When the stellar parameters are standardized in Run 2, the abundances determined for this line are 7.72 dex for ANU and 7.71 dex for Porto. While the ASU group measured 7.84 dex, their oscillator strength is smaller compared to ANU and Porto yet consistent with the Carnegie group. The Carnegie group also used MOOG but a different model atmosphere (MARCS). Carnegie's EW measurement was different by 1.3 m\aaa from ASU but the abundance is also 7.84 dex. Interestingly, while the Geneva group measured the same atomic parameters as ANU, their abundances (7.66 dex) were lower than ANU (7.72 dex). In Run 4, when the line list and atomic parameters are the same between all groups, the two sets of CoG groups (ANU and Porto; ASU and Carnegie) diverge from each other even though the total Fe abundances become more uniform. This may be an indication of how the differing methods and models effect the abundances on a line-by-line basis. Still, Run 4 is an improvement for almost every line, which is reflected in the overall agreement of most groups in Fig. \ref{elems2} (top-left) and discussed \S \ref{s.ironpeak}.

For line 5778.45\aaa, the ASU and Porto groups use slightly different atomic parameters as compared with the other groups, namely lower $\log (gf)$ values. While this generated expectedly larger iron abundances, the results were still within representative error. Geneva utilizes similar $\xi_l$ and $\log (gf)$ as the other groups for all runs, but their determination of A(Fe) during Run 3 was much lower. For line 5855.08\aaa, where again Porto has a $\log (gf)$ value that is 0.05 lower than the other groups, there is still good agreement among the abundance measurements. In comparison, the Geneva and Uppsala determinations are on the lower side for all analyses by $\sim$ 0.1 dex, despite the similar atomic parameters. Unlike the other selected lines, there is a large variation in the oscillator strengths for line 5856.08\aaa. The abundance results are diverse during Run 1, but become more copacetic through the implementation of standardized parameters and the line list. As expected, the results of all groups scale inversely with the oscillator strength, where Geneva and Uppsala have the largest $\log (gf)$ and the smallest abundances. On the other hand, Carnegie has the smallest $\log (gf)$ and largest abundance measurement. The range between these abundances is $\sim$ 0.3 dex for Runs 1 and 2, which is much larger than the $\pm$ 0.09 representative iron error. When the same line list is used in Run 3, the spread between groups decreases somewhat, with the most similar results occurring during Run 4. In this case it appears as though the adopted standardized line parameters are influencing the measured abundances more so than, for example, stellar atmospheric parameters or EWs.

Most groups use the same atomic parameters for line 6027.05\aaa, except Porto whose oscillator strength is relatively lower (4.07 eV) than the other groups (4.08 eV). However, this does not appear to have an impact on the abundances for Runs 1 and 2. Despite having similar atomic parameters, the Uppsala abundance determinations are notably larger than the other groups (+0.07 dex) for Runs 2--4, yet within representative error. Their overall iron abundances in Figure \ref{elems2} (top-left) is very similar to the other groups, so this pattern may not be expressed within all of their lines. Line 6151.62\aaa has the best agreement between groups for Run 2, with a range of only 0.04 dex. The atomic parameters are all similar and the EWs vary by a few m\aaa. This is one of the few lines where Run 2 has better agreement than Run 4. Line 6165.36\aaa shows varying oscillator strengths among the methods. However, the methods converge in Runs 3 and 4 for this line where the ranges are 0.08 dex and 0.06 dex, respectively.

\subsection{Appendix D.7: Other Elements Line Analysis}\label{a.otherlines}
This appendix offers a more a detailed discussion of the line-by-line analysis summarized in \S \ref{s.otherlines} for carbon, magnesium, sodium, aluminum, and barium and examines the trends seen in Table \ref{otherlines}.

{\it Carbon: }
For carbon, three lines were readily used between all groups, shown in Table \ref{otherlines}. Additionally, Carnegie had two extra lines in the 7100\aaa range. While all of the $\xi_l$ and $\log (gf)$ values were nearly identical, Porto tended to have larger abundance values ($\gtrsim$ 0.1 dex) during Runs 1 and 2 which may be due to their larger EW measurements. Both ASU and Uppsala had comparable values for 5052.15\aaa during Runs 1 and 2, but that similarity was not seen for either Runs 3 and 4. The same was true for line 5380.32\aaa. The 5052.15\aaa line showed the best agreement between groups during Run 1, while 6587.61\aaa had a slight improvement during Runs 3 and 4. In comparison, the abundance determinations by each group for line 5380.32\aaa were relatively similar for all of the analyses. The difference between Carnegie's 7111.47\aaa and 7113.18\aaa abundance determinations is noticeable, varying by $\sim$0.3 dex which is much higher than the carbon $\pm$0.07 dex representative error.

\vspace{1mm}

{\it Magnesium: }
For line 5711.09\aaa, both ASU and Uppsala determined abundances that were on the higher end of the scale by $\sim$0.1 dex. Interestingly, Uppsala was the only group who was able to measure this line in Runs 3 and 4. The contrary is true for the 6318.72\aaa magnesium line, where all groups produced results for Runs 3 and 4 and they agreed well, with the exception of Geneva. Despite having the same atomic parameters, the abundances by ASU and ANU were rather different during Runs 1 and 2.  Namely, ANU found significantly lower magnesium values for Runs 1 and 2: 7.68\aaa and 7.67\aaa, respectively, as compared to ASU: 7.82\aaa and 7.76\aaa, respectively.  This difference can be seen on the magnesium plot in Figure \ref{elems1} (middle-left), where ASU tends to yield higher A(Mg) results while ANU determines lower ones. We discuss the errors associated for an element with a single absorption line in \S \ref{s.error}.

\vspace{1mm}

{\it Sodium: }
For the two commonly measured sodium lines, namely 6154.23\aaa and 6160.75\aaa, the atomic parameters between groups are very similar. However, there are some variations in the oscillator strength, which varied from -1.622-- -1.5 and -1.363-- -1.2, respectively. ASU measured a larger EW and used a higher $\log (gf)$ for 6154.23\aaa, which yielded larger abundances in Runs 1 and 2. Porto continued to determine higher abundances for both Runs 1--4, which can be seen in Fig. \ref{elems1} (middle-left), although the spread decreased during Run 4. The results for the 6160.75\aaa are interesting in that the oscillator strengths are measured in decreasing order: Carnegie/Geneva, ANU, then Porto, while the EWs are Porto, Carnegie, then ANU. The abundance determinations for Runs 1--4 are Porto, Carnegie, and then ANU, in descending order. This implies that the EW measurement is more significant than the oscillator strength, although perhaps only for this line. Geneva produced abundances at the lower end of the range for all runs except Run 4. 

\vspace{1mm}

{\it Aluminum: }
For aluminum, three total lines were measured by the groups, where 5557.07\aaa was only measured by ANU and Uppsala and was not part of the set line list. The measurements of the 6696.02\aaa line used the same excitation potential but with oscillator strengths that varied by 0.3. Carnegie had both the highest $\log (gf)$ value and EW, which resulted in the lowest abundance values for Runs 1 and 2, as expected. ASU's low oscillator strength produced relatively higher abundance measurements for Runs 1 and 2, while the Runs 3 and 4 determinations were about average. The same was true for ASU, Porto, and Uppsala when measuring the 6698.67\aaa line: relatively low $\log (gf)$ values produced large abundances for Runs 1 and 2, but more average determinations for Runs 3 and 4. Geneva measured the highest A(Al) abundances for Runs 3 and 4.

It may be the case that the chosen $\log (gf)$ values are strongly affecting the overall Al abundance results. In Runs 1 and 2, the abundances for the two Al lines are consistent ( $< 0.1$ dex) with each other for all groups. However, for Runs 3 and 4, all of the groups get similar {\it average} abundances for Al (and quite similar EWs) but much higher errors because the two lines give discrepant results. For example, the bluer line gives significantly larger abundances.  

\vspace{1mm}

{\it Barium: }
As discussed in Appendix \ref{a.neutron}, singly ionized barium has three strong lines in our spectra which were measured by nearly all of our groups. In general, the atomic parameters are relatively similar between the groups for each line such that the small discrepancies do not always account for the varying results, see Table \ref{otherlines}. For example, a larger oscillator strength does not always produce a smaller abundance measurement. Therefore, we examine the EW measurements. For the 5853.69\aaa line, Porto measured the largest EW, followed by ANU and then ASU. In Run 1, ASU measures the largest abundance, along with the smallest $\log (gf)$. Porto determines the next largest abundance and then ANU. For Run 2, the descending order is Porto, ASU, and then ANU. Everything changes in Runs 3 and 4 where Uppsala and Porto measure the highest A(Ba II) content. For the 6141.73\aaa line, the ASU and Carnegie abundances are all much larger than the ANU and Uppsala determinations for Runs 1 and 2, although their EWs are not consistently smaller or larger. That pattern changes for the latter two analyses, where the measurements converge better during Run 4. The final Ba II line, 6496.91\aaa, has identical atomic parameters for all groups and a variety of EWs. Unfortunately, there does not appear to be any obvious pattern across the analyses.

\section{Appendix E: Comparison to Literature Abundances}\label{a.litcomp}

Expanding upon the discussion in \S \ref{s.litcomp} and relating to Table \ref{littable}, the abundance of Na is consistent across all analyses for HD~361, with average A(Na) = 6.06 - 6.10 dex, compared to a literature value of 6.15 dex. We find that HD~10700 is similarly consistent at 5.85 - 5.89 dex, compared to 5.88 dex in {\it Hypatia}. The HD~202206 range of 6.59 - 6.61 dex is in good agreement with the literature 6.59 dex. Spreads for HD~10700 are substantially lower than in the literature; in other stars spreads were larger for the Investigation, especially for Runs 3 and 4.

Again for Mg, the values are quite consistent across all analyses and with the literature for all three stars. 
The largest difference between the Investigation and a literature results was for HD~10700 with a range of 0.15 dex, which is less than the spread. Spreads are larger in the Investigation than the literature for HD~361 and HD~20226. For HD~10700 the spreads are comparable to {\it Hypatia}.

The abundances measured for Al are consistent with {\it Hypatia} for all stars. The exception was found in Runs 3 and 4 for HD~202206, where the values for the Investigation (6.65 dex and 6.63 dex, respectively) were lower by slightly more than the spread ({\it Hypatia} value of 6.75 dex with a range 0.12 dex). Differences were generally much smaller than this instance and were well within the spread. Spreads improved markedly with the introduction of the fixed line list in Runs 3 and 4 for all three stars, where the average spreads were 0.41, 0.43, 0.22, and 0.28 dex, respectively. HD~361 has an extremely small {\it Hypatia} spread and only three measurements in the literature. Otherwise literature spreads were somewhat smaller than the Investigation values.

Silicon was extremely consistent both across analyses and with the literature for all stars. In every case the literature values fell within the range spanned by the different analyses, and the largest difference between analyses was 0.06 dex, for HD~10700 Runs 2 and 3. The largest difference with the {\it Hypatia} A(Si) was 0.09 dex for HD10700 Run 3. Investigation ranges were always smaller than the {\it Hypatia} spreads.

Nickel behaves similarly to Fe, except that the systematically lower values in the Investigation are not seen. The Investigation spread are small and comparable to the literature for Runs 2 and 4, with averages of 0.09 and 0.12 dex, respectively. However, ranges are larger for Runs 1 and 3, which had averages of 0.22 and 0.24 dex, respectively. The {\it Hypatia} spread is smallest for HD~361 at 0.04 dex, but that can be attributed to the effect of all the literature values for HD~361 coming from groups with very similar observations and methodology.

Barium is measured by only one group for HD~361 and HD~202206 and by four groups for HD~10700. For HD~361 A(Ba II) differs from the value in \citet{GonzalezHernandez:2010p7714} by 0.015 dex and 0.025 dex in Runs 1 and 3, respectively. The abundance for A(Ba II) is much lower, 2.16 dex and 2.13 dex in Runs 2 and 4, when the stellar parameters are fixed. HD~10700 behaves similarly, save that the literature average,1.56 dex, lies roughly between the two extremes of fixed (1.50 dex) vs. free stellar parameters (1.69 dex). The spreads are large for HD~10700, with an average of 0.32 dex, and the low value reported by \citet{Takeda:2005p1589, Takeda:2007p1531} brings the {\it Hypatia} average down. The single value available for HD~202206 is 2.42 dex \citep{GonzalezHernandez:2010p7714} which falls between the higher values for the analyses that do not have fixed stellar parameters (average 2.47 dex) and the lower values for Runs 2 and 4 (average 2.30 dex).

Europium has a single value in the literature for HD~361 and HD~10700 and two for HD~202206, which differ by only 0.02 dex. The single measurement for HD~361 is lower than the Investigation values, which has an average of 0.62 dex, but falls within the average 0.18 dex spread. HD~10700 matches well. The Investigation averages, 0.50 dex, are much lower than {\it Hypatia} for HD~202206, but only two groups in the Investigation returned abundances. While it is difficult to draw conclusions from so few points, we can at least say that values in this study are broadly consistent with those in the literature.

\section{Appendix F: Supplemental Tables}\label{a.tables}
The first is the updated {\it Hypatia Catalog}, similar to the one seen in \citet{Hinkel14}, shown in Table \ref{update} including the specifics of the added data sets: such as the telescope(s) used, data resolution, S/N, wavelength range, model stellar atmosphere, EW technique (when applicable), whether spectral fitting or the CoG method was used, the solar normalization scale, the number of Fe I and Fe II lines implemented, and the total number of stars that were added into the {\it Hypatia Catalog}, respectively. We present the standard line lists used during Runs 3 and 4 for both the CoG methods (Table \ref{standardlines}) and the spectral fitting techniques (Table \ref{sfstandardlines}), with the appropriate references. The line lists used by the individual groups for Runs 1 and 2, as given to the lead author, are shown in Tables \ref{anulines}-\ref{uppsalalines}. It is our hope that providing all of this information will not only encourage others to share the details of their analysis but will help other groups/methods determine consistent results for future abundance determinations.

\renewcommand{\thetable}{F\arabic{table}}

\setcounter{table}{0}

\clearpage

\tiny    \LongTables
\begin{landscape}
\begin{deluxetable} {|p{2.0cm}p{4.0cm}p{1.0cm}p{1.0cm}p{1.7cm}p{1.3cm}p{1.3cm}p{2.0cm}p{1.8cm}p{1.0cm}p{1.0cm}|}
\tablecaption{\label{update} Hypatia 2.0 Update}
  \tablehead{
    \colhead{Catalog} &
    \colhead{Telescope} &
    \colhead{Resolution} &
    \colhead{S/N} &
    \colhead{$\lambda$ Range} &
    \colhead{Stellar} &
    \colhead{Eq.} &
    \colhead{CoG or} &
    \colhead{Solar} &
    \colhead{Num. of } &
    \colhead{Stars in } \\
    \colhead{} &
    \colhead{} &
    \colhead{($\Delta \lambda / \lambda$)} &
    \colhead{} &
    \colhead{(\aaa)} &
    \colhead{Atmo} &
    \colhead{Width} &
    \colhead{SF} &
    \colhead{Scale} &
    \colhead{FeI/II lines} &
    \colhead{Hypatia} \\
  }
\startdata
\hline
\hline
\citet{Adamow14} & Hobby-Ebberly Telescope (HET) at McDonald Observatory using High Resolution Spectrograph (HRS) & 60000 & $>$200-250 & 407-592  and 602-784  & ATLAS9 per \citet{Kurucz1993} & spectral fitting & SME & their own & 269 / 27 & 8 \\
\citet{Adibekyan12} & 3.6m telescope at ESO equipped with High Accuracy Radial Velocity Planet Searcher (HARPS) using CORALIE & 110000 & 70-2000 & 3800-6900  & ATLAS9 & ARES & MOOG & \citet{Anders:1989p3165} & 263 / 36 & 996 \\
\citet{Adi15} & La Silla Observatory using CORALIE and Ultraviolet and Visual Echelle Spectrograph (UVES) & 110000 & ~150 & 4780-6805  & ATLAS9 & ARES & MOOG & \citet{Anders:1989p3165} & 120 / 17 & 223 \\
\citet{Battistini15} & Fiber Fed Extended Range Optical Spectrograph (FEROS) on the ESO 1.5-m and 2.2-m telescopes; Soviet-Finnish Spectrograph (SOFIN) and FIbre-fed Echelle Spectrograph (FIES) on the Nordic Optical Telescope; HARPS on the ESO 3.6-m telescope; and Magellan Inamori Kyocera Echelle Spectrograph (MIKE) on the Magellan Clay telescope; UVES on the ESO Very Large Telescope (VLT) & 48000 / 80000 / 67000 / 120000 / 42000-80000 / 110000 & 150-300 & 3800-9200  / 4500-8800  / 3650-7300  / 3780-6910  / 3600-9300  / 5500-7500  & MARCS & IRAF {\sc SPLOT} & Uppsala EQWIDTH & \citet{Asplund:2009p3251} & 226 / 36 & 637 \\
\citet{Bensby14} & FEROS on the ESO 1.5-m and 2.2-m telescopes; SOFIN and FIES on the Nordic Optical Telescope; HARPS on the ESO 3.6-m telescope; and MIKE on the Magellan Clay telescope; UVES on the ESO Very Large Telescope & 48000 / 80000 / 67000 / 120000 / 42000-80000 / 110000 & 150-300 & 3800-9200  / 4500-8800  / 3650-7300  / 3780-6910  / 3600-9300  / 5500-7500  & MARCS & IRAF {\sc SPLOT} & Uppsala EQWIDTH & \citet{Asplund:2009p3251} & 226 / 36 & 689 \\
\enddata
\end{deluxetable}
\clearpage
\end{landscape}

\clearpage
\LongTables
\begin{deluxetable}{lllllll}
  \tablecaption{\label{standardlines} Standard Line List}
  \tablehead{
    \colhead{Ion} &
    \colhead{Wavelength} &
    \colhead{$\chi_l$} &
    \colhead{$\log (gf)$} &
    \colhead{$ \delta \Gamma^{5/2}_6$} &
    \colhead{$EW_{\odot}$} &
    \colhead{Reference} \\
    \colhead{} &
    \colhead{($\text{\AA}$)} &
    \colhead{(eV)} &
    \colhead{} &
    \colhead{} &
    \colhead{(m$\text{\AA}$)} &
    \colhead{} 
  }
 \startdata
CI & 5052.15 & 7.68 & -1.30 & 9.9 & 37.8 & \citet{Wiese96}\\
 & 5380.32 & 7.68 & -1.61 & 9.9 & 21.9 & \citet{Wiese96}\\
 & 6587.62 & 8.54 & -1.00 & 9.9 & 14.0 & \citet{Wiese96}\\
OI & 6156.80 & 10.74 & -0.43 & 2.5 & 4.1 & \citet{Feltzing:1998p886}\\
NaI & 6154.23 & 2.10 & -1.57 & 6.4 & 39.8 & \citet{Lambert:1968p5112}\\
 & 6160.75 & 2.10 & -1.27 & 6.4 & 58.4 & \citet{Lambert:1968p5112}\\
 MgI & 6318.71 & 5.11 & -1.97 & 9.9 & 39.8 & \citet{Lambert:1978p3182}\\
AlI & 6696.03 & 3.14 & -1.58 & 2.5 & 38.1 & \citet{Feltzing:1998p886}\\
\enddata
\end{deluxetable}

\clearpage
\LongTables
\begin{landscape}
\begin{deluxetable}{p{1.0cm}p{0.7cm}p{0.5cm}p{0.5cm}p{1.0cm}p{1.0cm}llp{1.9cm}p{1.9cm}p{1.9cm}p{1.9cm}p{2.2cm}}
  \tablecaption{\label{sfstandardlines} Spectral Fitting Standard Line List}
  \tablehead{
    \colhead{Wavelength} &
    \colhead{Elem.} &
    \colhead{Lower $\chi_l$} &
    \colhead{Upper $\chi_l$} &
    \colhead{$\log$(gf)} &
    \colhead{Rad.} &
    \colhead{Stark} &
    \colhead{Waals} &
    \colhead{Wavelength} &
    \colhead{Lande} &
    \colhead{Damp.} &
    \colhead{Stark} &
    \colhead{VDW} \\
    \colhead{(\aaa)} &
    \colhead{} &
    \colhead{(cm$^{-1}$)} &
    \colhead{(cm$^{-1}$)} &
    \colhead{} &
    \colhead{Damp.} &
    \colhead{} &
    \colhead{} &
    \colhead{Ref} &
    \colhead{Ref} &
    \colhead{Ref} &
    \colhead{Ref} &
    \colhead{Ref} 
  }
 \startdata 
5052.1443 & C & 61983 & 81768 & -1.3 & 8.69 & -4.69 & -7.31 & \citet{K10} & \citet{K10} & \citet{K10} & \citet{K10} & \citet{K10} \\ 
5082.3441 & Ni & 29503 & 49175 & -0.59 & 8.35 & -4.03 & 811.278 & \citet{K08} & \citet{K08} & \citet{K08} & \citet{K08} & \citet{BA-J,BPM} \\ 
5087.416 & Y II & 8743 & 28390 & -0.16 & 0 & 0 & 0 & \citet{Hannaford82} & \citet{Hannaford82} & \citet{Hannaford82} & \citet{Hannaford82} & \citet{K11} \\ 
5088.5377 & Ni & 31028 & 50675 & -1.04 & 8.31 & -5.58 & -7.51 & \citet{K08} & \citet{K08} & \citet{K08} & \citet{K08} & \citet{K08} \\ 
5088.9557 & Ni & 29673 & 49312 & -1.24 & 8.09 & -4.64 & 829.278 & \citet{K08} & \citet{K08} & \citet{K08} & \citet{K08} & \citet{BA-J,BPM} \\ 
5094.4107 & Ni & 30915 & 50538 & -1.07 & 8.18 & -4.51 & 788.277 & \citet{K08} & \citet{K08} & \citet{K08} & \citet{K08} & \citet{BA-J,BPM} \\ 
5115.3922 & Ni & 30923 & 50466 & -0.28 & 7.9 & -5.42 & 769.235 & \citet{K08} & \citet{K08} & \citet{K08} & \citet{K08} & \citet{BPM} \\ 
5141.7389 & Fe & 19550 & 38996 & -2.18 & 8.2 & -5.88 & 367.251 & \citet{K07} & \citet{K07} & \citet{K07} & \citet{K07} & \citet{BA-J,BPM} \\
\enddata
\end{deluxetable}
\clearpage
\end{landscape}

\clearpage
\LongTables
\begin{deluxetable}{lllllll}
  \tablecaption{\label{anulines} ANU Line List}
  \tablehead{
    \colhead{Wavelength} &
    \colhead{Species} &
    \colhead{$\chi_l$} &
    \colhead{$\log (gf)$} \\
    \colhead{($\text{\AA}$)} &
    \colhead{} &
    \colhead{(eV)} &
    \colhead{} \\
  }
 \startdata
5067.14 & 26 & 4.22 & -0.86 \\ 
5090.77 & 26 & 4.26 & -0.49 \\ 
5109.65 & 26 & 4.3 & -0.73 \\ 
5127.36 & 26 & 0.91 & -3.33 \\ 
5132.66 & 26.1 & 2.81 & -4.17 \\ 
5141.74 & 26 & 2.42 & -2.23 \\ 
5145.09 & 26 & 2.2 & -3.08 \\ 
5187.91 & 26 & 4.14 & -1.26 \\
\enddata
 \end{deluxetable}
 
 \clearpage
\LongTables
\begin{deluxetable}{lllllll}
  \tablecaption{\label{asulines} ASU Line List}
  \tablehead{
    \colhead{Wavelength} &
    \colhead{Species} &
    \colhead{$\chi_l$} &
    \colhead{$\log (gf)$} &
    \colhead{$ \delta \Gamma^{5/2}_6$} \\
    \colhead{($\text{\AA}$)} &
    \colhead{} &
    \colhead{(eV)} &
    \colhead{} &
    \colhead{} \\
  }
 \startdata
5380.34 & 6 & 7.68 & -1.61 & 9.9 \\ 
6587.62 & 6 & 8.54 & -1 & 9.9 \\ 
6156.8 & 8 & 10.7 & -0.43 & 2.5 \\ 
6154.23 & 11 & 2.1 & -1.57 & 6.4 \\ 
6318.72 & 12 & 5.11 & -1.97 & 9.9 \\ 
6696.03 & 13 & 3.14 & -1.58 & 2.5 \\ 
6698.67 & 13 & 3.14 & -1.63 & 9.9 \\ 
5793.08 & 14 & 4.93 & -2.06 & 1.9 \\
\enddata
 \end{deluxetable}

 \clearpage
\LongTables
\begin{deluxetable}{lllllll}
  \tablecaption{\label{carnegielines} Carnegie Line List}
  \tablehead{
    \colhead{Wavelength} &
    \colhead{Species} &
    \colhead{$\chi_l$} &
    \colhead{$\log (gf)$} \\
    \colhead{($\text{\AA}$)} &
    \colhead{} &
    \colhead{(eV)} &
    \colhead{} \\
  }
 \startdata
5234.625 & 26.1 & 3.221 & -2.23 \\ 
5284.098 & 26.1 & 2.891 & -3.195 \\ 
5307.361 & 26 & 1.608 & -2.987 \\ 
5322.041 & 26 & 2.279 & -2.803 \\ 
5325.559 & 26.1 & 3.221 & -3.324 \\ 
5412.784 & 26 & 4.435 & -1.89 \\ 
5414.046 & 26.1 & 3.221 & -3.645 \\ 
5425.247 & 26.1 & 3.199 & -3.39 \\
\enddata
 \end{deluxetable}

 \clearpage
\LongTables
\begin{deluxetable}{llrrrll}
  \tablecaption{\label{genevalines} Geneva Line List}
  \tablehead{
    \colhead{Wavelength} &
    \colhead{Species} &
    \colhead{Lower $\chi_l$} &
    \colhead{Upper $\chi_l$} &
    \colhead{$\log (gf)$} \\
    \colhead{($\text{\AA}$)} &
    \colhead{} &
    \colhead{(cm$^{-1}$)} &
    \colhead{(cm$^{-1}$)} &
    \colhead{} \\
  }
 \startdata
5052.1443 & 6 & 61985 & 81779 & -1.304 \\ 
6587.61 & 6 & 68857 & 84037 & -1.021 \\ 
6411.648 & 7 & 94796 & 110393 & -2.014 \\ 
5577.3386 & 8 & 15865 & 33795 & -8.241 \\ 
6158.1858 & 8 & 86634 & 102872 & -0.296 \\ 
6300.3038 & 8 & 0 & 15872 & -9.715 \\ 
6363.776 & 8 & 161 & 15875 & -10.19 \\ 
5688.205 & 11 & 16970 & 34550 & -0.404 \\
\enddata
 \end{deluxetable}

 \clearpage
\LongTables
\begin{deluxetable}{lllllll}
  \tablecaption{\label{portolines} Porto Line List}
  \tablehead{
    \colhead{Wavelength} &
    \colhead{Species}  &
    \colhead{$\chi_l$} &
    \colhead{$\log (gf)$} \\
    \colhead{($\text{\AA}$)} &
    \colhead{} &
    \colhead{(eV)} \\
  }
 \startdata
5247.06 & 26 & 0.09 & -4.941 \\
5855.08 & 26 & 4.61 & -1.531 \\
6056.01 & 26 & 4.73 & -0.489 \\
6200.32 & 26 & 2.61 & -2.397 \\
6733.15 & 26 & 4.64 & -1.429 \\
6750.16 & 26 & 2.42 & -2.615 \\
4523.40 & 26 & 3.65 & -1.871 \\
4531.62 & 26 & 3.21 & -1.801 \\
\enddata
 \end{deluxetable}

\clearpage
\LongTables
\begin{deluxetable}{lllllll}
  \tablecaption{\label{uppsalalines} Uppsala Line List}
  \tablehead{
    \colhead{Element} &
    \colhead{Wavelength}  \\
    \colhead{} &
    \colhead{($\text{\AA}$)} \\
  }
 \startdata
Al & 5557.063 \\
Al & 6696.023 \\
Al & 6698.673 \\
Ba & 5853.669 \\
Ba & 6141.715 \\
Ba & 6496.898 \\
C &  5052.145 \\
C &  5380.325 \\
\enddata
 \end{deluxetable}

\end{document}